\documentclass[preprint2]{aastex}

\slugcomment{Submitted to ApJS}

\shorttitle{Be Stars}
\shortauthors{McSwain \& Gies}

\begin{document}

\title{The Evolutionary Status of Be Stars: Results from a Photometric
Study of Southern Open Clusters}


\author{M. Virginia McSwain\altaffilmark{1,2}}
\affil{Department of Astronomy, Yale
University, P.O. Box 208101, New Haven, CT 06520-8101}
\email{mcswain@astro.yale.edu}

\author{Douglas R. Gies}
\affil{Department of Physics and Astronomy, Georgia State University,
P.O. Box 4106, Atlanta, GA 30302-4106}
\email{gies@chara.gsu.edu}

\altaffiltext{1}{Visiting Astronomer, Cerro Tololo Inter-American
Observatory.  CTIO is operated by AURA, Inc.\ under contract to the
National Science Foundation.}
\altaffiltext{2}{NSF Astronomy and Astrophysics Postdoctoral Fellow}


\begin{abstract} 

Be stars are a class of rapidly rotating B stars with circumstellar disks
that cause Balmer and other line emission.  There are three possible
reasons for the rapid rotation of Be stars: they may have been born as
rapid rotators, spun up by binary mass transfer, or spun up during the
main-sequence (MS) evolution of B stars.  To test the various formation
scenarios, we have conducted a photometric survey of 55 open clusters in
the southern sky.  Of these, five clusters are probably not physically
associated groups and our results for two other clusters are not reliable,
but we identify 52 definite Be stars and an additional 129 Be candidates
in the remaining clusters.  We use our results to examine the age and
evolutionary dependence of the Be phenomenon.  We find an overall increase
in the fraction of Be stars with age until 100 Myr, and Be stars are most
common among the brightest, most massive B-type stars above the zero-age
MS (ZAMS).  We show that a spin-up phase at the terminal-age MS (TAMS)
cannot produce the observed distribution of Be stars, but up to 73\% of
the Be stars detected may have been spun-up by binary mass transfer.  
Most of the remaining Be stars were likely rapid rotators at birth.

Previous studies have suggested that low metallicity and high cluster
density may also favor Be star formation.  Our results indicate a possible
increase in the fraction of Be stars with increasing cluster distance from
the Galactic center (in environments of decreasing metallicity).  
However, the trend is not significant and could be ruled out due to the
intrinsic scatter in our data.  We also find no relationship between the
fraction of Be stars and cluster density.

\end{abstract}



\section{Introduction}

Be stars are well known to be a class of rapidly rotating stars
\citep{slettebak1949, slettebak1966,slettebak1992}. \citet{slettebak1966}
showed that the observed equatorial rotational velocities, $v_{rot}$, of
Be stars are 70\%$-$80\% of the critical breakup velocity, $v_{crit}$,
although \citet*{townsend2004} and \citet{fremat2005} showed that observed
Be star rotational velocities may be systematically underestimated due to
the effects of gravitational darkening on spectral lines.  Therefore Be
stars may rotate much closer to their critical velocities than the
available observations suggest.  This rapid rotation may be combined with
weaker processes, such as non-radial pulsations or magnetic fields
\citep{porter2003}, to move material from the stellar surface into a disk
\citep{townsend2004}.

There are three possible reasons for the rapid rotation of Be stars: they
may have been born as rapid rotators, spun up by binary mass transfer, or
spun up during the MS evolution of B stars.  Many authors
\citep{mermilliod1982, grebel1997, zorec1997, fabregat2000} have found the
highest fractions of classical Be stars among spectral types B0$-$B2,
appearing most often between the ages $13-25$ Myr \citep{fabregat2000}.  
\citeauthor{fabregat2000} use these results to argue that the Be
phenomenon is due to an evolutionary spin-up among B-type stars towards
the end of the MS lifetime \citep{meynet2000, keller2001}.  On the other
hand, \citet{abt1979} found no dependence on the Be frequency with age,
and \citet{mermilliod1982} and \citet{slettebak1985} noted that Be stars
are found throughout the entire MS band, from the ZAMS through the TAMS,
indicating that Be stars can occur at any age.  Therefore it is also
possible that the evolution of close binary systems may lead to mass
transfer onto B stars, thereby increasing their angular momentum and
inducing disks \citep{pols1991}.  Finally, \citet{zorec1997} find constant
fractions of Be stars with luminosity class, and they argue that the Be
phenomenon is not associated with a particular evolutionary phase.  
Rather, it is more likely due to rapid rotation in B stars at the time of
their formation.  Because these various studies disagree about the
relationship between the Be frequency and evolution, further
investigations are necessary to resolve the issue.

Therefore we present here an extensive photometric survey of open clusters
to study their Be star populations.  Our detection technique is
demonstrated for the cluster NGC 3766 in \citet{mcswain2005}, hereafter
referred to as Paper 1.  In this paper we present photometric data for 55
open clusters, and we identify a total of 52 definite Be stars and 129 Be
star candidates in the 48 reliable open clusters.  We use our sample of Be
stars to conduct a statistical analysis of the Be phenomenon and explore
the reasons for their rapid rotation.


\section{Observations and Data Analysis}

\setcounter{footnote}{2}
We selected the target open clusters for this study using the WEBDA
database\footnote{The WEBDA database is maintained by J.-C. Mermilliod and
is available online at obswww.unige.ch/webda/navigation.html.}.  We chose
clusters with ages between 1-32 Myr, $V-M_V < 13$, and angular diameters
between 5--20\arcsec.  However, we also included some older clusters to
improve our statistics, and many of the ages and distance moduli from
WEBDA were later revised using other references.  We restricted our
targets to clusters in the southern sky; all have declination $<
0^{\circ}$.  To eliminate clusters with obvious associated nebulosity, we
used the Space Telescope Science Institute (STScI) Digitized Sky
Survey\footnote{The STScI Digitized Sky Survey is available online at
http://stdatu.stsci.edu/cgi-bin/dss\_form.} to view a red image of each
cluster.  While the requirement of little to no nebulosity reduced the 
difficulty of the photometry, it did eliminate many very young clusters 
from our sample.

We made photometric observations of the clusters over 14 nights between
2002 March 27 -- April 2 and 2002 June 19 -- 25 with the CTIO 0.9m
telescope and SITe 2048 CCD.  The images were binned using a CCD summing
factor of 2 pixels $\times$ 2 pixels due to the slow readout time of the
chip.  Without binning, the chip has a plate scale of
$0\farcs401$~pixel$^{-1}$, but with binning the plate scale increased by a
factor of 2.

Most clusters were observed at least twice in each of the Str\"{o}mgren
$b$ and $y$ filters as well as a narrow band H$\alpha$ filter, and the
exposure times usually varied from 5--120 s duration depending on the
brightness of the cluster.  During the 2002 March run, we also observed
five standard stars from the list of \citet{cousins1987}: HD 79039, HD
80484, HD 104664, HD 105498, and HD 128726.  We observed the standard
stars HD 105498, HD 128726, HD 156623, HD 157795, HD 167321, and HD 216743
from the lists of \citet{cousins1987} and \citet{clausen1997} during the
2002 June nights.  Unless the weather prohibited observations, each
standard star was observed in each band at a minimum of three different
airmasses.  All of the images were processed in IRAF\footnote{IRAF is
distributed by the National Optical Astronomy Observatory, which is
operated by the Association of Universities for Research in Astronomy,
Inc., under cooperative agreement with the National Science Foundation.}
using nightly bias and dome flat frames.

The photometry of each standard star and cluster was measured and
calibrated according to the the methods discussed in Paper 1.  We measured
the errors in the magnitudes by combining the estimated instrumental
errors, the standard deviation of the aperture correction, and the errors
in the transformation coefficients, and these quantities vary for each
cluster due to nightly variations in atmospheric conditions and the
exposure time used.  From our images, we also performed astrometry of the
clusters according to the technique described in Paper 1, and the errors
are typically less than $0\farcs10$ in both $\alpha \cos \delta$ and
$\delta$.  However, several clusters did not fit into a single field of
view, and they were observed in three or four fields.  These clusters are
Collinder 272, Hogg 16, IC 2395, NGC 2343, NGC 6425, NGC 6531, Stock 13,
and Trumpler 18.  The astrometry errors of these clusters are larger, but
usually less than $0\farcs50$ in both $\alpha \cos \delta$ and $\delta$.

Our photometry and astrometry of each star in the 55 clusters are listed in
Table \ref{table1}, available online (see Note to Table \ref{table1}).  
Although we often exclude stars near the edge of the CCD or near bad
columns, the photometry of each open cluster is usually complete enough to
include all of its (unsaturated) OB stars.  It is complete to $y = 13$ for
the brightest clusters, but we obtained deeper images and measured
magnitudes up to $y = 17$ in some cases.

\placetable{table1}

From the calibrated photometry, we created ($b-y$, $y$) color-magnitude
and $(b-y, y-\rm H\alpha)$ color-color diagrams like those described in
Paper 1 for NGC 3766.  Because the MS of the ($b-y$, $y$) diagram is
nearly vertical in the region of massive OB stars, it was very difficult
to constrain the distance modulus and the age of each cluster.  
Therefore, to determine the cluster parameters ($E(B-V)$, $V_0-M_V$, and
age), we searched the available literature for the various values for each
cluster.  We used these values to plot the theoretical isochrones from 
\citet{lejeune2001} and theoretical color-color curve using colors derived 
from \citet{kurucz1979} model spectra.  This technique is described for 
NGC 3766 in Paper 1.  The \citet{lejeune2001} isochrone models do not 
provide Str\"omgren magnitudes, so we transformed the Johnson $V$ and $B$ 
magnitudes to the Str\"omgren system using the transformations
\begin{equation}
 y = V - 0.038(B-V)
\end{equation}
\citep{cousins1985} and 
\begin{equation}
B-V = 1.584(b-y) + 0.681m_1 - 0.116
\end{equation}
\citep{turner1990}, as described in Paper 1.  We also transformed $E(B-V)$ 
to $E(b-y)$ using the relationship
\begin{equation}
E(b-y) = 0.745 \; E(B-V)
\end{equation}
from \citet{fitzpatrick1999}, assuming $R \equiv A(V)/E(B-V) = 3.1$.  In 
very few cases, a cluster did not have published parameters available, or 
the published values did not agree well with our observations, so we fit 
our data to the theoretical curves using a grid of values for $E(b-y)$,
$V_0-M_V$, and log age.  For all clusters, we retained the parameters that 
resulted in the best fit between our data and the theoretical curves, and 
the resulting parameters are listed in Table \ref{clusters}.  Our 
technique to select definite and possible Be stars from the normal B 
stars in our images according to their excess brightness in the narrow 
band H$\alpha$ filter is described in Paper 1, and Table \ref{clusters} 
also includes the numbers of Be and B stars identified in each cluster.

\placetable{clusters}

 
\section{Results from Photometric Study}

Before we consider the results from this photometric survey, it is
important to review the errors inherent in this study.  Saturated stars
affect the number of very bright B and Be stars that we detect, but this
problem is ameliorated by the availability of data on most bright cluster
members in the literature.  However, we do not include saturated stars in
the number of Be and B stars in Table \ref{clusters}.  A second problem is
that our photometry technique only detects those Be stars that had active
H$\alpha$ emission at the time of our observations, so we detect only the
lower limit of the true Be population.  We showed in Paper 1 that the true
fraction of Be stars is probably between $1.2-2.3$ times the number
detected.  Assuming that the ratio of active to inactive Be stars is
similar among different clusters, the neglect of inactive Be stars will
have no impact on our statistical comparison of the Be star frequency
between clusters.  The largest source of errors in our study is due to the
difficulty detecting weakly emitting Be stars since the errors in their
photometry are comparable to the strength of their emission.  There is
also a possibility that unreddened foreground stars or, less often,
background supergiants, contaminate the MS and appear as weak emitters in
the color-color diagram.  Field stars cause special problems for the five
targets that are not truely bound clusters, and the ages, reddening, and
distance moduli for these groups are not representative of the entire
stellar population.  Finally, one of our requirements for identifying
B-type stars (see Paper 1) is that $b-y < E(b-y)$, which neglects the
inherent width of the MS, and we may omit some B and Be stars redward of
this cutoff.  This is especially a problem at the faint end of the B star
MS.

Our photometry of six clusters, excluding the less reliable clusters
Bochum 13 and Trumpler 27, did not cover the entire range of B star
magnitudes so the total number of B stars in these clusters is
underestimated in Table \ref{clusters}.  We show below that the number of 
Be stars omitted is negligible, in agreement with the results of 
\citet{keller2001}.  However, we account for the missing late B-type stars 
by computing the mass function, $\xi(\log M)$.
The \citet{lejeune2001} isochrones assign a mass to each point along the 
curve, and we interpolated to convert the observed $y$ magnitudes of the B 
stars to $M$.  For the allowed mass range of B-type stars, we binned the 
stars according to $\log M$ and counted the unnormalized number, $N$, in 
each bin.  The slope of the linear fit between $\log N$ and $\log M$ 
provides $\Gamma$, where
\begin{equation}
\Gamma = d \log \xi(\log M) / d \log M.
\end{equation}
In the fit, each value of $N$ was weighted according to $N^{1/2}$, which 
decreases the dependence on the less common high mass stars, and the 
incomplete lower mass bins were excluded entirely.  The average $\log M$ 
in each bin was used to compute the total number of B-type stars expected, 
$N_{tot}$.  We calculated $\Gamma$ and $N_{tot}$ using $6-15$ $\log M$ 
bins, equally spaced along the MS curve in the color-magnitude diagram.  
Table \ref{massfcn} provides $\Gamma$, $N_{tot}$, $N_{missing}$, and the 
observed mass range of B-type stars, $\Delta M$, for the six clusters 
analyzed.  This mass range varies depending on the cluster age, and the 
upper mass limit is comparable to the observed maximum B star mass in each 
cluster.  The minimum mass of B stars is 2.3 $M_\odot$ from the 
\citet{lejeune2001} models.  We found a slight dependence between $\Gamma$ 
and the number of bins used, and the errors for $\Gamma$ and $N_{tot}$ 
reflect the $1\sigma$ variance.  Our values for $\Gamma$ are consistent 
with the values for massive stars in open clusters found by 
\citet{massey1995}.

\placetable{massfcn}

In this study, we find a total of 52 definite Be stars and an additional
129 possible Be stars in 48 clusters.  (Hereafter we ignore the regions
that are not true open clusters as well as Bochum 13 and Trumpler 27,
discussed in Appendix A.)  We also detect 2210 B-type stars, including the
Be stars, and we estimate from the clusters' mass distributions that 328
B-type stars are missing from our sample.  Considering only the definite
Be stars in these clusters, the mean percentage of Be stars is only 2.0\%.  
Including the possible Be stars raises the mean percentage to 7.1\%.  
This is much lower than expected; \citet{abt1987} found that 18\% of field
B0 to B7 III-V stars in the Bright Star Catalogue (BSC;
\citealt{hoffleit1991}) were classified as Be stars in the literature.  
However, our study accurately accounts for the total number of B stars
within the cluster volume, while Abt's magnitude-limited survey is biased
towards early B stars that are probably more rich in Be stars (e.g.\
\citealt{mermilliod1982}).  Furthermore, many of the B stars in the BSC
are members of the Gould Belt, which has $7.4 < \log \rm{age} < 7.8$
\citep{perrot2003} and coincides with a maximum in the Be star fraction
(see Figure \ref{befreq} below).  This may explain the unusual richness of 
Be stars in the BSC.

Using our extensive sample of Be stars, we investigate several factors
that may influence the population of Be stars in open clusters, beginning
with metallicity.  Using clusters within the same age interval in the
interior and exterior of the Galaxy, the Large Magellanic Cloud (LMC), and
the Small Magellanic Cloud (SMC), \citet*{maeder1999} found that the ratio
of Be stars decreases sharply with increasing metallicity.  However,
\citet*{keller1999} found no difference in the fraction of Be stars in the
LMC and SMC, although there is a factor of two difference in their
metallicities.  To look for this influence in our data, we investigated
the distribution of the target clusters in the Galactic plane.  We
calculated the distance of each cluster from the Galactic center,
$d_{cent}$, using $R_0 = 8000$~pc as the Sun's distance from the Galactic
center and assuming that each cluster lies within the plane of the Galaxy
(valid since all have $\mid b \mid < 6^\circ$).  The Galactic longitude,
$\ell$, latitude, $b$, distance from the Sun, $d_\odot$, and $d_{cent}$
for each cluster are listed in Table \ref{distribution}.  The spatial
distribution of clusters with respect to the Sun and the Galactic center
is illustrated in Figure \ref{galaxy}.

\placetable{distribution}
\placefigure{galaxy}

\begin{figure}
\includegraphics[angle=90,scale=0.3]{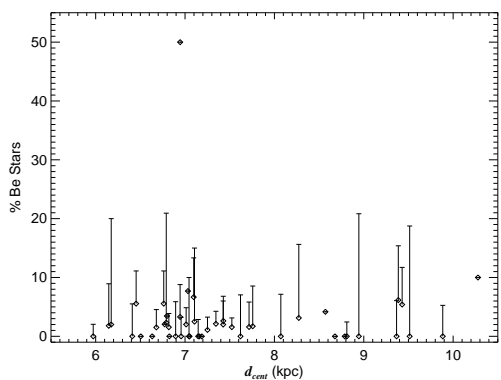}
\caption{
The percentage of Be stars (\textit{diamonds}) is shown as a function of 
their distance from the Galactic center, $d_{cent}$.  The upper limit for 
each cluster, the percentage of definite and possible Be stars, is also 
shown.
\label{distancefcn}
}
\end{figure}

The numbers of definite and possible Be star detections from our
photometry are included in Table \ref{clusters}, and we plot 
the percentage of Be stars as a function of distance from the
Galactic center in Figure \ref{distancefcn}.  The diamonds are the lower
limits of the Be star fraction, determined from the number of definite
detections, and the upper limit assumes that all of the possible
candidates are in fact Be stars.  Note that this upper limit does not
represent the formal error in our measurement, which can be significant in
clusters with few B-type stars (especially NGC 6268, in which we find that
one of its two B stars is a Be star).  For the 35 clusters with $d_{cent}
< R_0$, the mean Be star population is 1.8\%--6.5\%.  The outer 13
clusters have a mean population of 3.3\%--10.5\%.  Using the Galactic
metallicity gradient $\Delta Z/$kpc $= -0.0019$ to $-0.0029$
\citep{maeder1999}, over the 4.3 kpc range in distance from the Galactic
center, $\Delta Z = .0082-.0125$.  Interpolating from
\citeauthor{maeder1999}, the fraction of Be stars in our clusters should
increase by a factor of $\approx 2$ over this range in metallicity.  The
mean Be star populations do increase slightly over this distance, but
Figure \ref{distancefcn} does not provide compelling evidence supporting a
dependence on the Galactic Be star fraction with metallicity.

\placefigure{distancefcn}

\begin{figure}
\includegraphics[angle=90,scale=0.3]{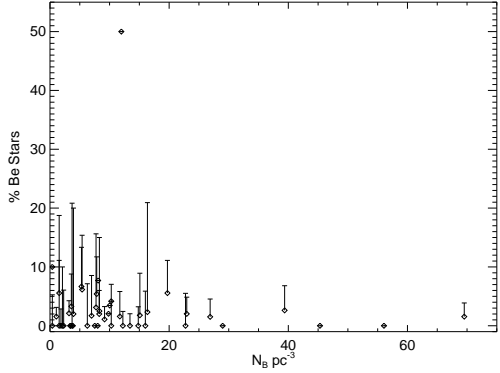}
\caption{
The percentage of Be stars is shown as a function of the cluster density 
in the same format as Figure \ref{distancefcn}.
\label{density}
}
\end{figure}

The number of Be stars may also be related to the stellar density in the
cluster.  \citet*{strom2005} suggest that the average rotational velocity
is greater for higher density clusters, presumably allowing more Be stars
to form. They show that the average rotational velocity of unevolved B
stars in the dense clusters $h$ and $\chi$ Persei is more than $2\times$
that of field stars of comparable ages.  To investigate the role of the
stellar density in each cluster, we calculated the mean B star density of
the clusters. We determined the cluster center using the mean positions of
the B-type stars, and the standard deviation of the B star positions
relative to the cluster center defines the radius of the assumed
spherical volume. However, the cluster NGC 6268 has only two B stars, so 
we used the mean distance from the cluster center to define the radius in
this case. We plot the percentage of Be stars against the number of B
stars per cubic parsec in Figure \ref{density}.  Our data show no
correlation between the two.

\placefigure{density}

We also consider the relationship between the number of Be stars in open
clusters and the cluster age, a possible signature of evolutionary effects
involved in the origin of Be star disks.  Past studies have disagreed on
the relationship between the ages and numbers of Be stars.  
\citet{fabregat2000} found that classical Be stars are most abundant
between $7.1 < \log$~age $< 7.4$.  On the other hand,
\citet{mermilliod1982} and \citet{slettebak1985} found a weaker dependence
on age.  Instead, they noted that Be stars occupy the entire MS band, from
the ZAMS through the TAMS, indicating that Be stars occur at any
evolutionary stage.  \citet{mermilliod1982} found that the frequency of Be
stars depends instead on spectral type; the abundance peaks for types
B1--B2 and B7--B8.  However, \citet{zorec1997} also found no distinct peak
the Be star frequency with luminosity class, and they argue that the Be
phenomenon is more likely due to the characteristics of B stars at the
time of their formation.  This would imply a higher fraction of Be stars
among young clusters.

\begin{figure}
\includegraphics[angle=90,scale=0.3]{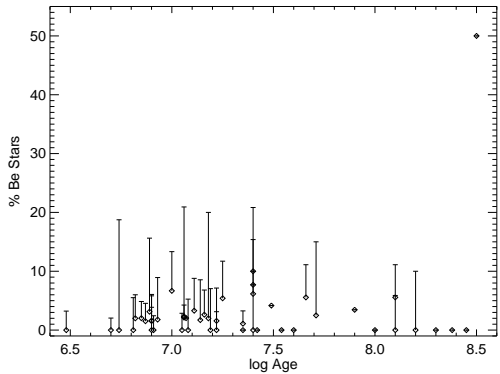}
\caption{
The percentage of definite Be stars is shown as a function of cluster age 
in the same format as Figure \ref{distancefcn}.  
\label{agefunction}
}
\end{figure}

We present a plot of the Be star percentage as a function of cluster age
in Figure \ref{agefunction}.  While we did not compute the formal error in
the ages, the agreement among different ages provided in the literature
and the good match between the theoretical isochrones and our
color-magnitude diagrams suggest that the cluster ages are accurate to
$\pm 0.2$ dex in most cases.  Because no clear relationship between the
percentage of Be stars and the cluster age emerges from Figure
\ref{agefunction}, we plot the average Be frequency in four age bins in
Figure \ref{befreq}.  The bins correspond to the ages $\log \rm age <
7.0$, $7.0 \le \log \rm age < 7.4$, $7.4 \le \log \rm age < 8.0$, and
$\log \rm age \ge 8.0$.  Our definite Be star frequency and the Poisson
errors are plotted using solid lines, and the total of definite and
possible Be stars, also with the Poisson errors, are represented with
dashed lines.  The plotted ages represent the mean age observed within
each bin, and the age error bars represent the maximum and minimum ages
from our sample.  We are surprised to see that the fraction of Be stars is
generally highest among clusters with $7.4 \le \log \rm age < 8.0$, unlike
previous studies that found a peak among clusters with $7.0 \le \log \rm
age < 7.4$ \citep{fabregat2000}.  Figure \ref{befreq} reveals that the
population of Be stars generally increases with time until $7.4 \le \log
\rm age < 8.0$, then declines among older clusters.

\placefigure{agefunction}
\placefigure{befreq}

\begin{figure}
\includegraphics[angle=90,scale=0.3]{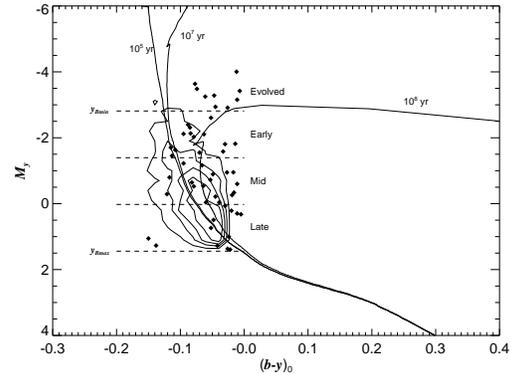}
\caption{
The definite Be star detections are shown on an absolute
color-magnitude diagram.  Three isochrones with ages $10^5$, $10^7$, and
$10^8$~yr are shown for comparison.  The B stars are divided into four 
regions: evolved, early, mid, and late spectral types.  The contours 
represent the distribution of normal B-type stars detected in our sample 
(29\%, 43\%, 57\%, and 76\%, respectively). 
\label{abscolor}
}
\end{figure}

Finally, we plot the 52 definite Be stars in Figure \ref{abscolor} using 
their intrinsic colors and absolute magnitudes.  For comparison, the 
normal B-type stars are plotted as contours overlying the absolute
color-magnitude diagram.  To derive the absolute magnitudes, we used the
relation 
\begin{equation}
A(y) = 3.034 \: E(B-V)
\end{equation}
interpolated from \citet{fitzpatrick1999}.  The horizontal dashed lines in
Figure \ref{abscolor} divide the Be population into four bins: evolved
($M_y < y_{Bmin}$ at the ZAMS), early B spectral types, mid B stars, and
late B stars.  The isochrone for $10^5$~yr represents a very young age,
close to the ZAMS, and the isochrones for a distribution of stars with
ages of $10^7$ and $10^8$~yr are also plotted.  Although the Be stars are 
generally a redder population than the B-type stars (the two most
significant outliers to the lower left of the ZAMS are No.\ 12 in NGC 6531 
and No.\ 355 in Trumpler 34; see discussion in Appendix A), it is unlikely 
that the Be stars in a given cluster are a more evolved population than 
their neighboring B stars, and we believe that other effects are largely 
responsible for their position relative to the ZAMS.  The circumstellar 
disk can impart additional reddening, rotationally induced gravitational 
darkening may cause them to appear cooler than a non-rotating star of the 
same mass, or light may be preferentially scattered by the disks and cause 
the stars to appear brighter and more evolved \citep{slettebak1985, 
zorec1997}.

\placefigure{abscolor}

\begin{figure}
\includegraphics[angle=90,scale=0.3]{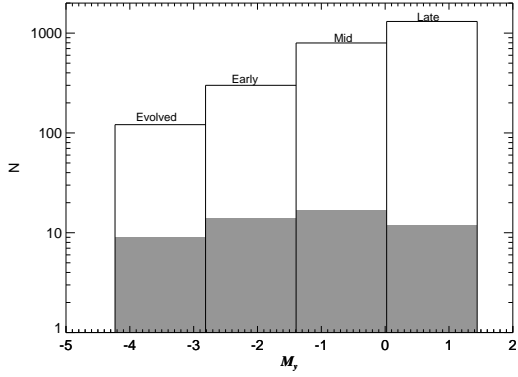}
\caption{
A histogram plot of the total Be and B star populations of all
clusters.  The bins are divided into evolved, early, mid, and late B 
spectral types.
\label{histogram}
}
\end{figure}

Although Figure \ref{abscolor} reveals that the definite Be stars are
rather uniformly distributed in $M_y$, the relative fraction of Be and B
stars decreases with later spectral tyes.  Be stars comprise a respective
7.4\%, 4.7\%, 2.1\%, and 0.9\% of the evolved, early, mid, and late type B
stars in our sample.  A histogram plot illustrating the relative
populations for each bin is shown in Figure \ref{histogram}.  These
results indicate that the relative fraction of Be stars attains a maximum
among the evolved stars that are approaching the TAMS.  (Note that these
evolved stars do not correspond to the oldest age bin defined above, where
a decline in the Be population is observed.)  \citet{fabregat2000} also
found that the Be phenomenon occurs during the second half of the MS,
which they argue is the result of an evolutionary spin-up as the stars
approach the TAMS.

\placefigure{histogram}


\section{Conclusions}

Our results indicate that the Be phenomenon is not strongly dependent on 
metallicity or cluster density, but it is clearly influenced by the
evolutionary state of B stars.  We observe a higher fraction of evolved Be
stars, but if Be stars were born as rapid rotators as believed by
\citet{zorec1997}, the opposite trend would be observed; more Be stars
would be observed in unevolved stars, in the youngest clusters, since the
rotation rate decreases during the MS lifetime \citep{meynet2000}.  Also,
Be stars born as rapid rotators would have a much different distribution
along the MS.  A rotationally distorted star with mass $M_B$ at the 
critical breakup velocity will have an equatorial radius 
\begin{equation}
R_e = 1.5 \left (0.914 + \frac{0.143}{(M_B/M_\odot)^2} \right ) R_B
\end{equation}
\citep{fremat2005}, where $R_B$ is the radius of a non-rotating B star.  
The resulting critical breakup velocity is
\begin{equation}
v_{crit} = 436.7  \left ( \frac{M_B/M_\odot}{R_e/R_\odot} \right )^{1/2} 
\rm{km~s^{-1}}.
\end{equation}
Table \ref{critrot} compares this $v_{crit}$ with the mean projected 
rotational velocity, $\langle v \sin i \rangle$, of B stars from 
\citet*{abt2002}.  Masses and radii are from \citet{harmanec1988}.  Later 
B-type stars have a mean rotational velocity closer to their critical 
value, so that if these velocities reflect their birth spins, then we 
should expect to see a higher fraction of Be stars among late B stars.  We 
show in Figure \ref{histogram} that such a distribution is not observed, 
hence we doubt that many Be stars are born as fast rotators.

\placetable{critrot}

\citet{meynet2000} predicted that stars with $M > 12 M_\odot$ experience a
short spin-up phase at the end of their MS lifetimes, and this spin-up
phase could be responsible for the high fraction of evolved Be stars
observed in this study.  We do not expect an age dependence with this
scenerio; as long as a cluster has TAMS B stars, Be stars would be
expected. However, the duration of this spin-up phase is expected to be
short, only about 1\% of the MS lifetime, so we should observe a
comparable Be star fraction if the TAMS spin-up phase is the origin of the
rapid rotation.  While lower mass B stars are not expected to experience
such a spin-up, \citet{meynet2000} predict that these stars may remain
rapid rotators and in fact increase their fraction of the breakup velocity
as they evolve along the MS.  Once again, this would support a larger
fraction of late Be stars, which is not observed.

We find it more likely that Be stars are spun-up by mass transfer in close
binary systems.  Although there is a perceived deficiency of short period
Be star binaries (e.g.\ \citealt{abt1978}), recent work has led to the
discovery of 55 Be star binaries with periods $< 100$ d \citep{gies2000,
raguzova2005}, and Be stars have been observed in nearly all stages of the
proposed binary evolution scenario.  The initially more massive star
expands to fill its critical Roche lobe during the H shell burning phase,
triggering Case B mass transfer and causing the spin-up of the mass gainer
\citep{gies2000}.  The massive Algol binary RY Per is an example of such a
spin-up process in action; the F7: II-III donor star is engaged in active
mass transfer via Roche lobe overflow onto the B4: V companion, and the
mass gainer has been spun up to more than seven times faster than the
synchronous rate \citep{barai2004}.  Commonly, the mass ratio becomes
inverted during the mass transfer and the orbital separation $a$
increases, leading to a dramatic increase in the timescale for tidal
synchronization ($\sim a^{17/2}$; \citealt{hilditch2001}).  This timescale
becomes much longer than the H shell burning phase, so the mass gainer
will retain its rapid rotation long after the mass transfer is complete.  
At the conclusion of the mass transfer phase, the donor star is left as a
stripped down, helium star remnant -- a small but hot subdwarf
\citep{gies2000}.  $\phi$ Per is the only confirmed Be+sdO system
\citep{thaller1995, gies1998}, but three additional candidates have been
identified: 59 Cyg \citep{maintz2004}, HR 2142 \citep*{waters1991}, and FY
CMa \citep{rivinius2004}.  Depending on its mass, the sdO star will
eventually become a white dwarf (WD) or explode as a supernova to form a
Be + neutron star (NS) binary or an unbound system \citep{gies2000}.  To
date, no Be + WD systems are known, but 130 Be/X-ray binaries (presumably
with NS companions) have now been identified \citep{raguzova2005}.

\citet{pols1991} and \citet{vanbever1997} predicted that the fraction of
mass gainers is highest among B0--B3 stars, decreasing with later spectral
types, consistent with the distribution of Be stars that we observe.  If
the mass transfer is conservative (as expected for systems with a high
initial mass ratio, $q > 0.6$), \citet{pols1991} showed that the mass
gainers would be observed as rejuvenated, bright MS stars, including blue
stragglers. However, Be stars formed this way may not appear near the ZAMS
due to disk reddening and gravitational darkening.  On the other hand,
\citeauthor{pols1991} showed that there is a minimum $q$ (between
$0.3-0.5$, depending on their model) for the production of Be stars from
close binary evolution with non-conservative mass transfer. Therefore, Be
stars formed by mass transfer would be found preferentially in the upper
part of the MS.

Previous studies have estimated the percentage of the Be stars that are
products of close binary evolution.  \citet{pols1991} predicted that
1.3\%$-$3.9\% of B-type stars should be observed as rapid rotators due to
angular momentum received during case B mass transfer, and
\citet{vanbever1997} predicted a similar number (0.9\%$-$4.8\%).  While
the predicted total number of mass gainers is comparable to the number of
Be stars that we observe, not all of these mass gainers are expected to
become Be stars.  \citet{pols1991} showed that the predicted number of
spun-up B stars in the BSC is 40\%$-$60\% of the observed number of Be
stars in that sample.  However, because not all rapidly rotating B-type
stars exhibit the Be phenomenon, \citet{vanbever1997} argued that Be stars
represent about half of the total number of rapidly rotating B stars, so
the fraction of Be stars in the BSC formed by mass transfer is closer to
20\%.  We show above that the fraction of Be stars in the BSC is not
representative of the fraction in open clusters, so the estimate of
\citet{vanbever1997} cannot be applied to our sample, and there is no
spectroscopic survey in the literature that provides relative numbers of
Be stars and normal, rapidly rotating B stars.

\begin{figure}
\includegraphics[angle=90,scale=0.3]{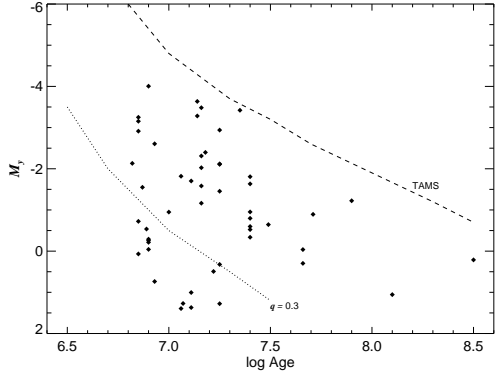}
\caption{
$M_y$ of the definite Be stars is plotted as a function of cluster 
age (\textit{diamonds}).  The absolute magnitude of a B star at the 
TAMS is also shown (\textit{dashed curve}; \citealt{lejeune2001}).  For 
a primary at the TAMS and system mass ratio $q=0.3$, $M_y$ of the 
secondary is also shown (\textit{dotted curve}).  
\label{binaryevol}
}
\end{figure}

In lieu of an accurate theoretical prediction, we can estimate the
fraction of mass gainer Be stars from our results.  We plot the observed
$M_y$ of our definite Be stars versus their age in Figure
\ref{binaryevol}.  Using the isochrone models of \citet{lejeune2001}, we
determined the mass of a primary at the TAMS and its $M_y$ (dashed curve)
over a range of cluster ages.  For a close binary system with $q \approx
1$ and the primary at the TAMS, the mass transfer will be conservative
\citep{pols1991} and the rejuvenated secondary will appear brighter than
the TAMS $M_y$. Assuming a minimum $q=0.3$ for the production of Be stars
\citep{pols1991}, we also identified the $M_y$ of the mass gainer, shown
as a dotted curve.  (Note that for log age $> 7.5$, the limiting $M_y$ for
the secondary is greater than $y_{Bmax}$.)  Figure \ref{binaryevol}
reveals that only one of our definite Be stars is found at the TAMS and
thus may have acquired its rapid rotation during a spin-up phase at this
stage of its evolution.  In contrast, we find that 38 Be stars, or 73\%,
fall in the mass range associated with binary mass ratios of $0.3 \le q <
1$ and are good candidates for spin-up by binary mass transfer.  (Note
that two Be stars in NGC 2439 have nearly the same $M_y \approx -2.1$, so
these points overlap in Figure \ref{binaryevol}.)  There are 14 Be stars
below the binary mass ratio line of $q < 0.3$, and some of these were
probaby born as rapid rotators.  However, four of these (members of the
clusters NGC 6231, NGC 6322, and Stock 13) are likely Herbig Be stars, and
their disks may be due either to rapid rotation or the remnants of star
formation.  We also caution that the $M_y$ relations in Figure
\ref{binaryevol} neglect gravitational darkening and disk reddening in Be
stars, so our comparison of the predictions of the three Be formation
scenarios is not exact.  Nevertheless, our results clearly emphasize the
relative dominance of the binary spin-up process, while neither a TAMS
spin-up phase nor rapid rotation at the time of formation appear to play
a significant role in the formation of Be stars.

\placefigure{binaryevol}

\acknowledgments

We thank Charles Bailyn, Juan Zorec, and the referee, Helmut Abt, for
their helpful comments on our manuscript.  We are grateful to Charlie
Finch and John Helsel for their help reducing the large amount of
photometry.  MVM thanks NOAO for travel support to observe with the CTIO
0.9m telescope, and she is supported by an NSF Astronomy and Astrophysics
Postdoctoral Fellowship under award AST$-$0401460.  Financial support also
was provided by the NSF through grant AST$-$0205297 (DRG).  Institutional
support has been provided from the GSU College of Arts and Sciences and
from the Research Program Enhancement fund of the Board of Regents of the
University System of Georgia, administered through the GSU Office of the
Vice President for Research.

Facilities: \facility{CTIO}


\appendix

\section{Notes on Individual Clusters}

While the results from each cluster are summarized by Table
\ref{clusters}, further comments regarding each cluster are presented
below.  The definite and possible Be stars in each cluster are identified
using their assigned number from Table \ref{table1}.  (For cross
referencing purposes, note that Table \ref{table1} lists other identifiers
and WEBDA numbers for these stars whenever possible.)  Particular sources
of error, including saturated stars and contaminating foreground stars,
are also discussed.  Unless otherwise noted, spectral types and other
information about individual stars are from the SIMBAD database.  

\begin{figure}
\includegraphics[angle=0,scale=0.4]{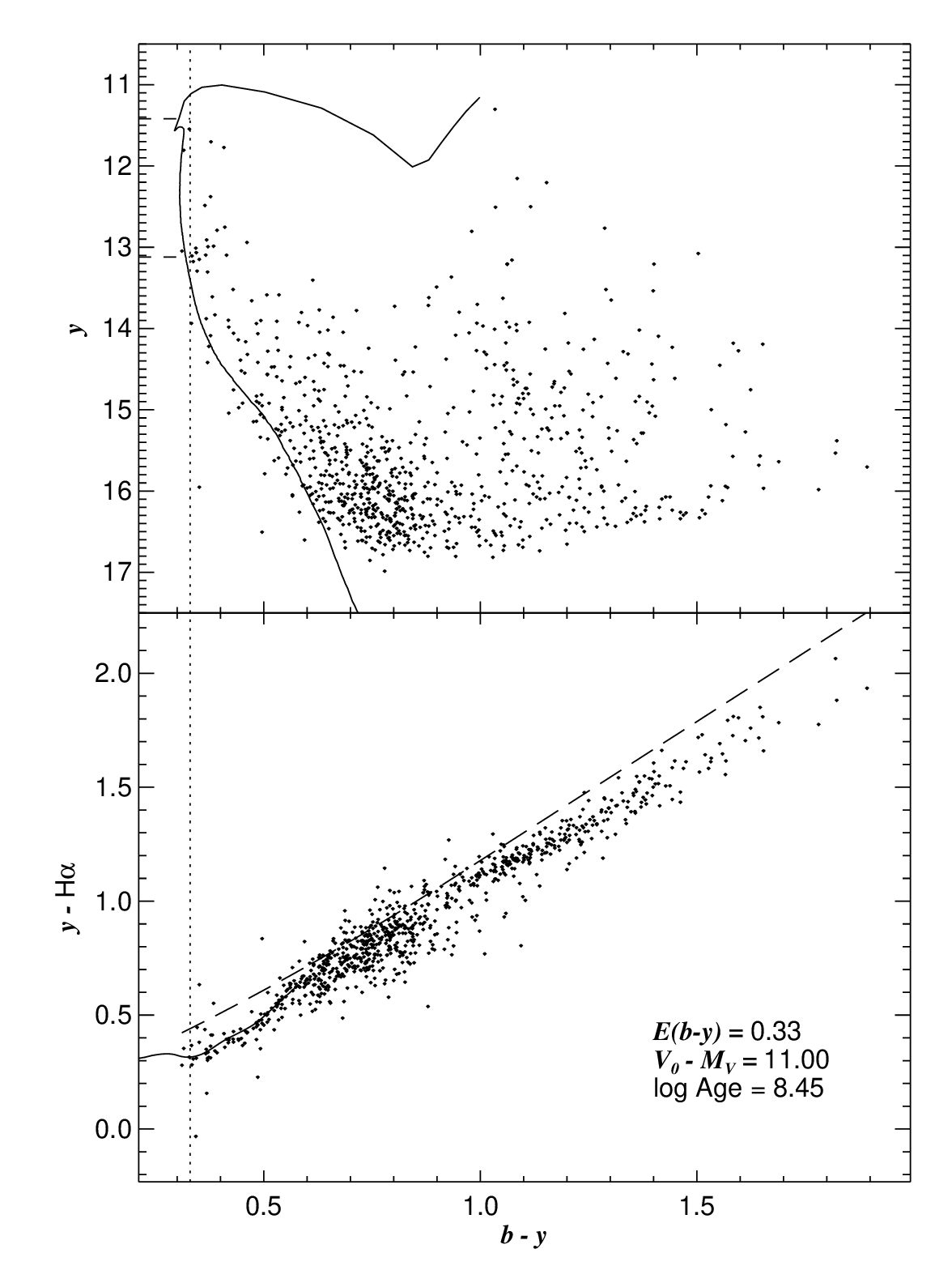}
\caption{
Color-magnitude (\textit{top}) and color-color (\textit{bottom}) diagrams
of the cluster Basel 1.  The isochrone fit (\textit{solid line in top}),
color-color fit (\textit{solid line in bottom}), reddening value
(\textit{dotted lines}), $y_{Bmin}$ and $y_{Bmax}$ (\textit{dashed lines
in top}), and parabolic fit of unreddened stars in the 
atlas of \citet{jacoby1984} (\textit{dashed line in bottom}) are also 
shown.  Be stars (\textit{large}) are distinguished from all other stars 
(\textit{small}) in both diagrams (although in this cluster no Be stars 
are present).  Potential foreground stars or white dwarfs are also marked 
($\times$), although none are present in our data for this cluster.
\label{Basel1}
}
\end{figure}

\begin{figure}
\includegraphics[angle=0,scale=0.4]{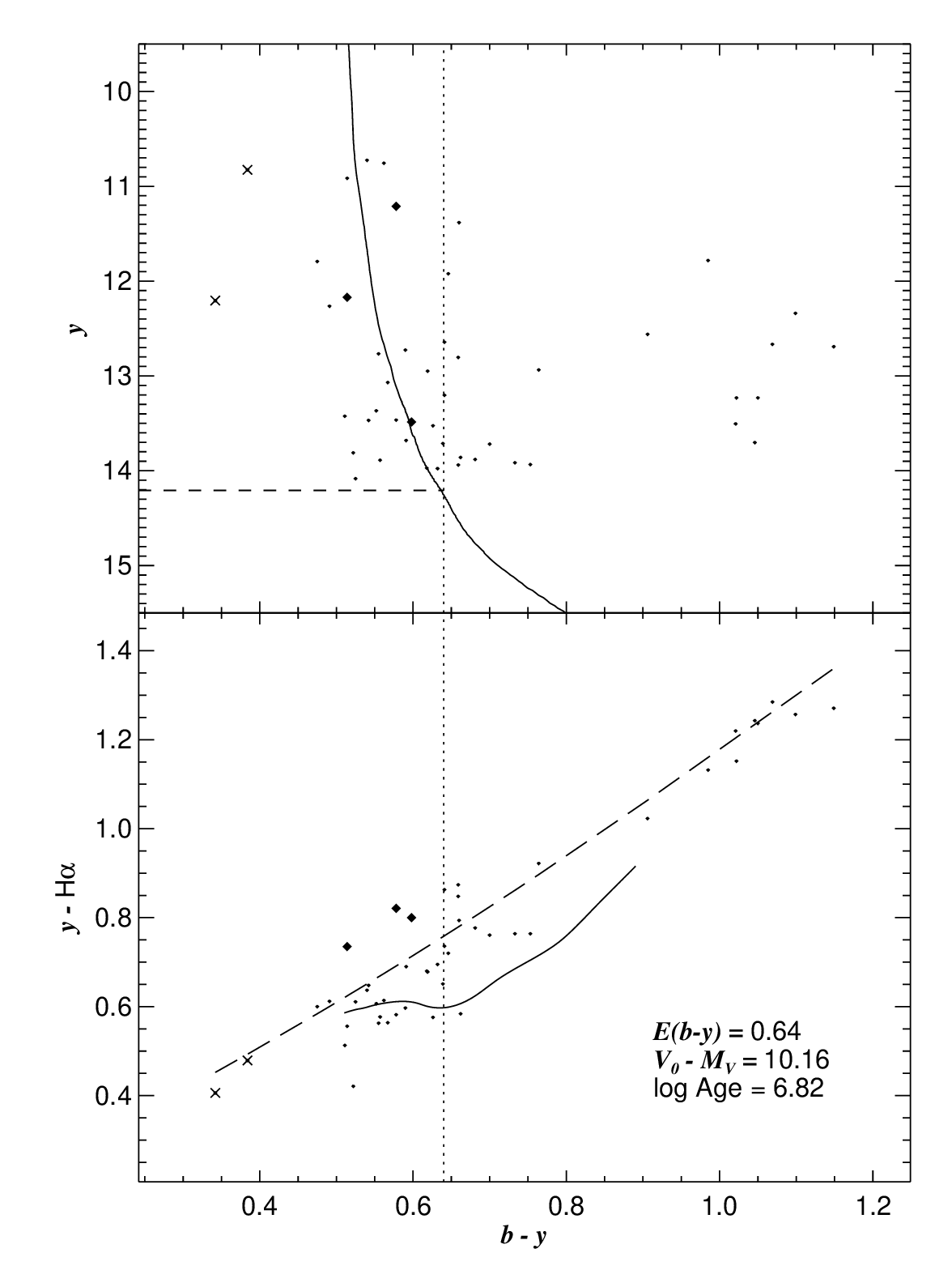}
\caption{
Color-magnitude (\textit{top}) and color-color 
(\textit{bottom}) diagrams of the cluster Bochum 13
in the same format as Figure \ref{Basel1}.
\label{Bochum13}
}
\end{figure}

\begin{figure}
\includegraphics[angle=0,scale=0.4]{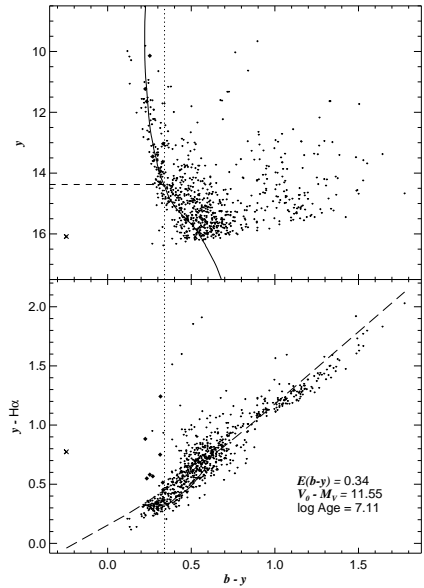}
\caption{
Color-magnitude (\textit{top}) and color-color 
(\textit{bottom}) diagrams of the cluster Collinder 272
in the same format as Figure \ref{Basel1}.
\label{Collinder272}
}
\end{figure}

\begin{figure}
\includegraphics[angle=0,scale=0.4]{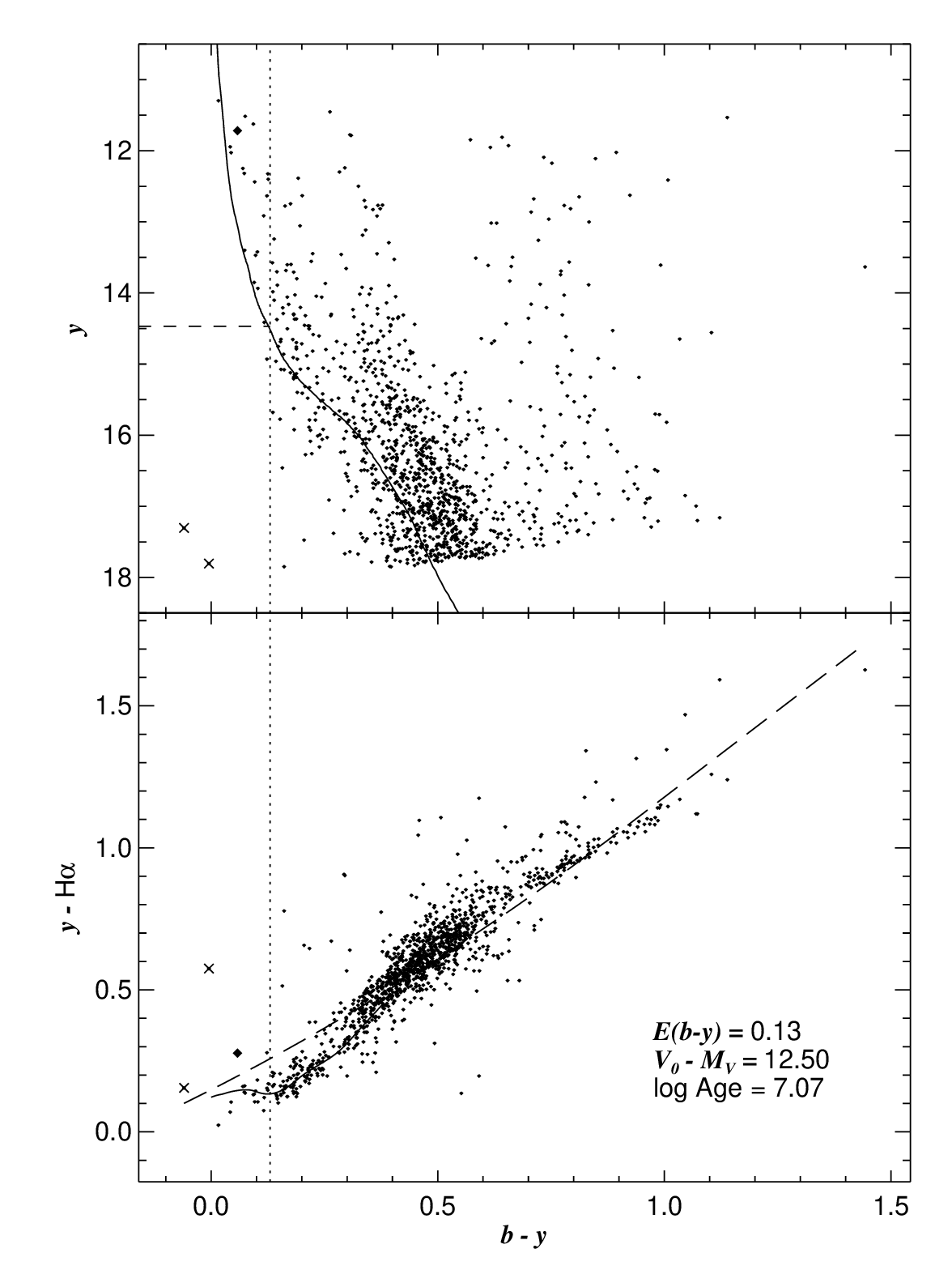}
\caption{
Color-magnitude (\textit{top}) and color-color 
(\textit{bottom}) diagrams of the cluster Haffner 16
in the same format as Figure \ref{Basel1}.
\label{Haffner16}
}
\end{figure}

\begin{figure}
\includegraphics[angle=0,scale=0.4]{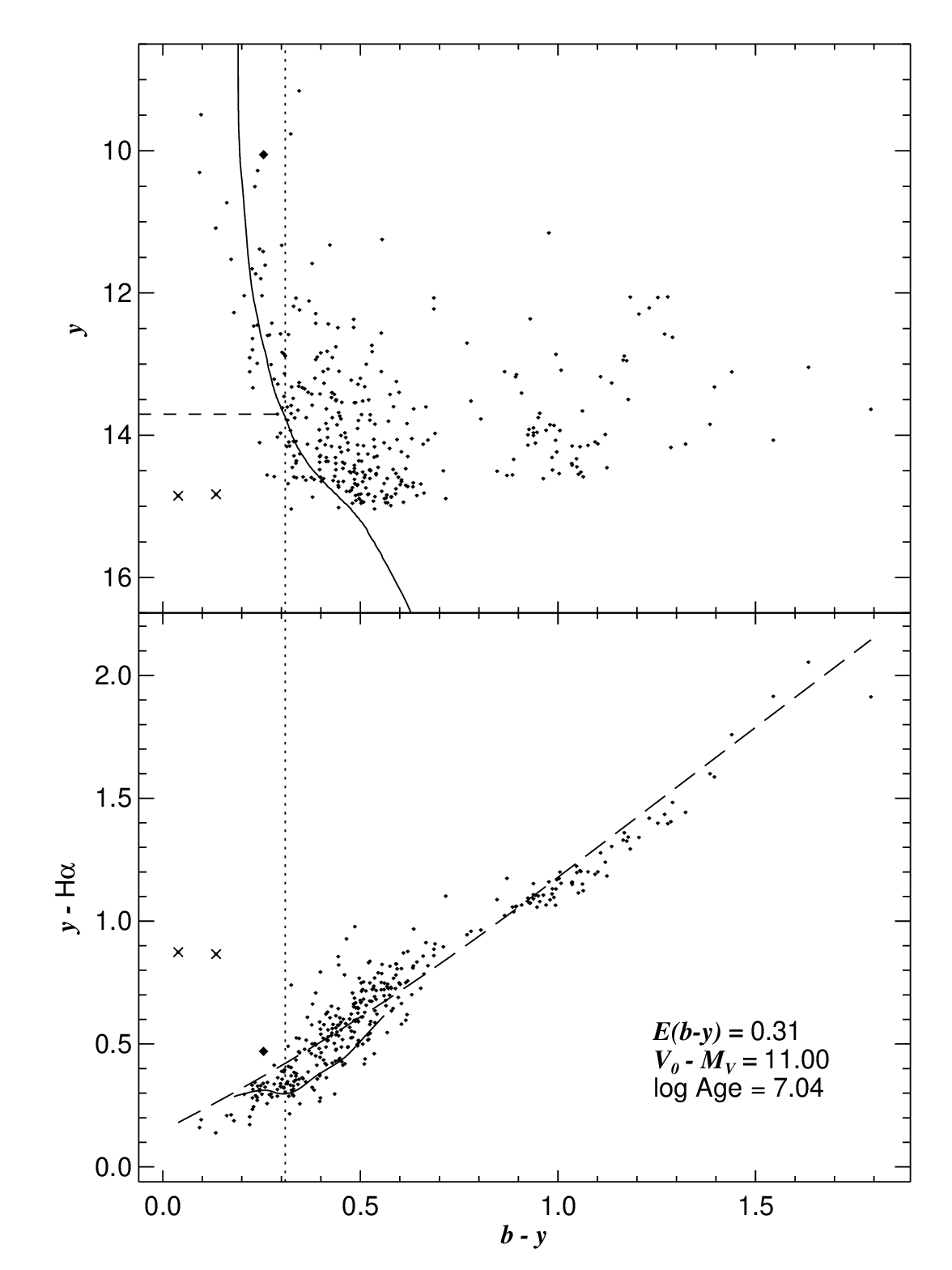}
\caption{
Color-magnitude (\textit{top}) and color-color 
(\textit{bottom}) diagrams of the cluster Hogg 16
in the same format as Figure \ref{Basel1}.
\label{Hogg16}
}
\end{figure}

\begin{figure}
\includegraphics[angle=0,scale=0.4]{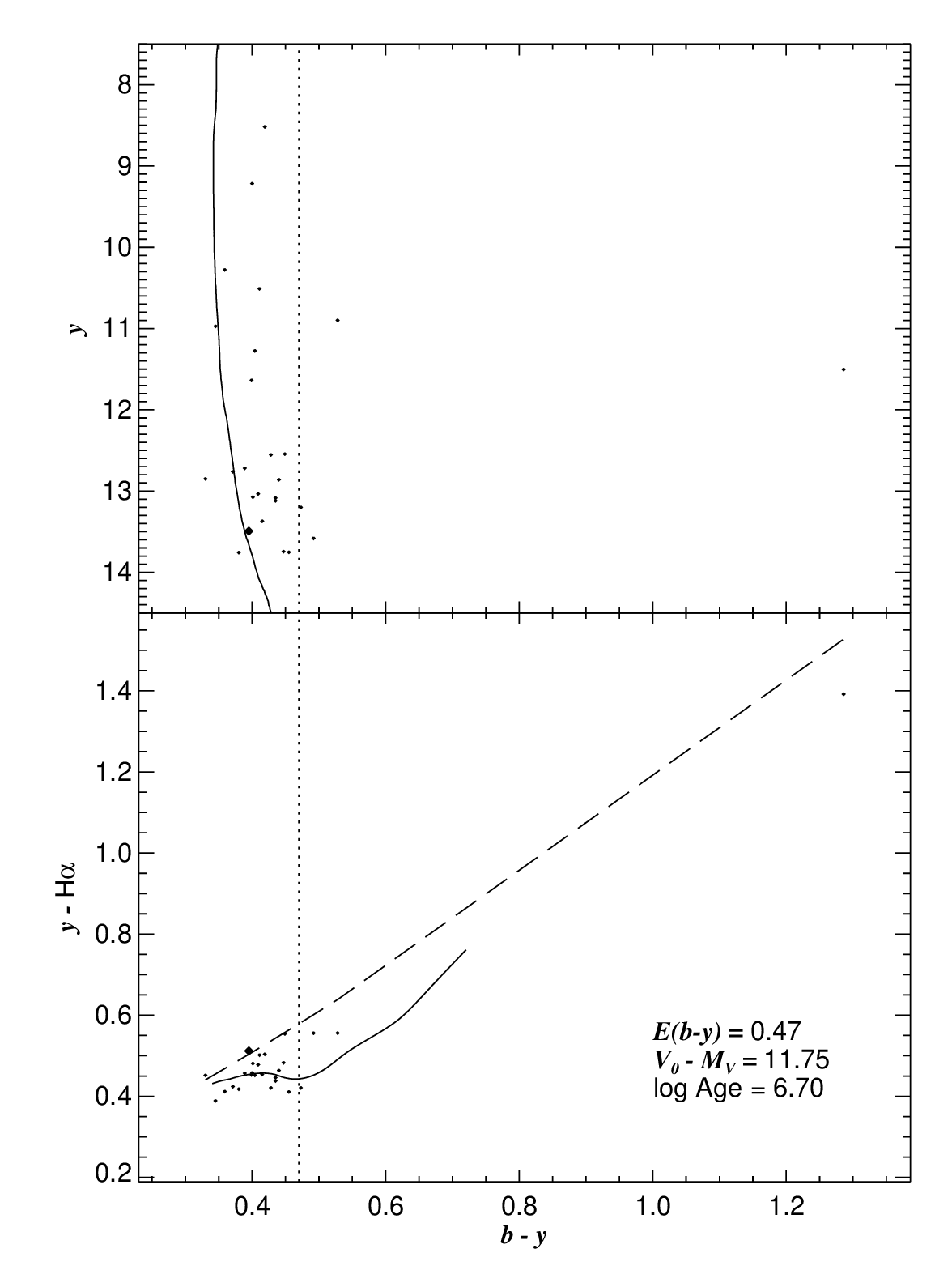}
\caption{
Color-magnitude (\textit{top}) and color-color 
(\textit{bottom}) diagrams of the cluster Hogg 22
in the same format as Figure \ref{Basel1}.
\label{Hogg22}
}
\end{figure}

\begin{figure}
\includegraphics[angle=0,scale=0.4]{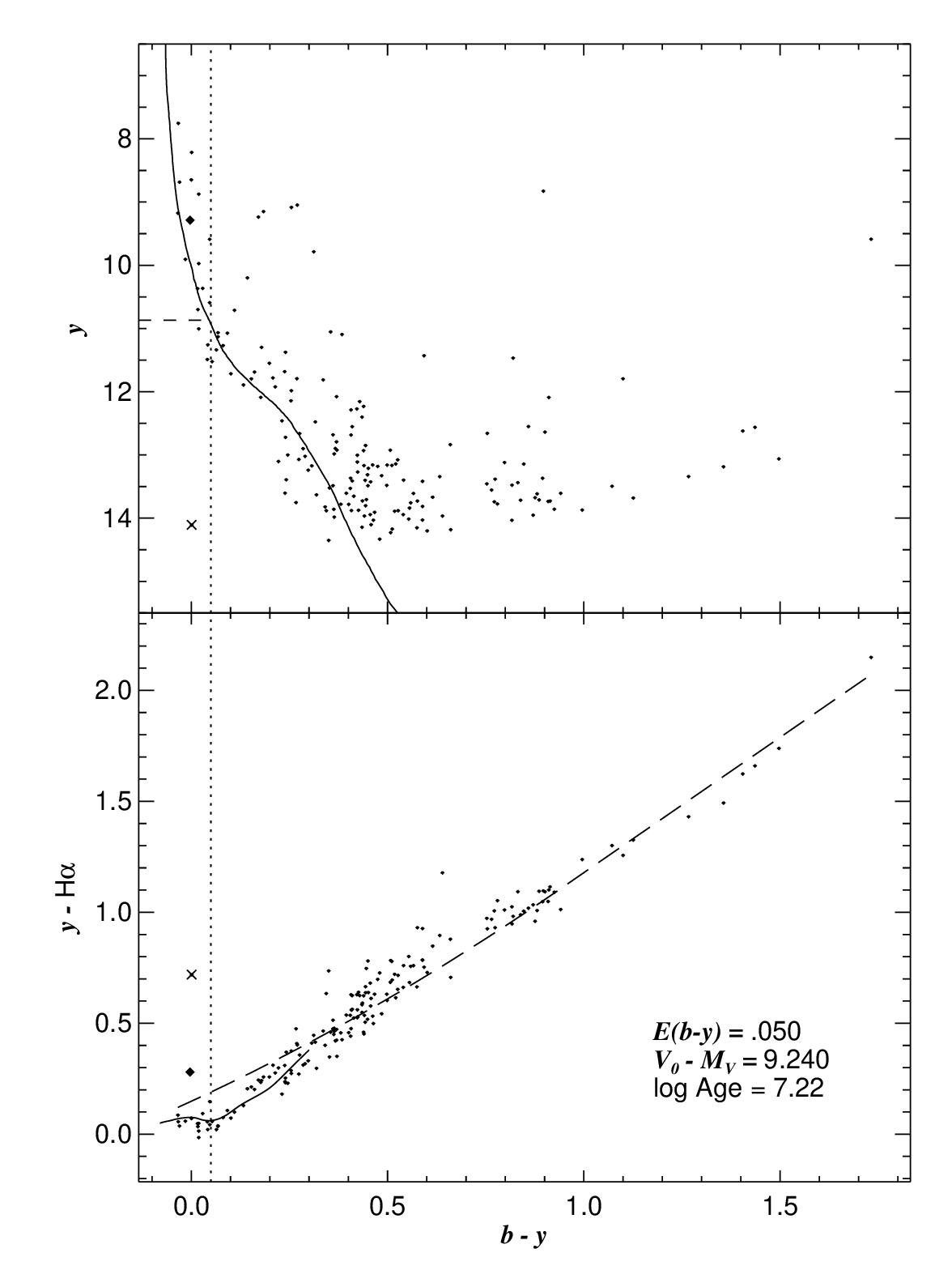}
\caption{
Color-magnitude (\textit{top}) and color-color 
(\textit{bottom}) diagrams of the cluster IC 2395
in the same format as Figure \ref{Basel1}.
\label{IC2395}
}
\end{figure}

\begin{figure}
\includegraphics[angle=0,scale=0.4]{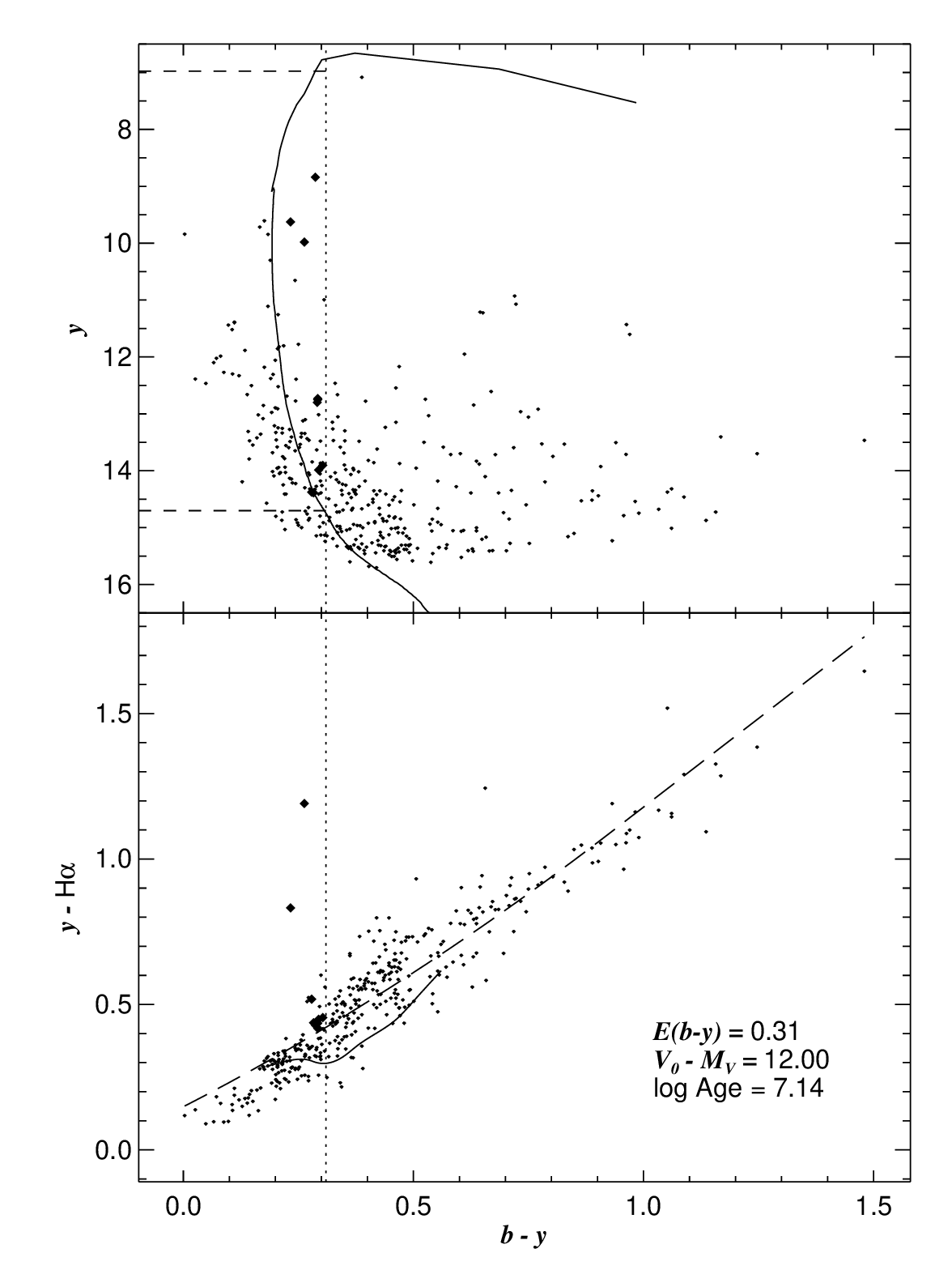}
\caption{
Color-magnitude (\textit{top}) and color-color 
(\textit{bottom}) diagrams of the cluster IC 2581
in the same format as Figure \ref{Basel1}.
\label{IC2581}
}
\end{figure}

\begin{figure}
\includegraphics[angle=0,scale=0.4]{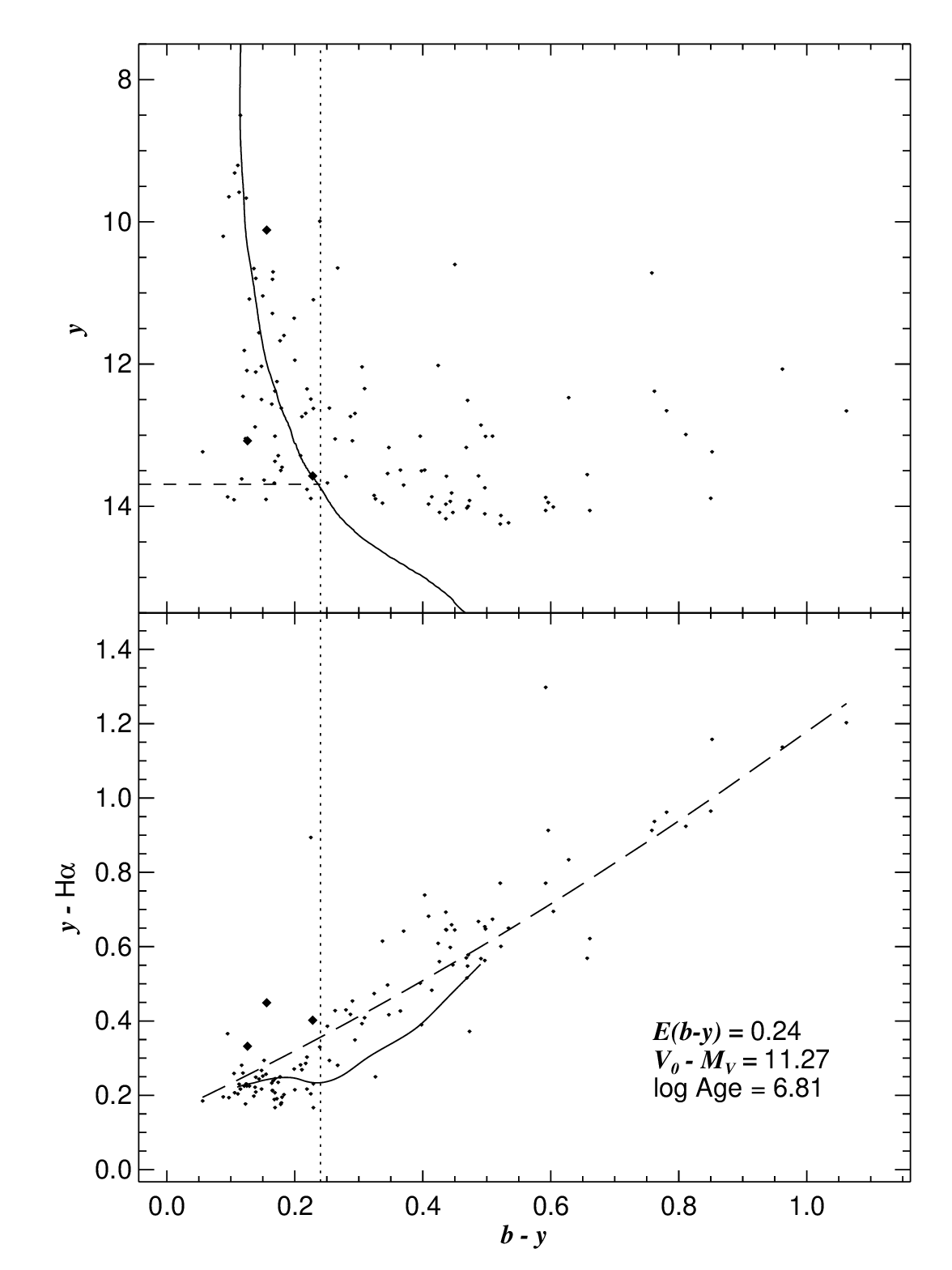}
\caption{
Color-magnitude (\textit{top}) and color-color 
(\textit{bottom}) diagrams of the cluster IC 2944
in the same format as Figure \ref{Basel1}.
\label{IC2944}
}
\end{figure}

\begin{figure}
\includegraphics[angle=0,scale=0.4]{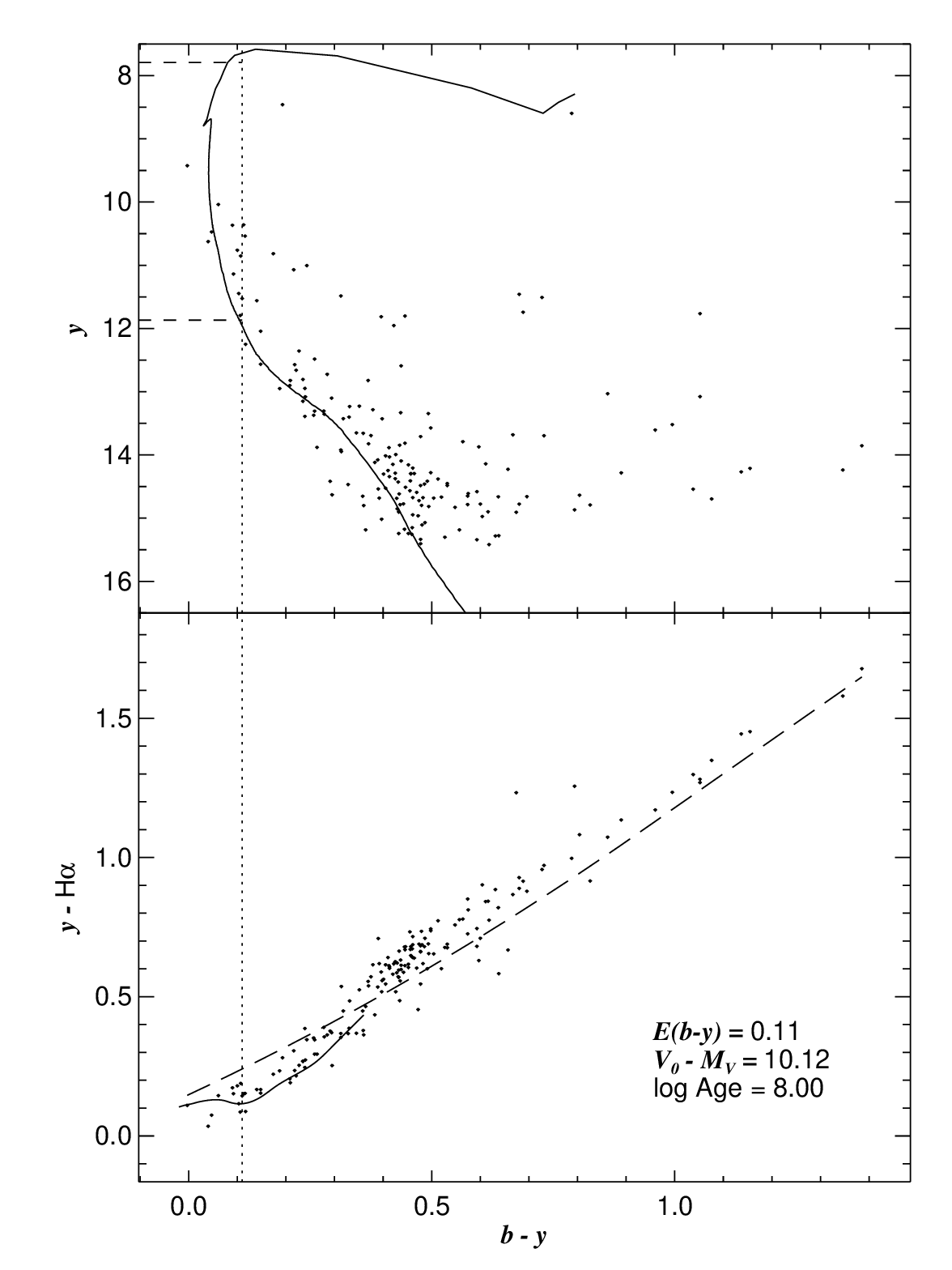}
\caption{
Color-magnitude (\textit{top}) and color-color 
(\textit{bottom}) diagrams of the cluster NGC 2343
in the same format as Figure \ref{Basel1}.
\label{NGC2343}
}
\end{figure}

\begin{figure}
\includegraphics[angle=0,scale=0.4]{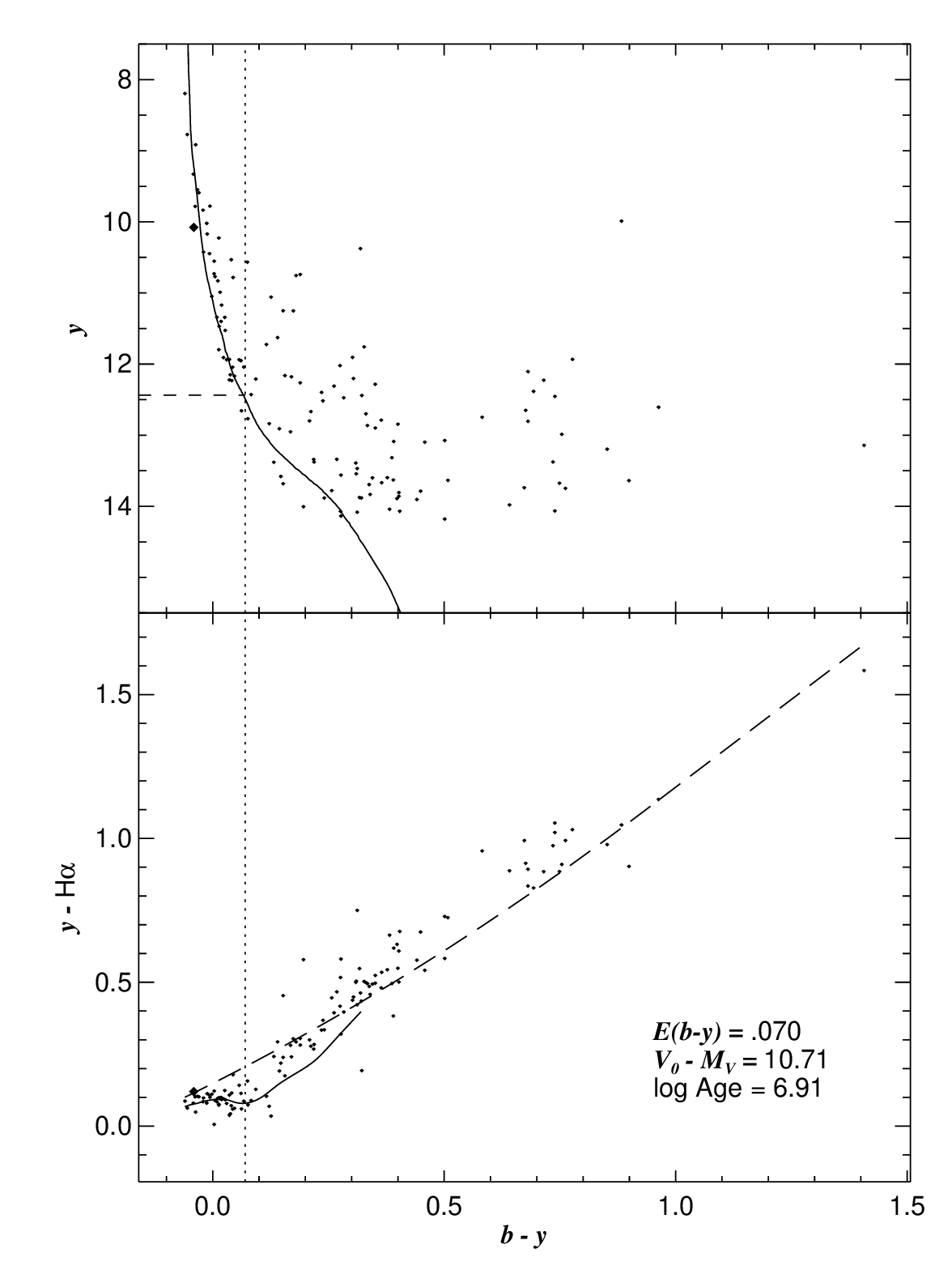}
\caption{
Color-magnitude (\textit{top}) and color-color 
(\textit{bottom}) diagrams of the cluster NGC 2362
in the same format as Figure \ref{Basel1}.
\label{NGC2362}
}
\end{figure}

\begin{figure}
\includegraphics[angle=0,scale=0.4]{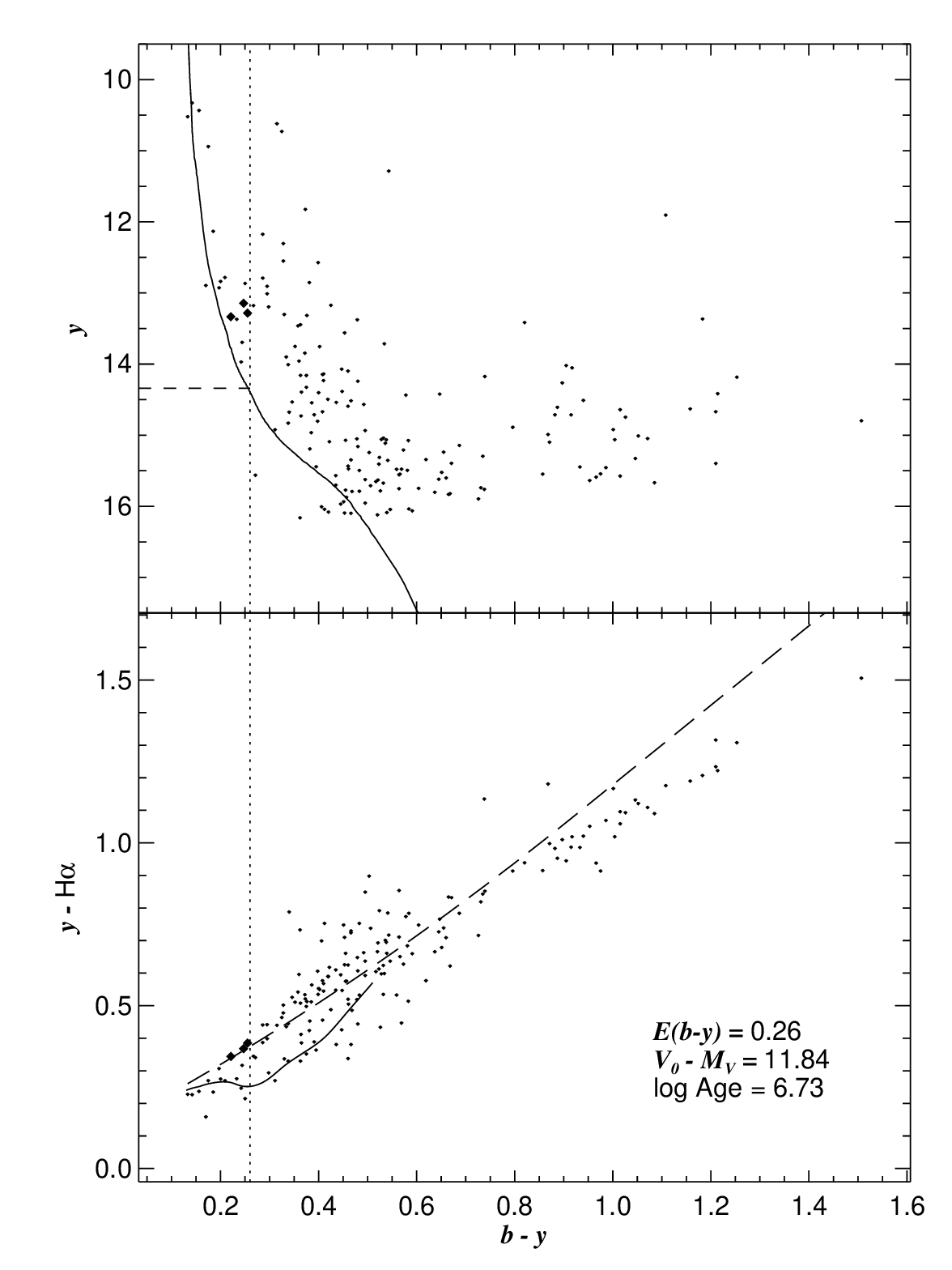}
\caption{
Color-magnitude (\textit{top}) and color-color 
(\textit{bottom}) diagrams of the cluster NGC 2367
in the same format as Figure \ref{Basel1}.
\label{NGC2367}
}
\end{figure}

\begin{figure}
\includegraphics[angle=0,scale=0.4]{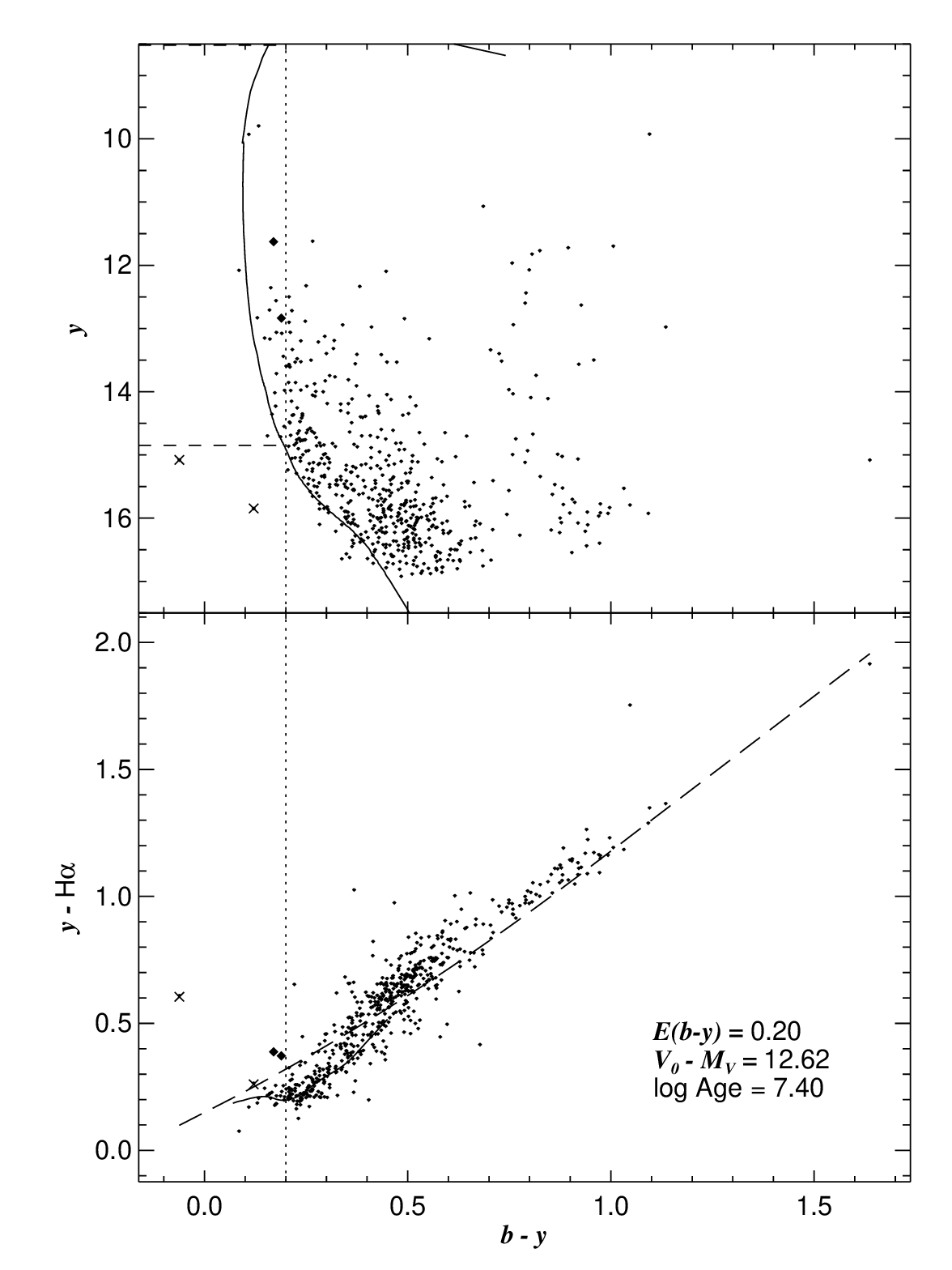}
\caption{
Color-magnitude (\textit{top}) and color-color 
(\textit{bottom}) diagrams of the cluster NGC 2383
in the same format as Figure \ref{Basel1}.
\label{NGC2383}
}
\end{figure}

\begin{figure}
\includegraphics[angle=0,scale=0.4]{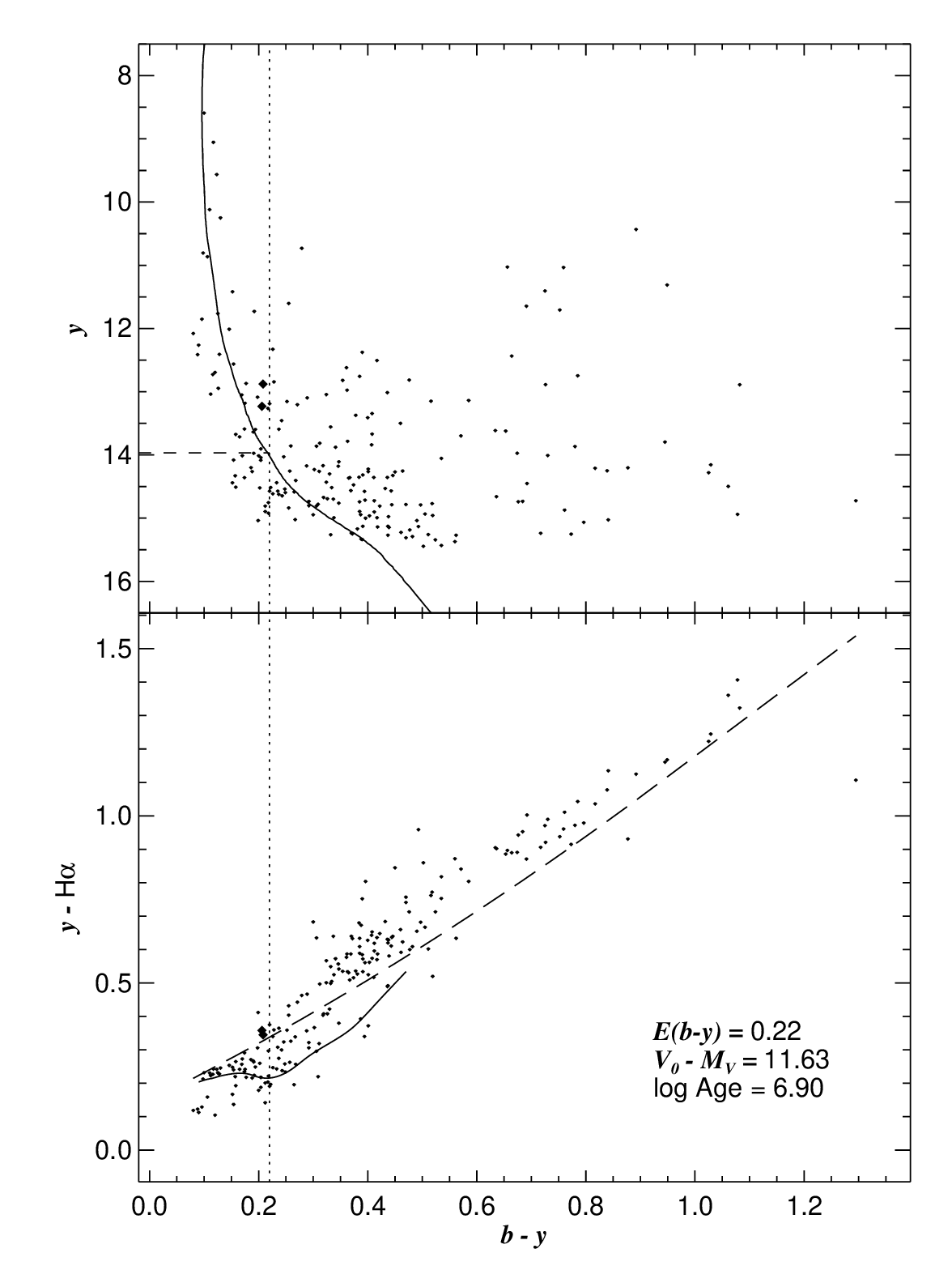}
\caption{
Color-magnitude (\textit{top}) and color-color 
(\textit{bottom}) diagrams of the cluster NGC 2384
in the same format as Figure \ref{Basel1}.
\label{NGC2384}
}
\end{figure}

\begin{figure}
\includegraphics[angle=0,scale=0.4]{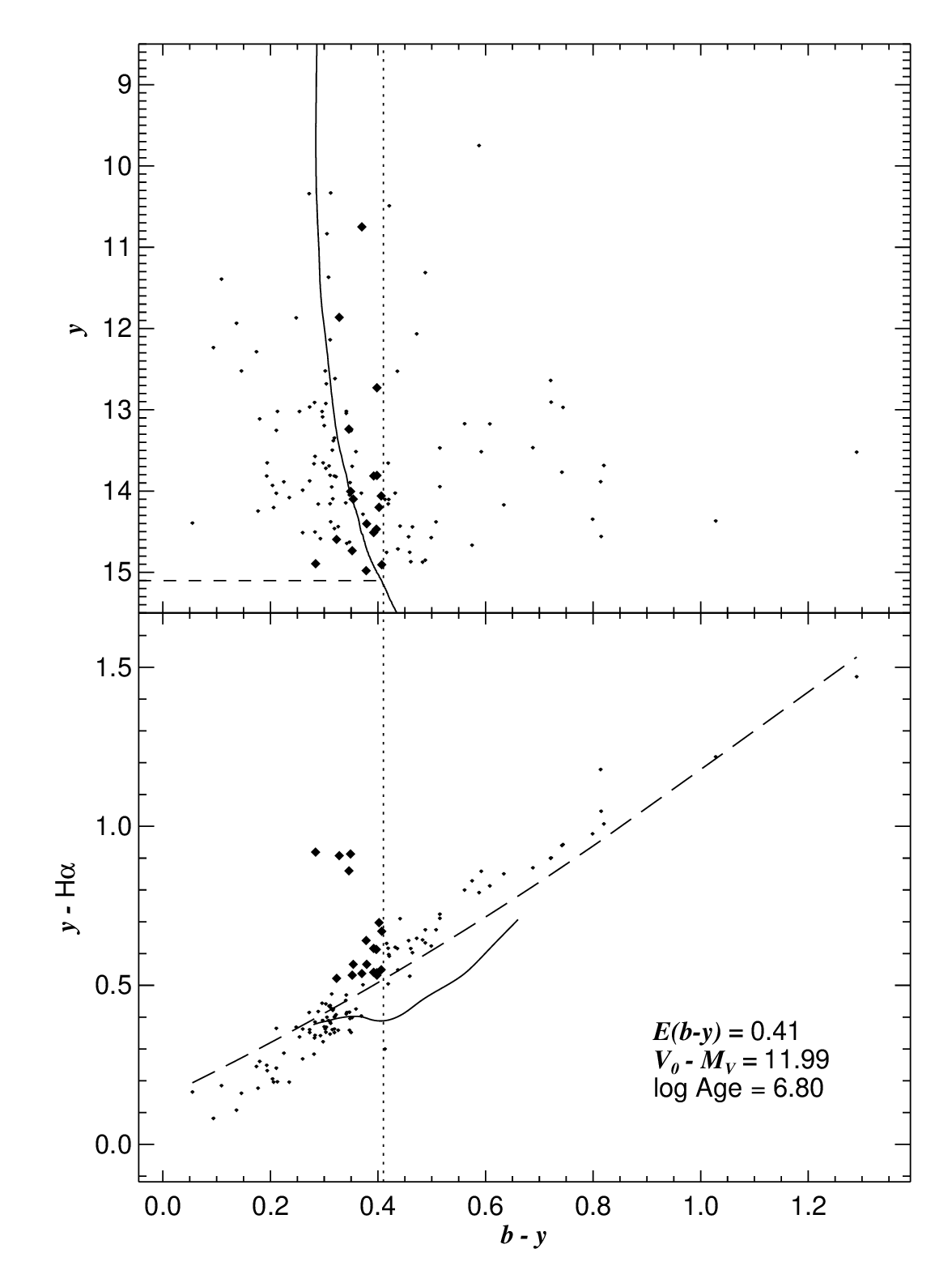}
\caption{
Color-magnitude (\textit{top}) and color-color 
(\textit{bottom}) diagrams of the cluster NGC 2414
in the same format as Figure \ref{Basel1}.
\label{NGC2414}
}
\end{figure}

\begin{figure}
\includegraphics[angle=0,scale=0.4]{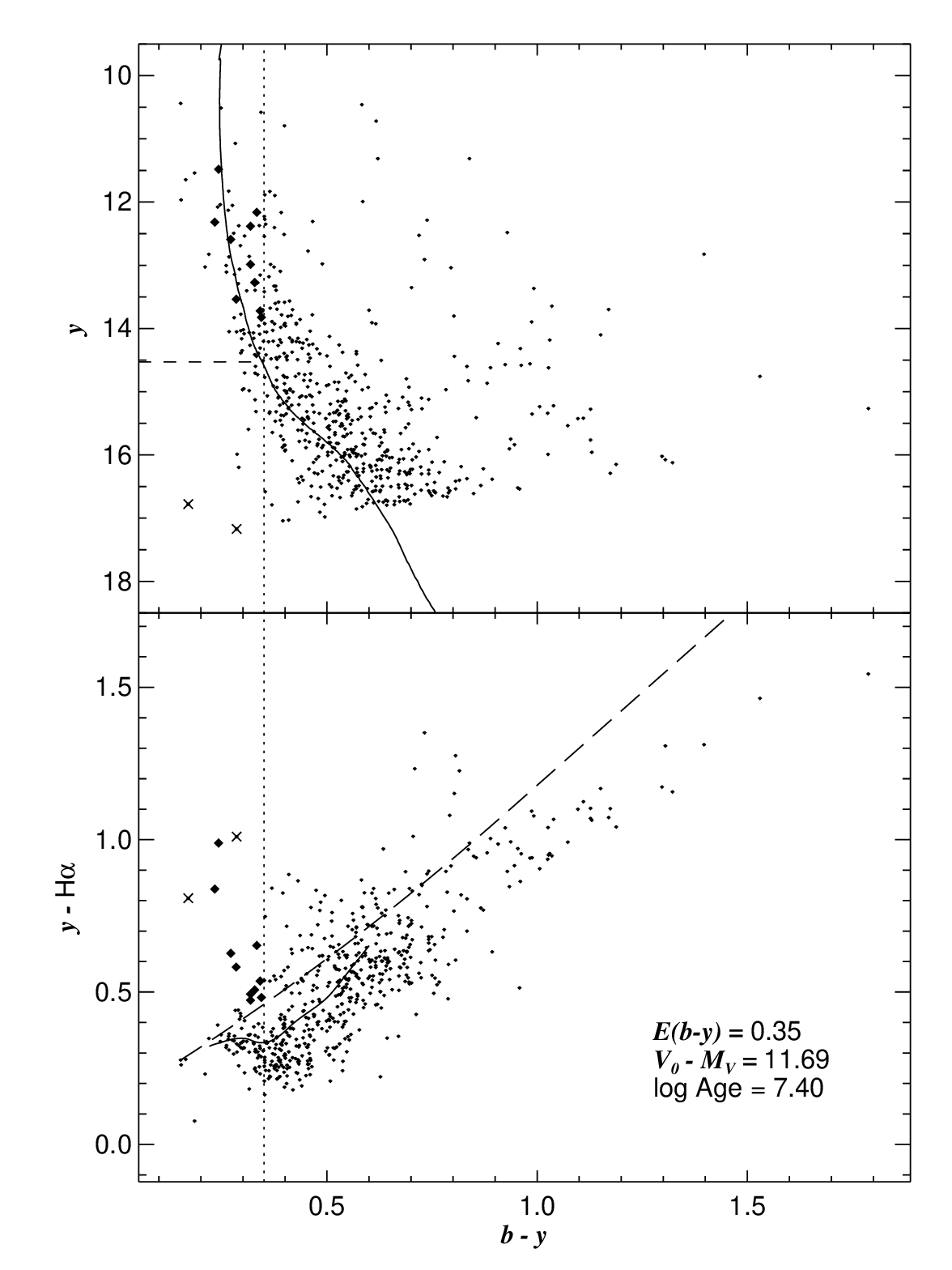}
\caption{
Color-magnitude (\textit{top}) and color-color 
(\textit{bottom}) diagrams of the cluster NGC 2421
in the same format as Figure \ref{Basel1}.
\label{NGC2421}
}
\end{figure}

\begin{figure}
\includegraphics[angle=0,scale=0.4]{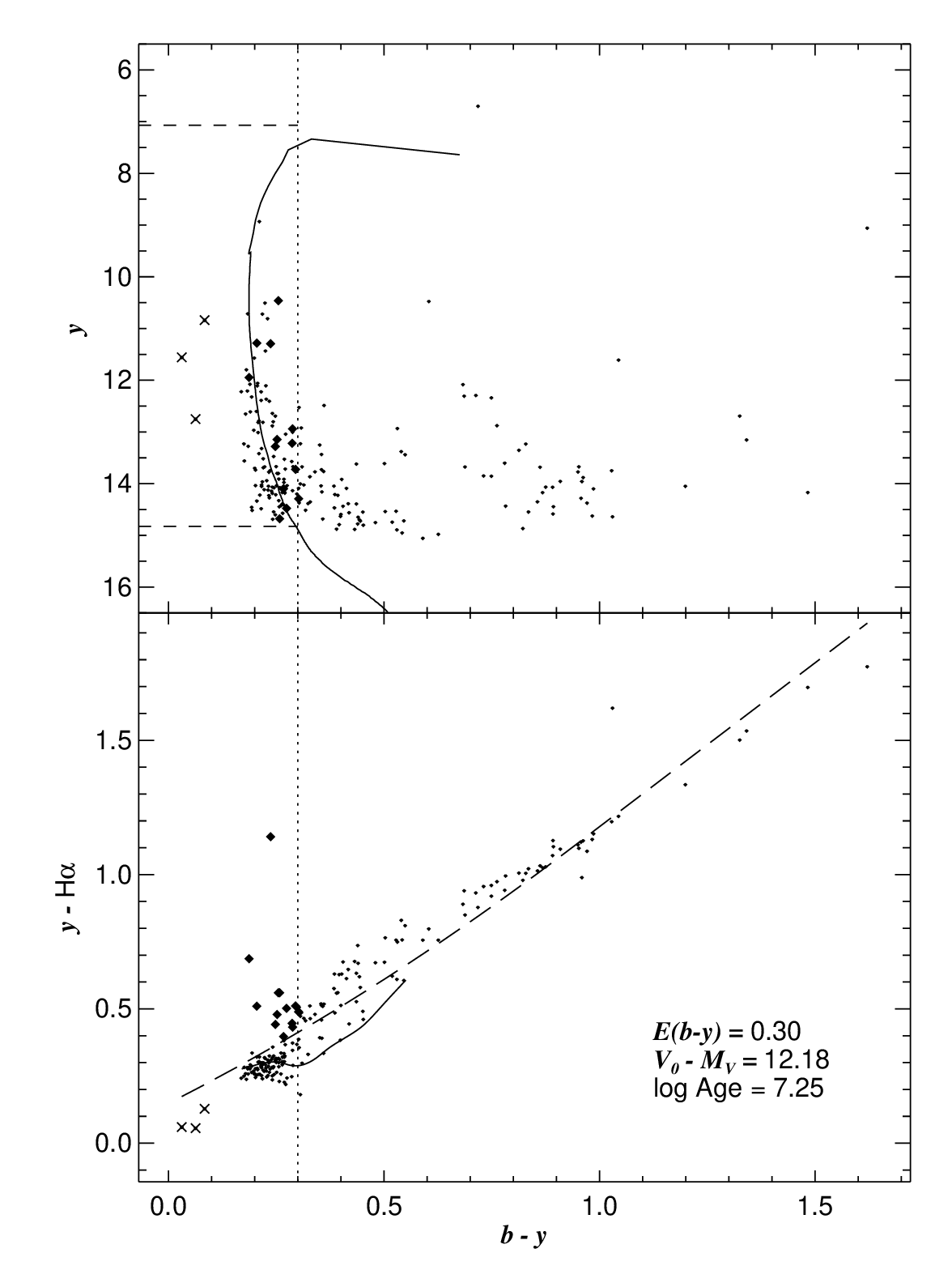}
\caption{
Color-magnitude (\textit{top}) and color-color 
(\textit{bottom}) diagrams of the cluster NGC 2439
in the same format as Figure \ref{Basel1}.
\label{NGC2439}
}
\end{figure}

\begin{figure}
\includegraphics[angle=0,scale=0.4]{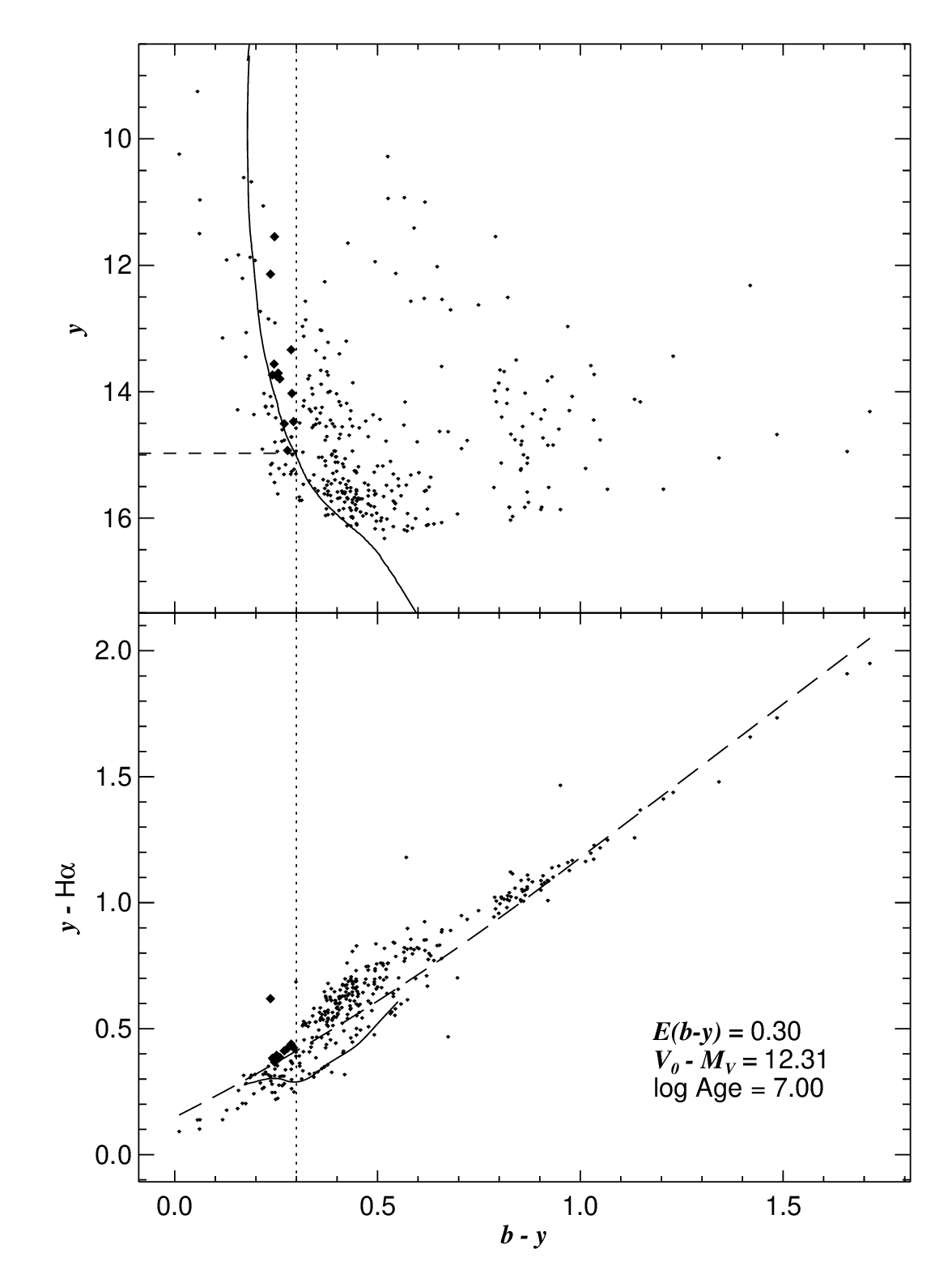}
\caption{
Color-magnitude (\textit{top}) and color-color 
(\textit{bottom}) diagrams of the cluster NGC 2483
in the same format as Figure \ref{Basel1}.
\label{NGC2483}
}
\end{figure}

\begin{figure}
\includegraphics[angle=0,scale=0.4]{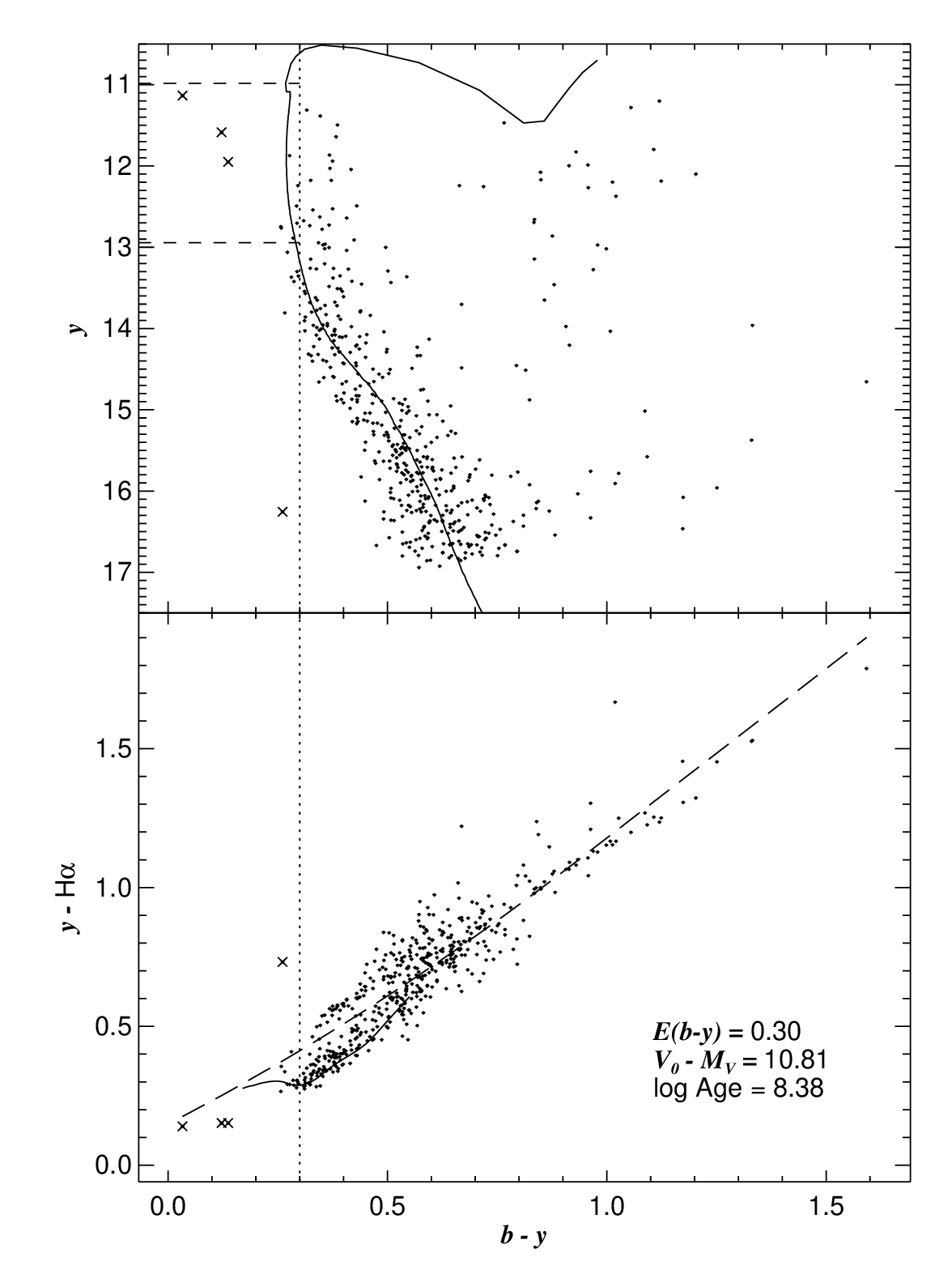}
\caption{
Color-magnitude (\textit{top}) and color-color 
(\textit{bottom}) diagrams of the cluster NGC 2489
in the same format as Figure \ref{Basel1}.
\label{NGC2489}
}
\end{figure}

\begin{figure}
\includegraphics[angle=0,scale=0.4]{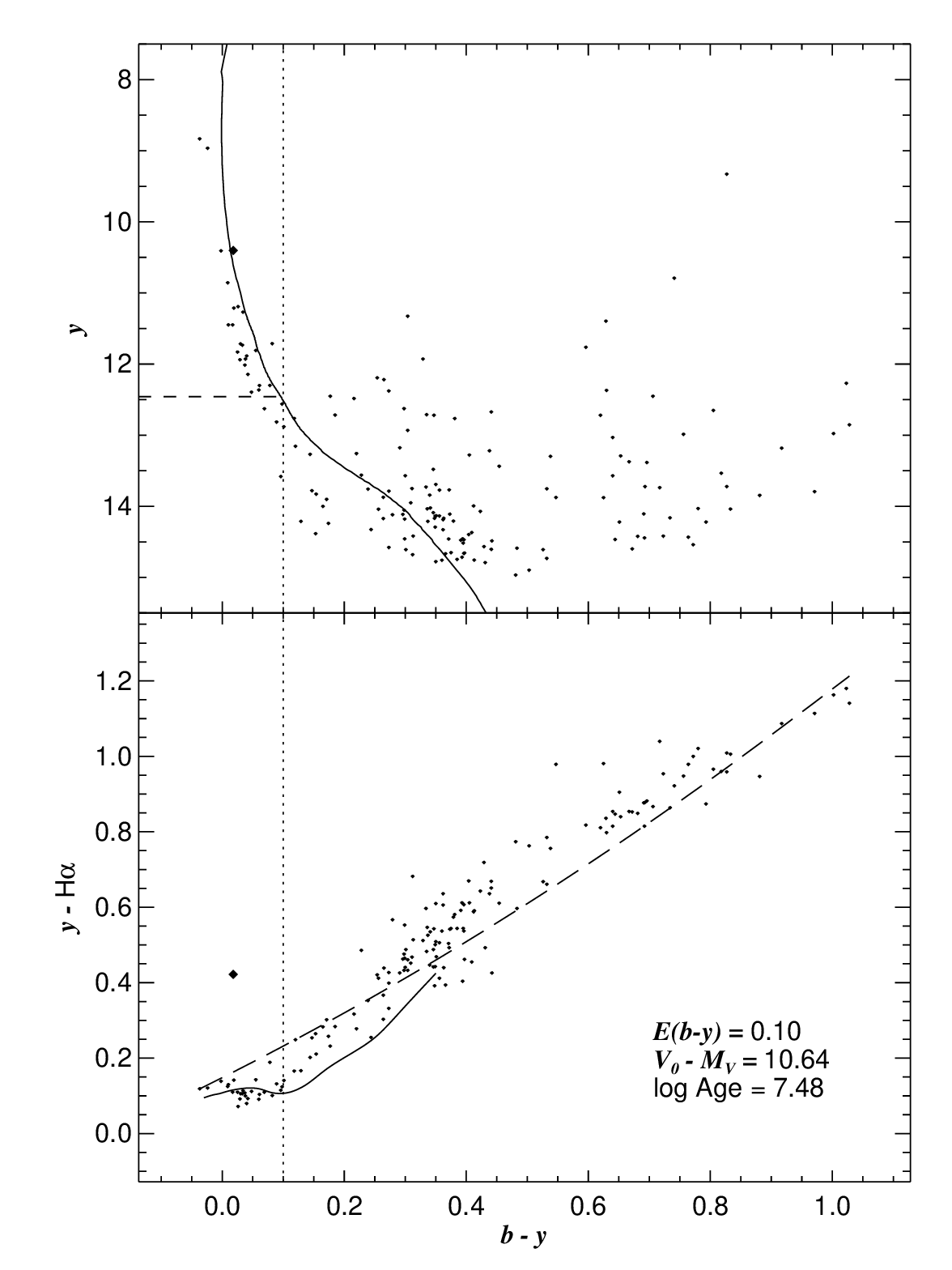}
\caption{
Color-magnitude (\textit{top}) and color-color 
(\textit{bottom}) diagrams of the cluster NGC 2571
in the same format as Figure \ref{Basel1}.
\label{NGC2571}
}
\end{figure}

\begin{figure}
\includegraphics[angle=0,scale=0.4]{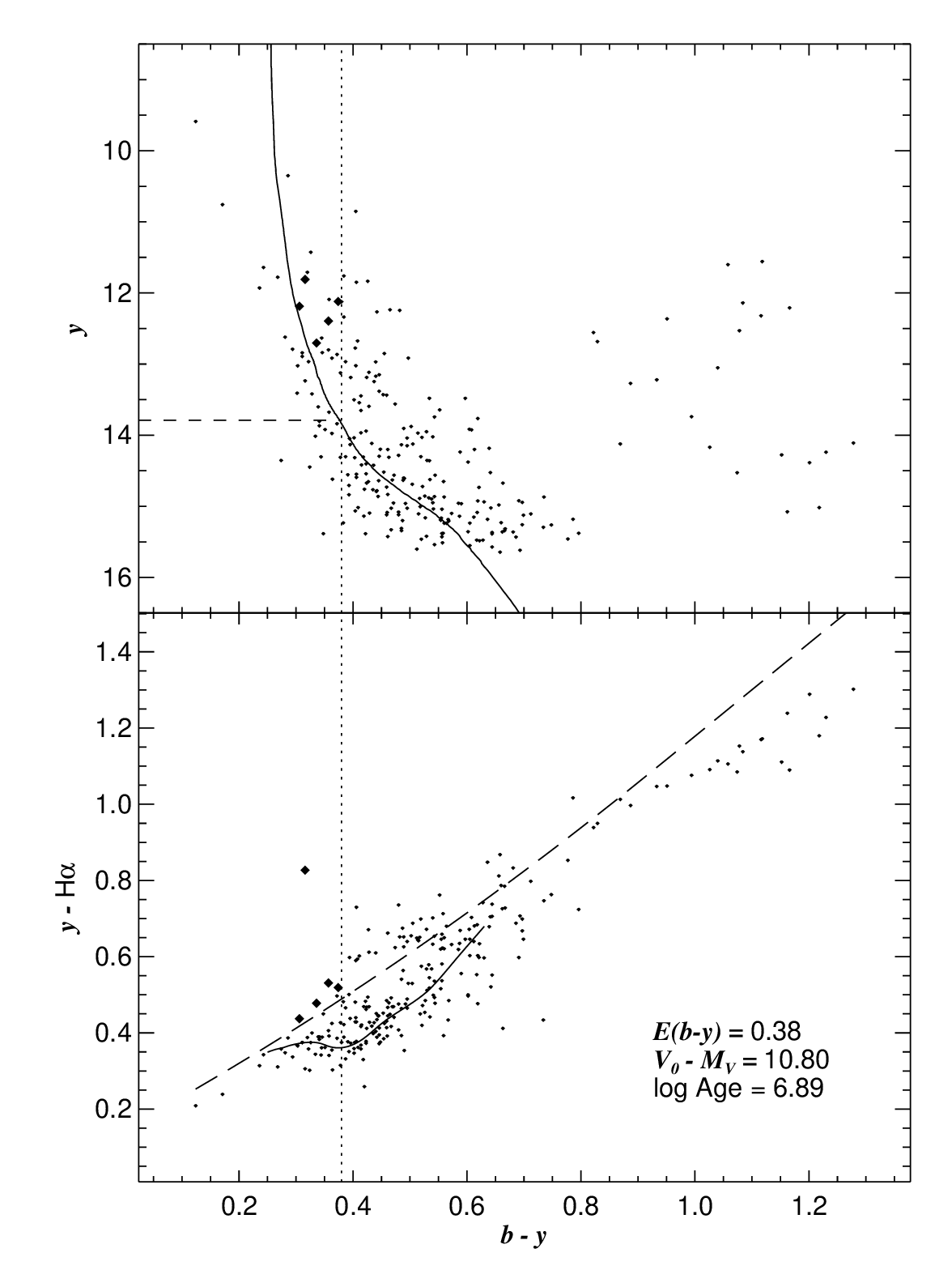}
\caption{
Color-magnitude (\textit{top}) and color-color 
(\textit{bottom}) diagrams of the cluster NGC 2659
in the same format as Figure \ref{Basel1}.
\label{NGC2659}
}
\end{figure}

\begin{figure}
\includegraphics[angle=0,scale=0.4]{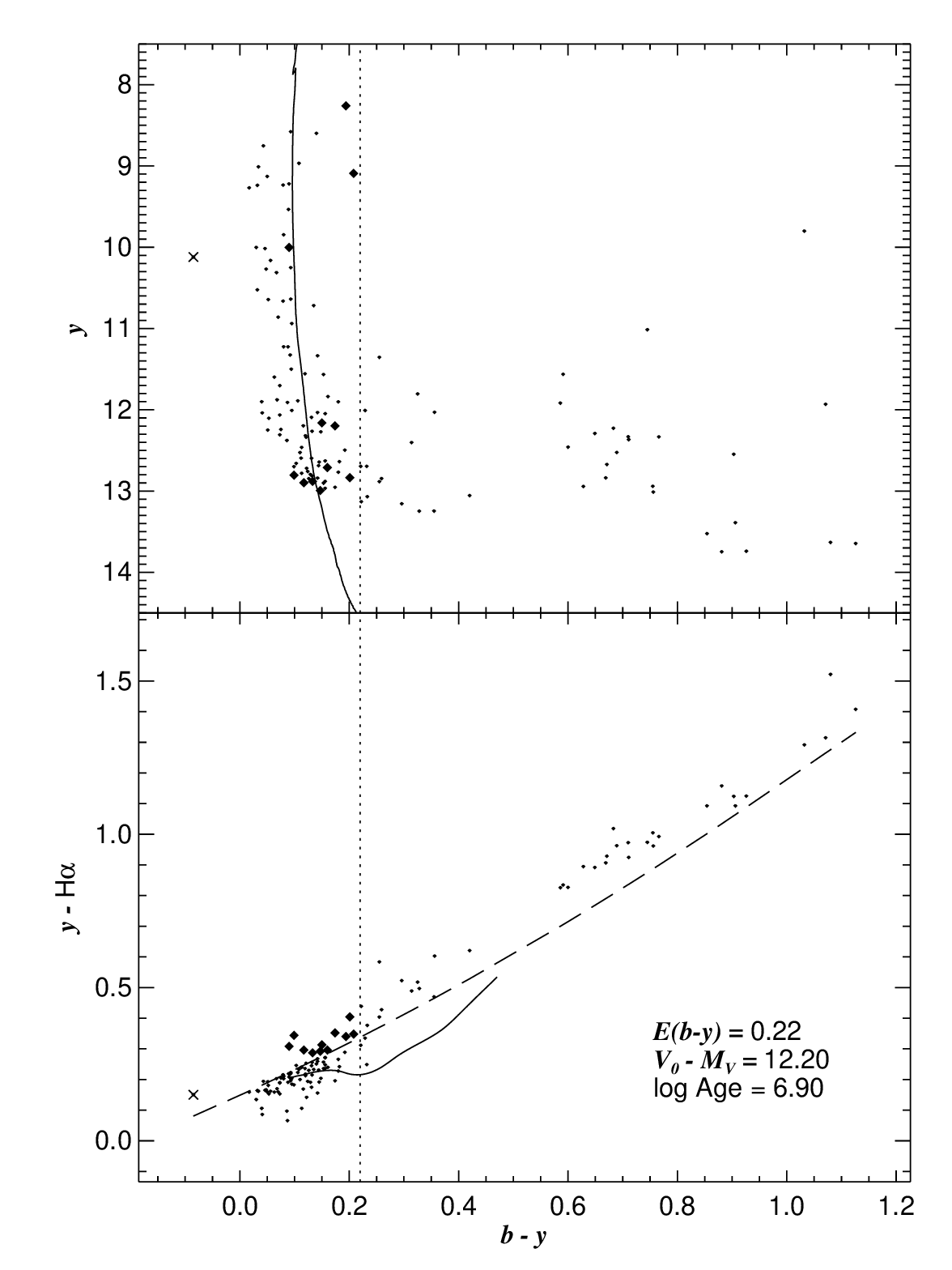}
\caption{
Color-magnitude (\textit{top}) and color-color 
(\textit{bottom}) diagrams of the cluster NGC 3293
in the same format as Figure \ref{Basel1}.
\label{NGC3293}
}
\end{figure}

\begin{figure}
\includegraphics[angle=0,scale=0.4]{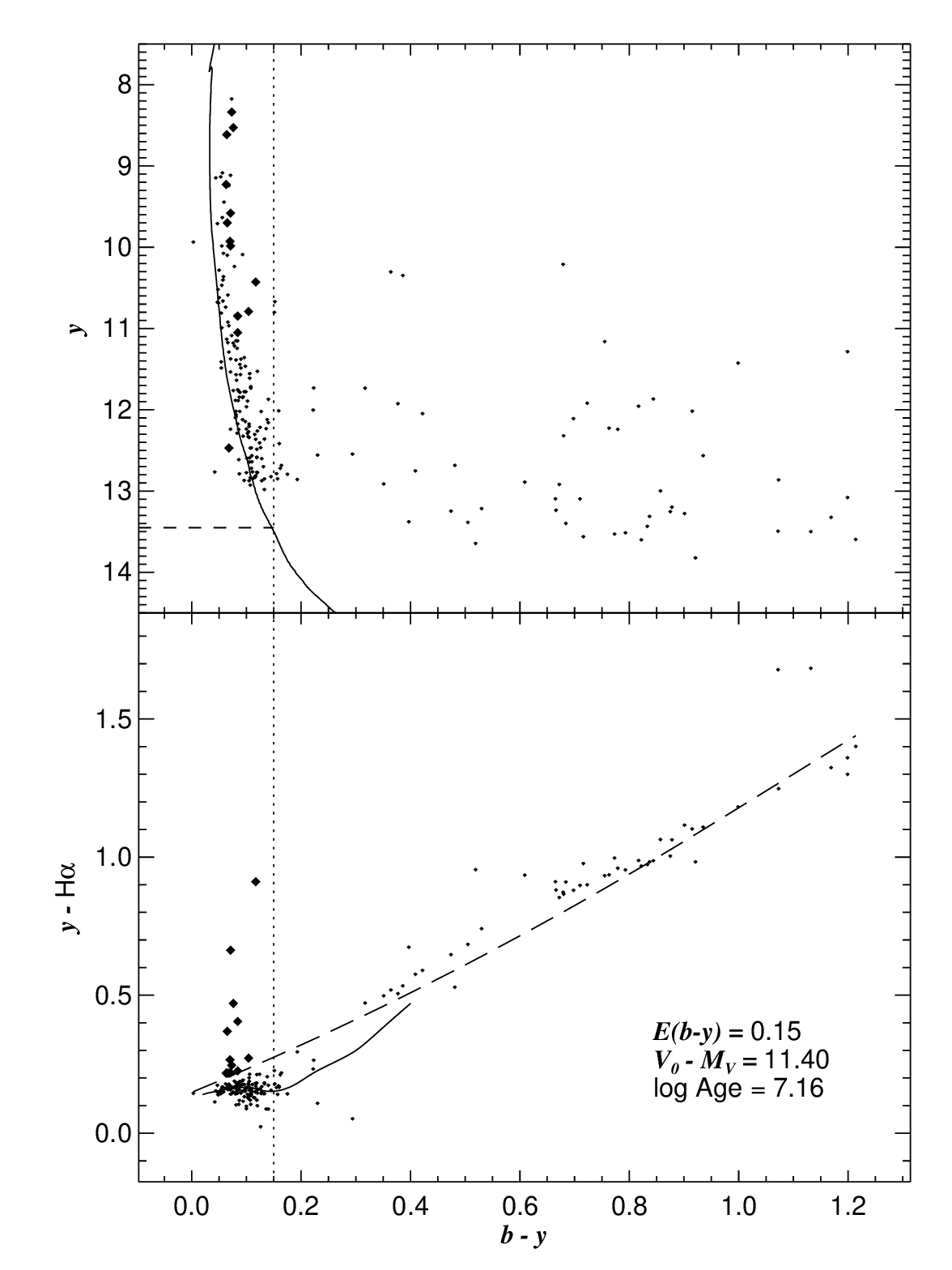}
\caption{
Color-magnitude (\textit{top}) and color-color 
(\textit{bottom}) diagrams of the cluster NGC 3766
in the same format as Figure \ref{Basel1}.
\label{NGC3766}
}
\end{figure}

\begin{figure}
\includegraphics[angle=0,scale=0.4]{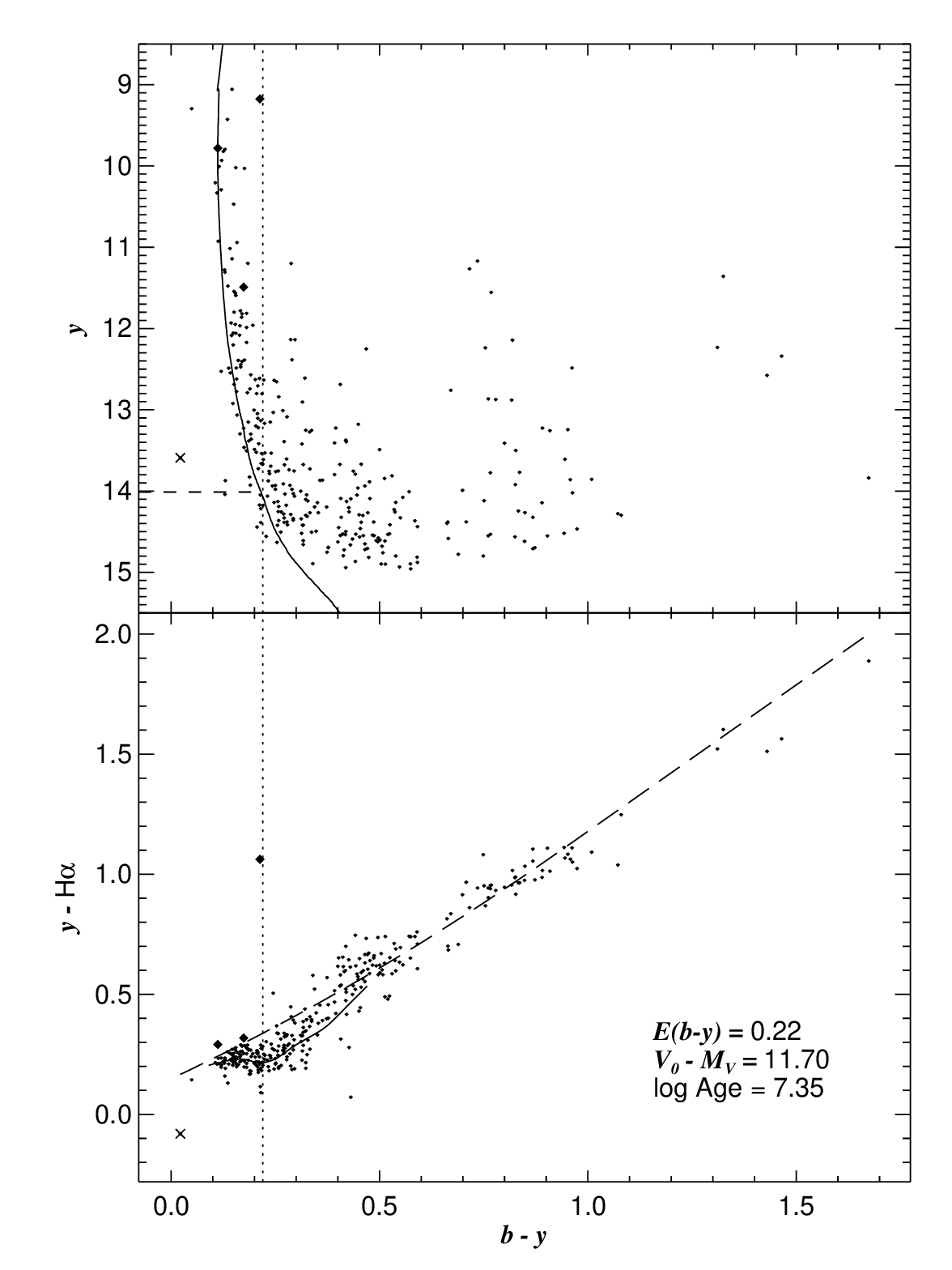}
\caption{
Color-magnitude (\textit{top}) and color-color 
(\textit{bottom}) diagrams of the cluster NGC 4103
in the same format as Figure \ref{Basel1}.
\label{NGC4103}
}
\end{figure}

\begin{figure}
\includegraphics[angle=0,scale=0.4]{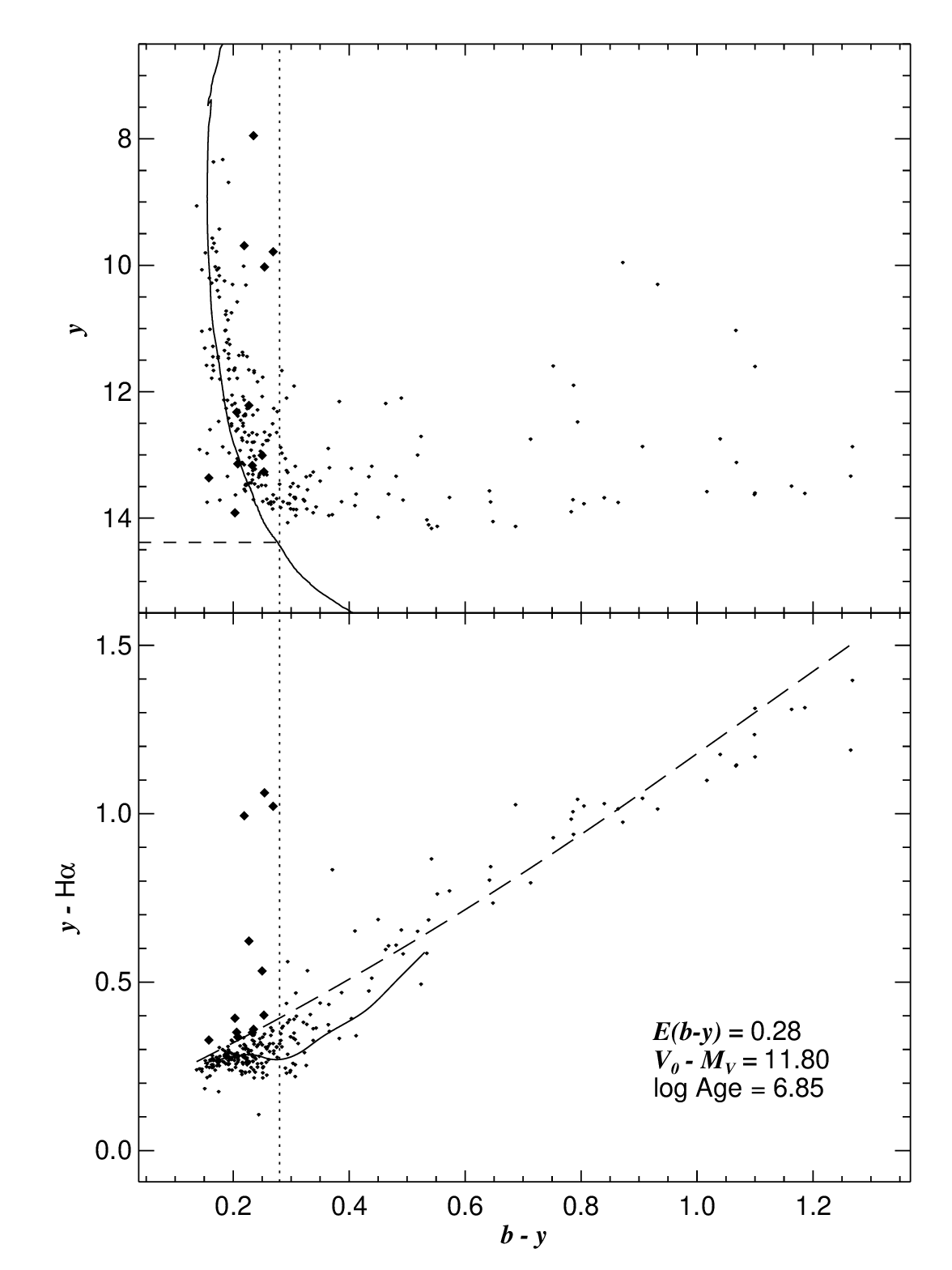}
\caption{
Color-magnitude (\textit{top}) and color-color 
(\textit{bottom}) diagrams of the cluster NGC 4755
in the same format as Figure \ref{Basel1}.
\label{NGC4755}
}
\end{figure}

\begin{figure}
\includegraphics[angle=0,scale=0.4]{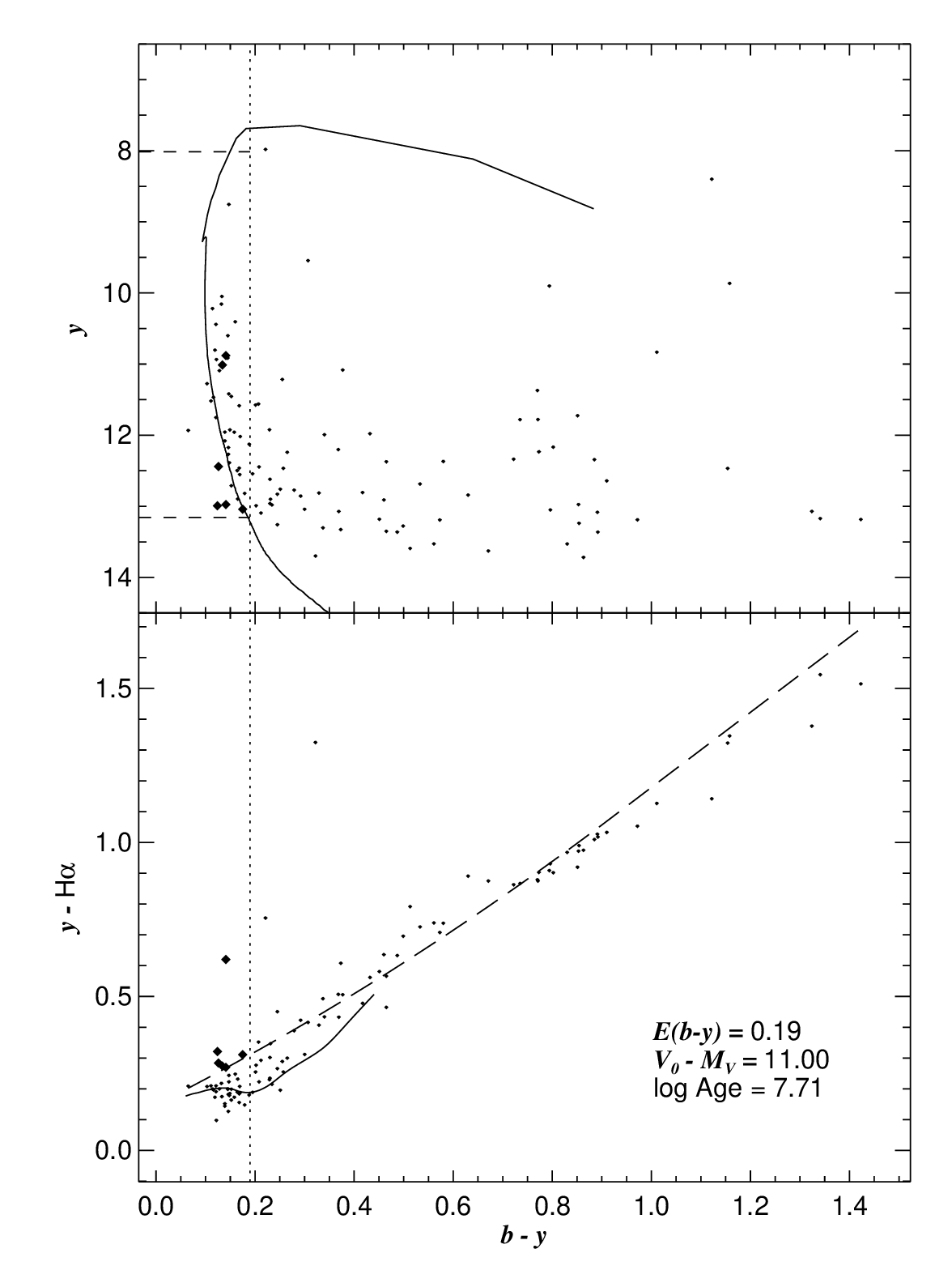}
\caption{
Color-magnitude (\textit{top}) and color-color 
(\textit{bottom}) diagrams of the cluster NGC 5281
in the same format as Figure \ref{Basel1}.
\label{NGC5281}
}
\end{figure}

\begin{figure}
\includegraphics[angle=0,scale=0.4]{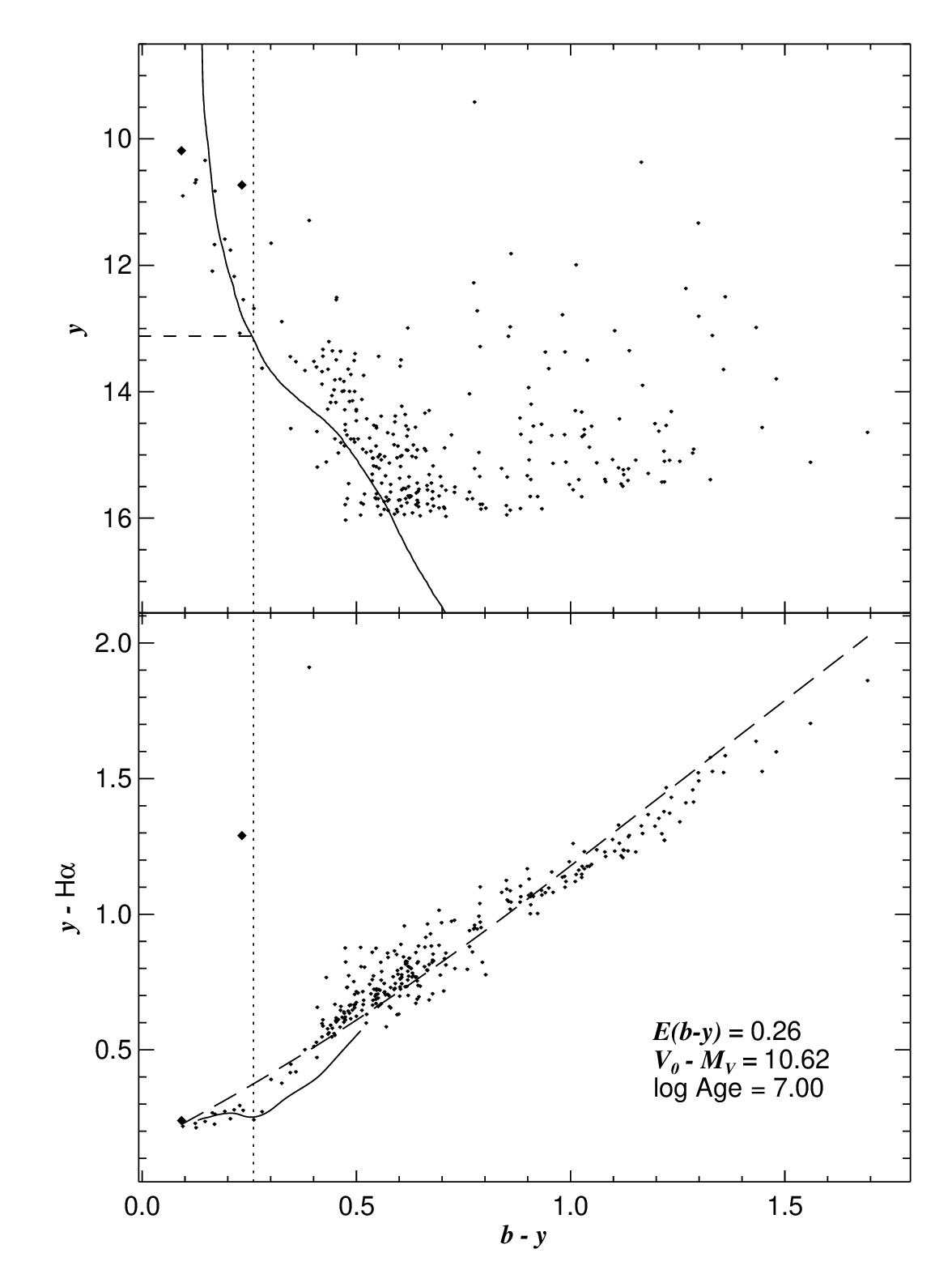}
\caption{
Color-magnitude (\textit{top}) and color-color 
(\textit{bottom}) diagrams of the cluster NGC 5593
in the same format as Figure \ref{Basel1}.
\label{NGC5593}
}
\end{figure}

\begin{figure}
\includegraphics[angle=0,scale=0.4]{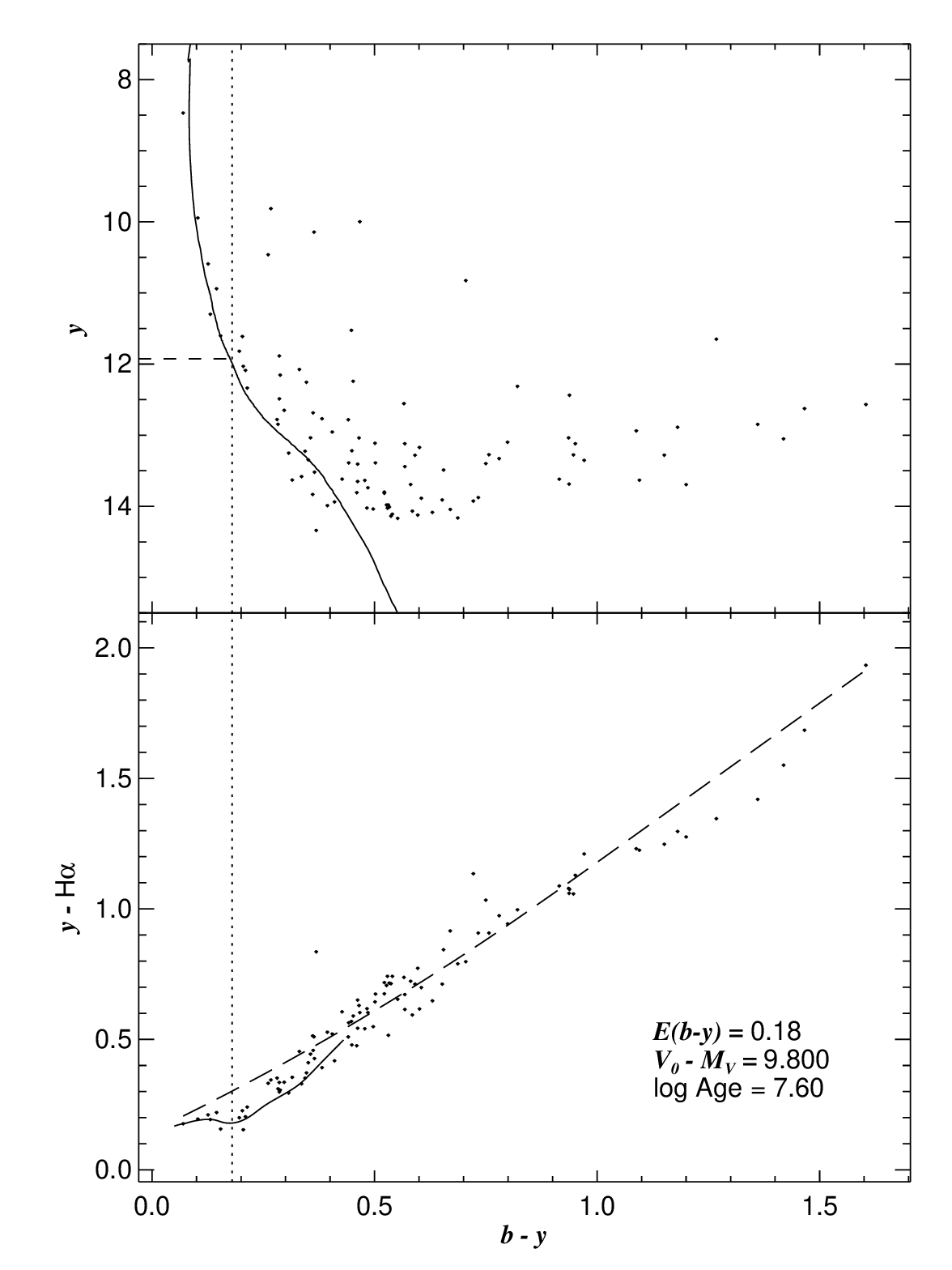}
\caption{
Color-magnitude (\textit{top}) and color-color 
(\textit{bottom}) diagrams of the cluster NGC 6178
in the same format as Figure \ref{Basel1}.
\label{NGC6178}
}
\end{figure}

\begin{figure}
\includegraphics[angle=0,scale=0.4]{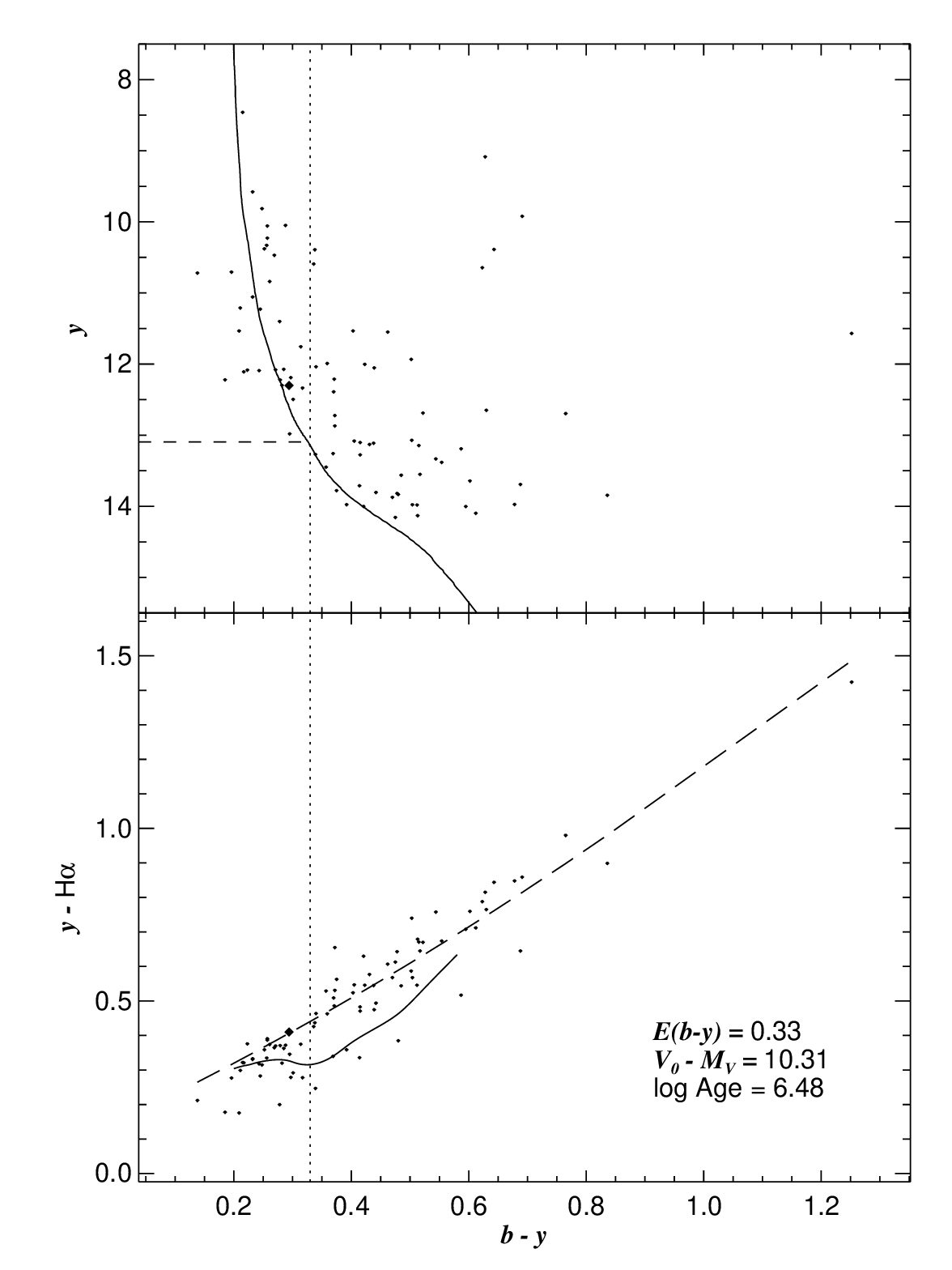}
\caption{
Color-magnitude (\textit{top}) and color-color 
(\textit{bottom}) diagrams of the cluster NGC 6193
in the same format as Figure \ref{Basel1}.
\label{NGC6193}
}
\end{figure}

\begin{figure}
\includegraphics[angle=0,scale=0.4]{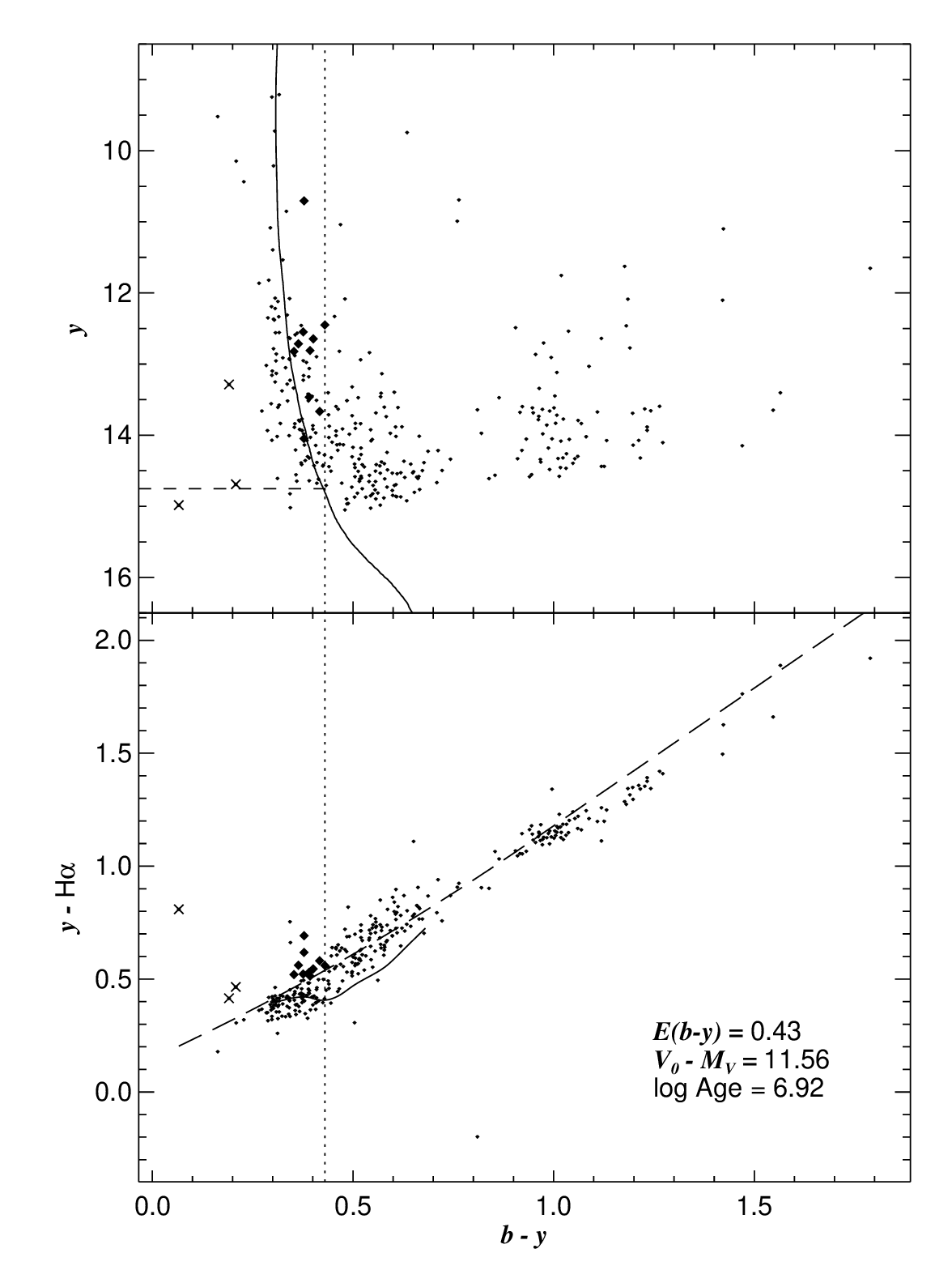}
\caption{
Color-magnitude (\textit{top}) and color-color 
(\textit{bottom}) diagrams of the cluster NGC 6200
in the same format as Figure \ref{Basel1}.
\label{NGC6200}
}
\end{figure}

\begin{figure}
\includegraphics[angle=0,scale=0.4]{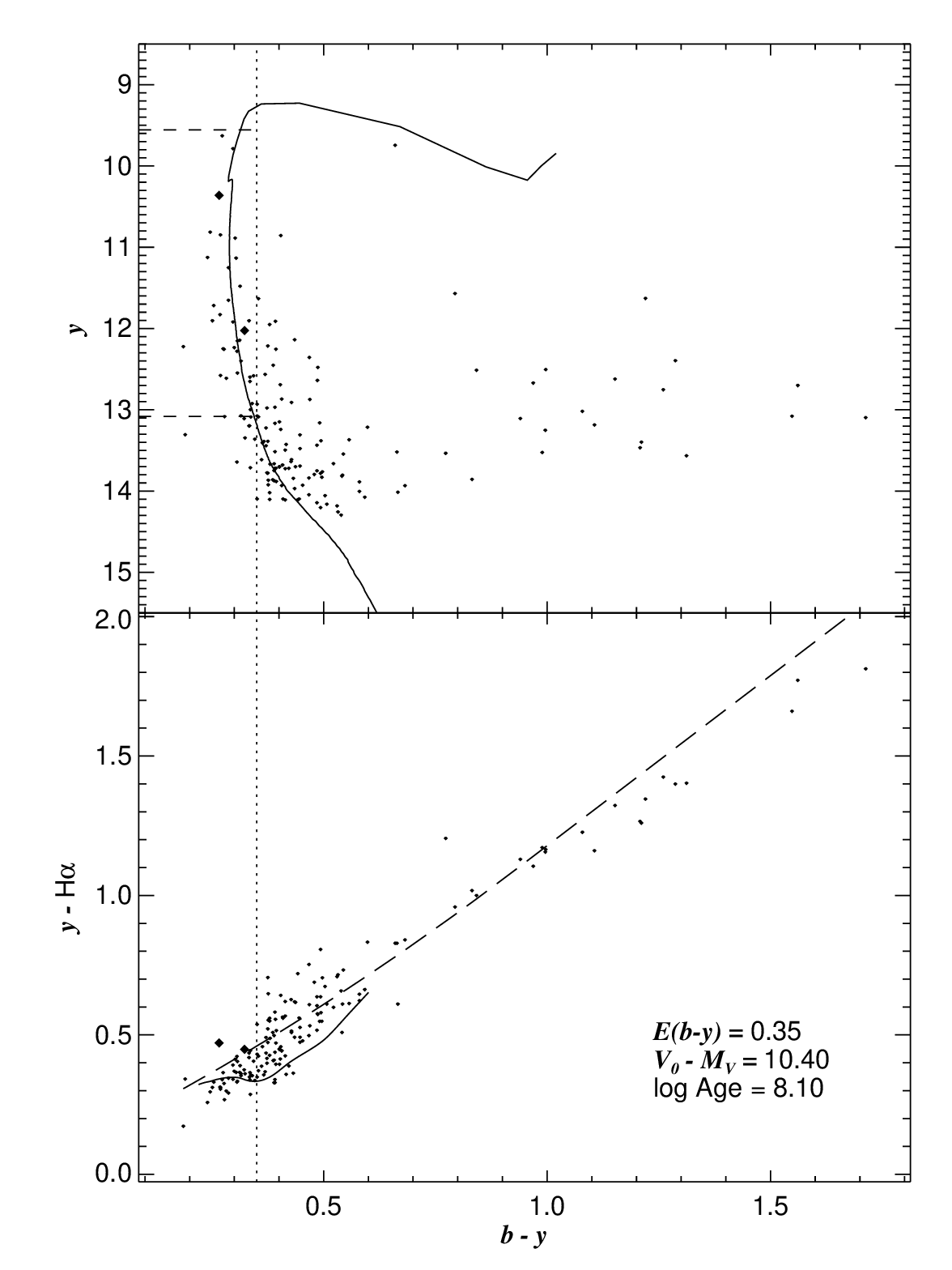}
\caption{
Color-magnitude (\textit{top}) and color-color 
(\textit{bottom}) diagrams of the cluster NGC 6204
in the same format as Figure \ref{Basel1}.
\label{NGC6204}
}
\end{figure}

\begin{figure}
\includegraphics[angle=0,scale=0.4]{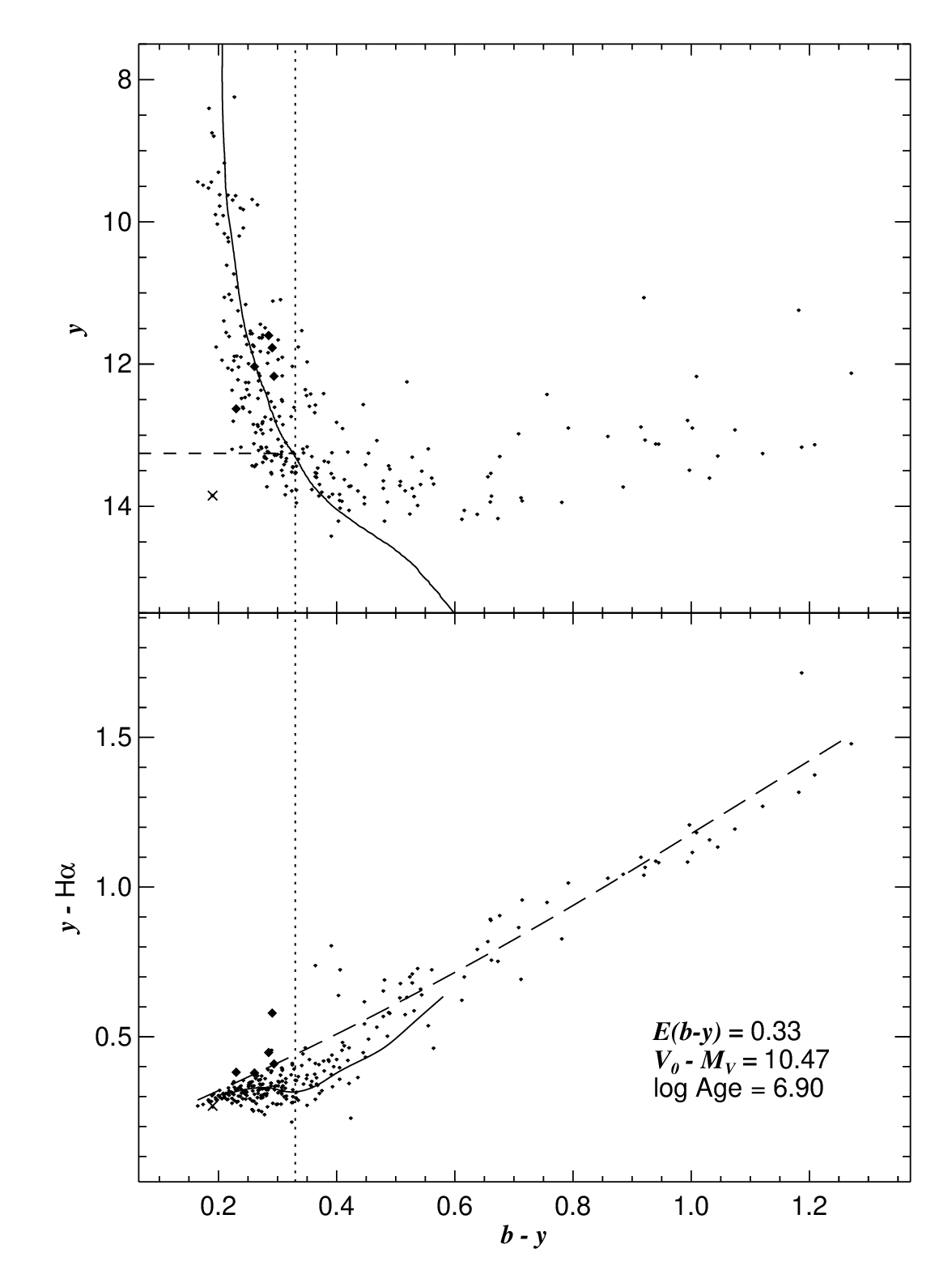}
\caption{
Color-magnitude (\textit{top}) and color-color 
(\textit{bottom}) diagrams of the cluster NGC 6231
in the same format as Figure \ref{Basel1}.
\label{NGC6231}
}
\end{figure}

\begin{figure}
\includegraphics[angle=0,scale=0.4]{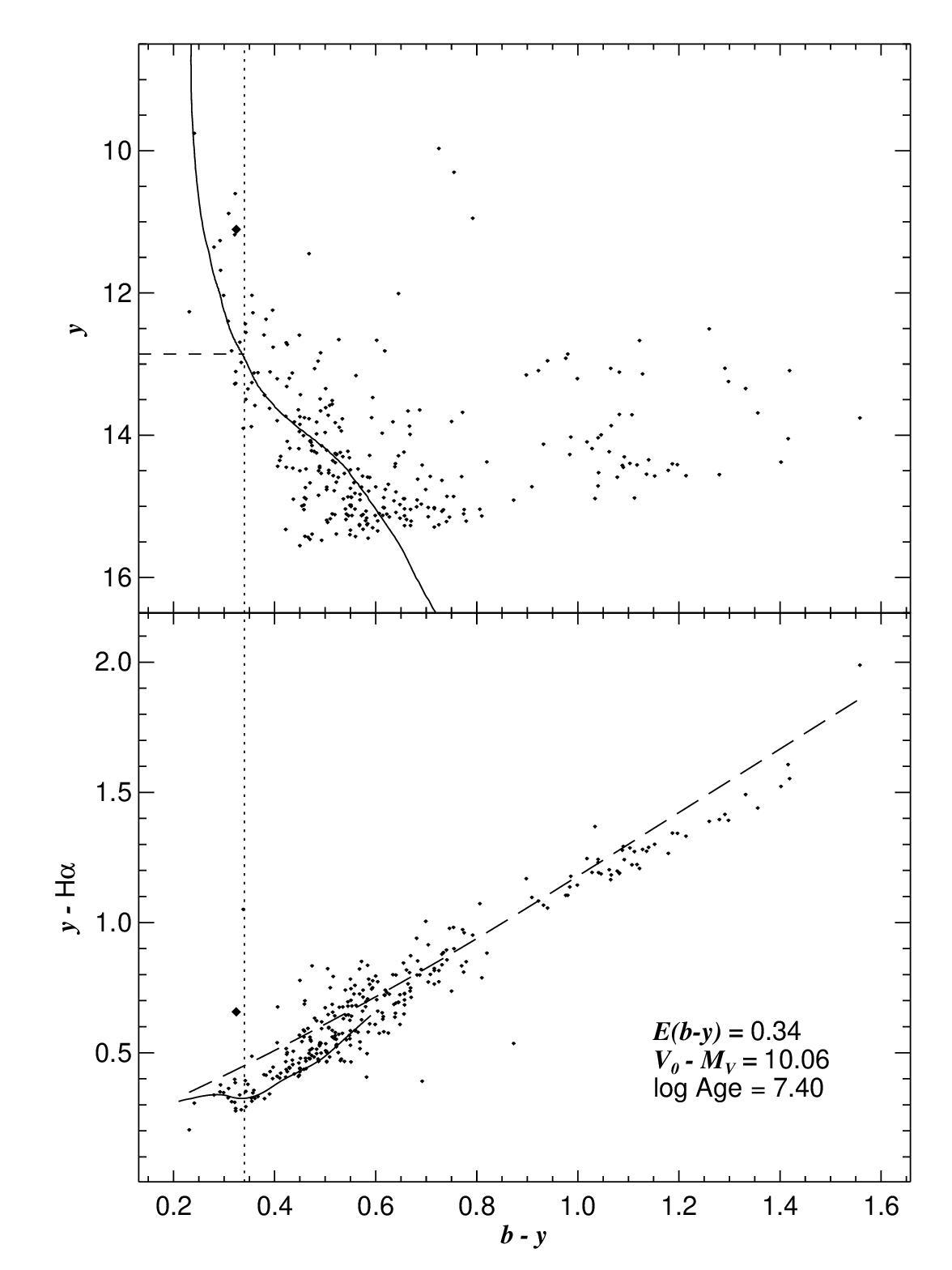}
\caption{
Color-magnitude (\textit{top}) and color-color 
(\textit{bottom}) diagrams of the cluster NGC 6249
in the same format as Figure \ref{Basel1}.
\label{NGC6249}
}
\end{figure}

\begin{figure}
\includegraphics[angle=0,scale=0.4]{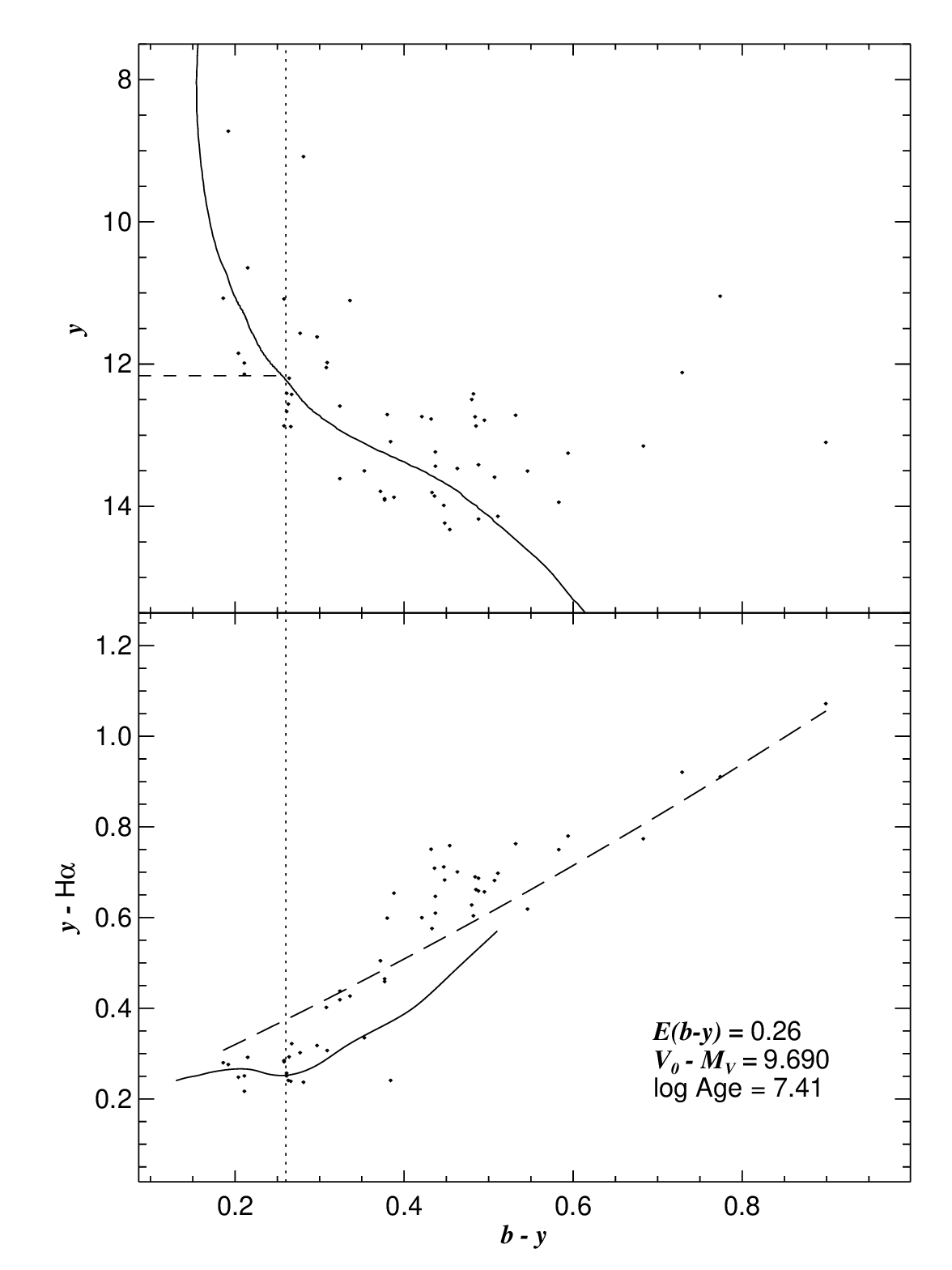}
\caption{
Color-magnitude (\textit{top}) and color-color 
(\textit{bottom}) diagrams of the cluster NGC 6250
in the same format as Figure \ref{Basel1}.
\label{NGC6250}
}
\end{figure}

\begin{figure}
\includegraphics[angle=0,scale=0.4]{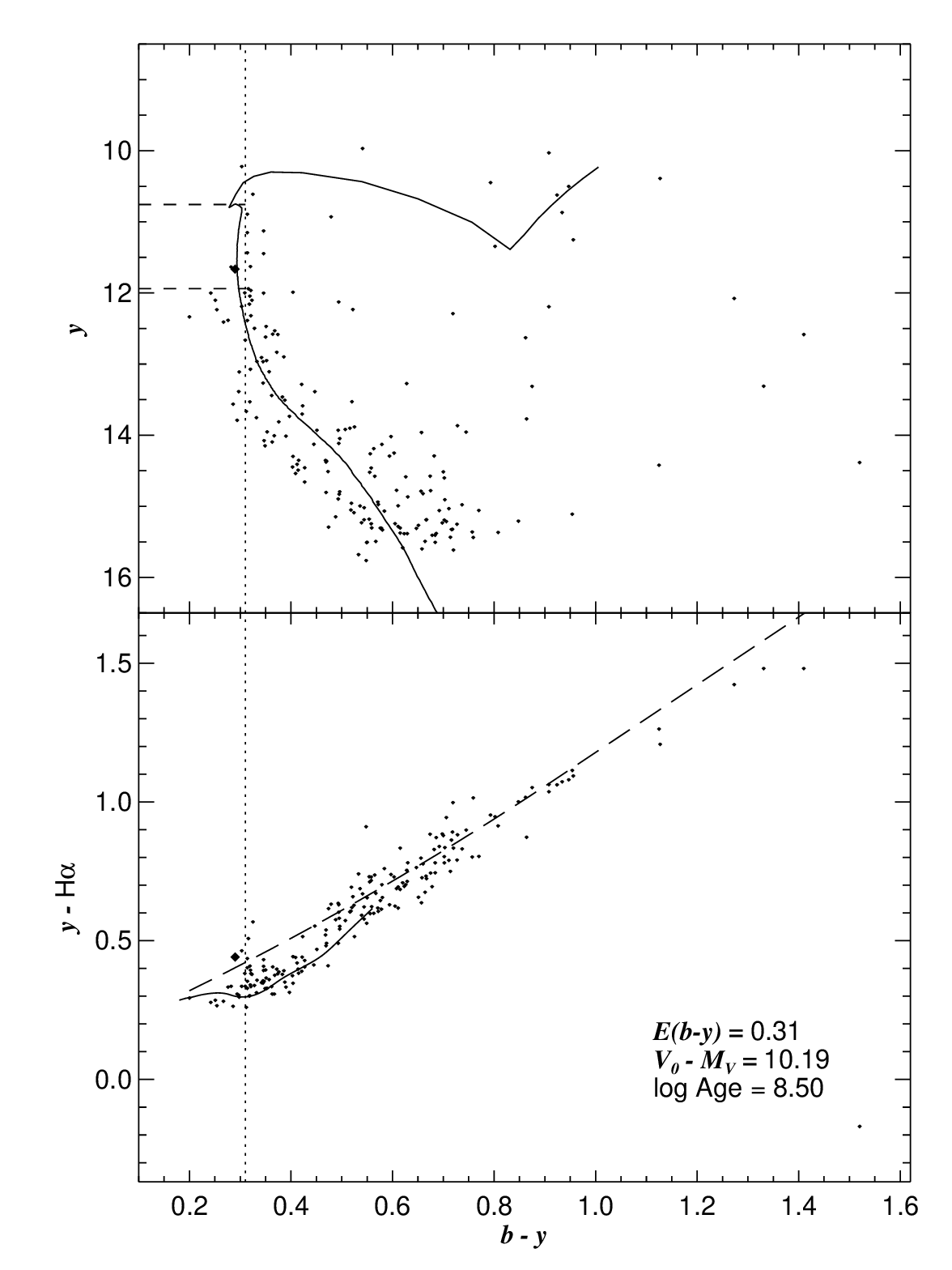}
\caption{
Color-magnitude (\textit{top}) and color-color 
(\textit{bottom}) diagrams of the cluster NGC 6268
in the same format as Figure \ref{Basel1}.
\label{NGC6268}
}
\end{figure}

\begin{figure}
\includegraphics[angle=0,scale=0.4]{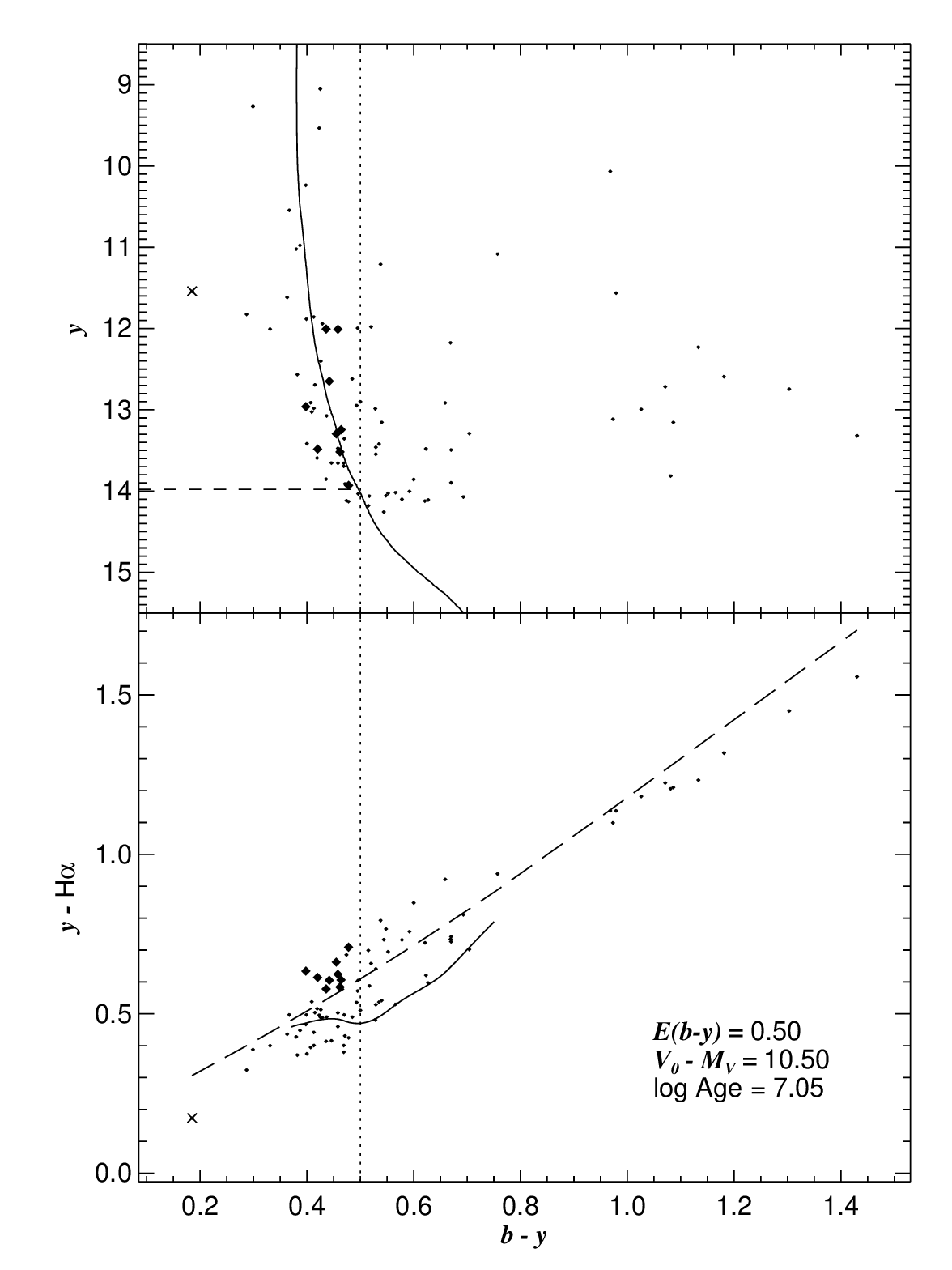}
\caption{
Color-magnitude (\textit{top}) and color-color 
(\textit{bottom}) diagrams of the cluster NGC 6322
in the same format as Figure \ref{Basel1}.
\label{NGC6322}
}
\end{figure}

\begin{figure}
\includegraphics[angle=0,scale=0.4]{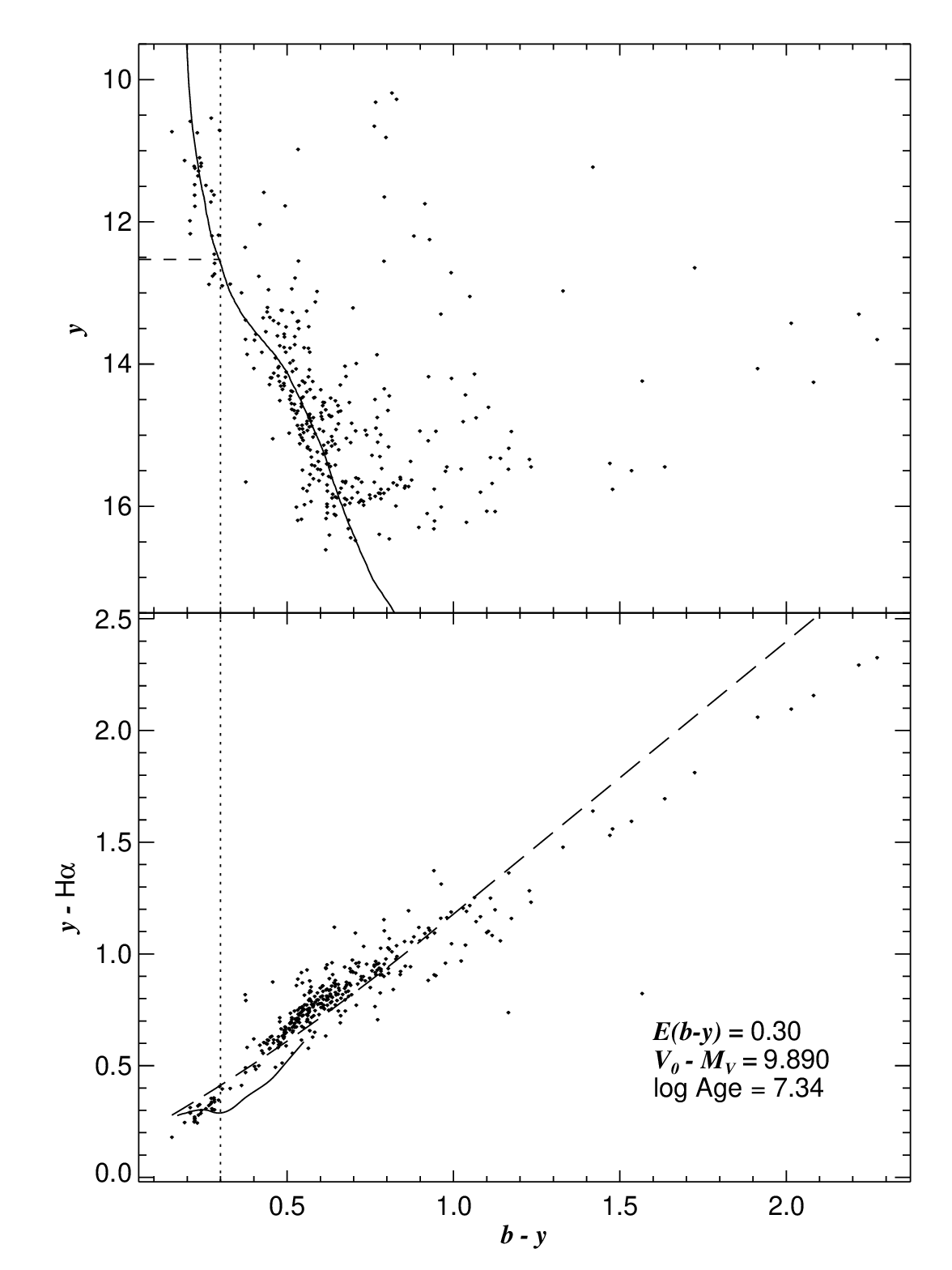}
\caption{
Color-magnitude (\textit{top}) and color-color 
(\textit{bottom}) diagrams of the cluster NGC 6425
in the same format as Figure \ref{Basel1}.
\label{NGC6425}
}
\end{figure}

\begin{figure}
\includegraphics[angle=0,scale=0.4]{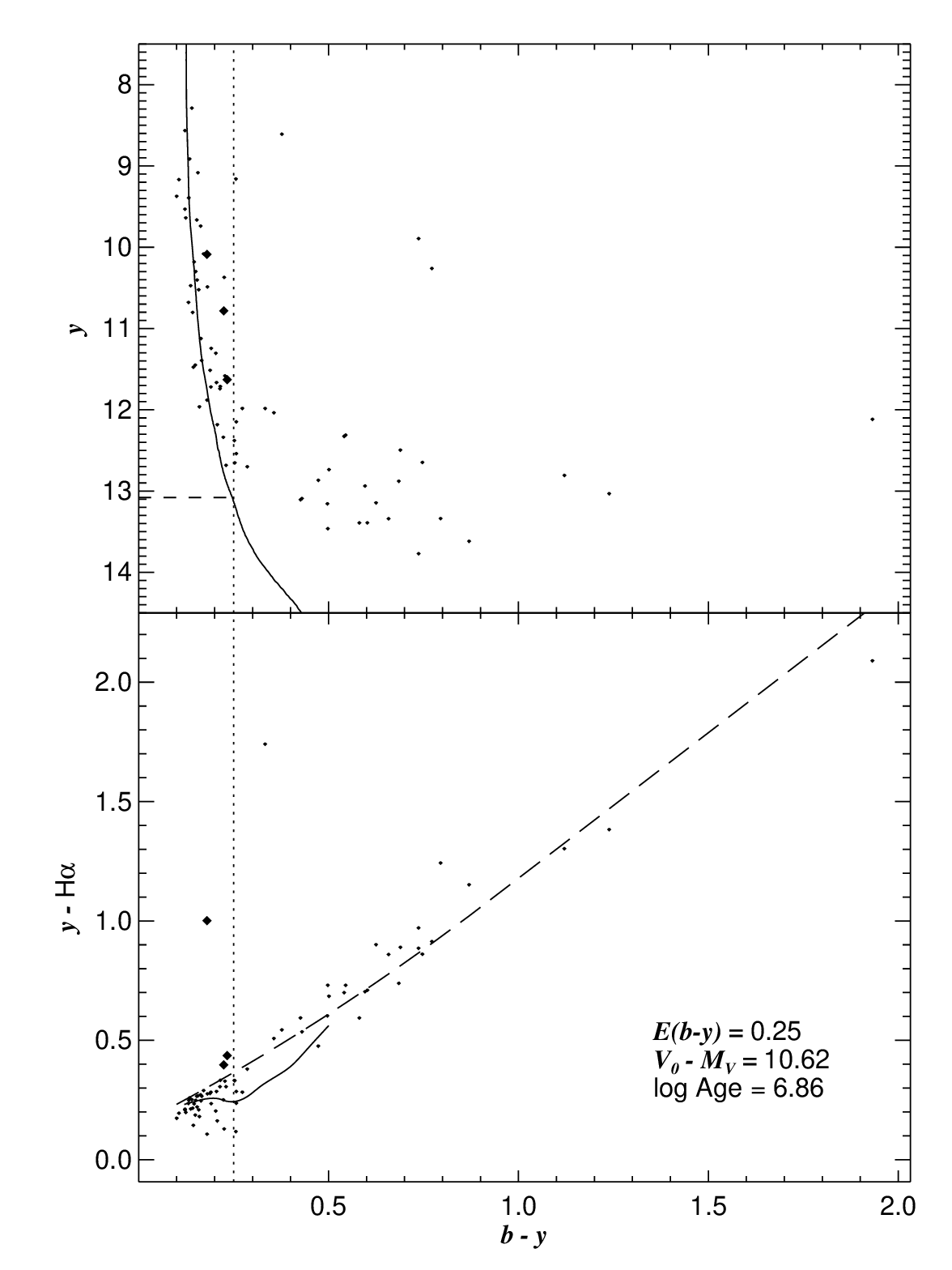}
\caption{
Color-magnitude (\textit{top}) and color-color 
(\textit{bottom}) diagrams of the cluster NGC 6530
in the same format as Figure \ref{Basel1}.
\label{NGC6530}
}
\end{figure}

\begin{figure}
\includegraphics[angle=0,scale=0.4]{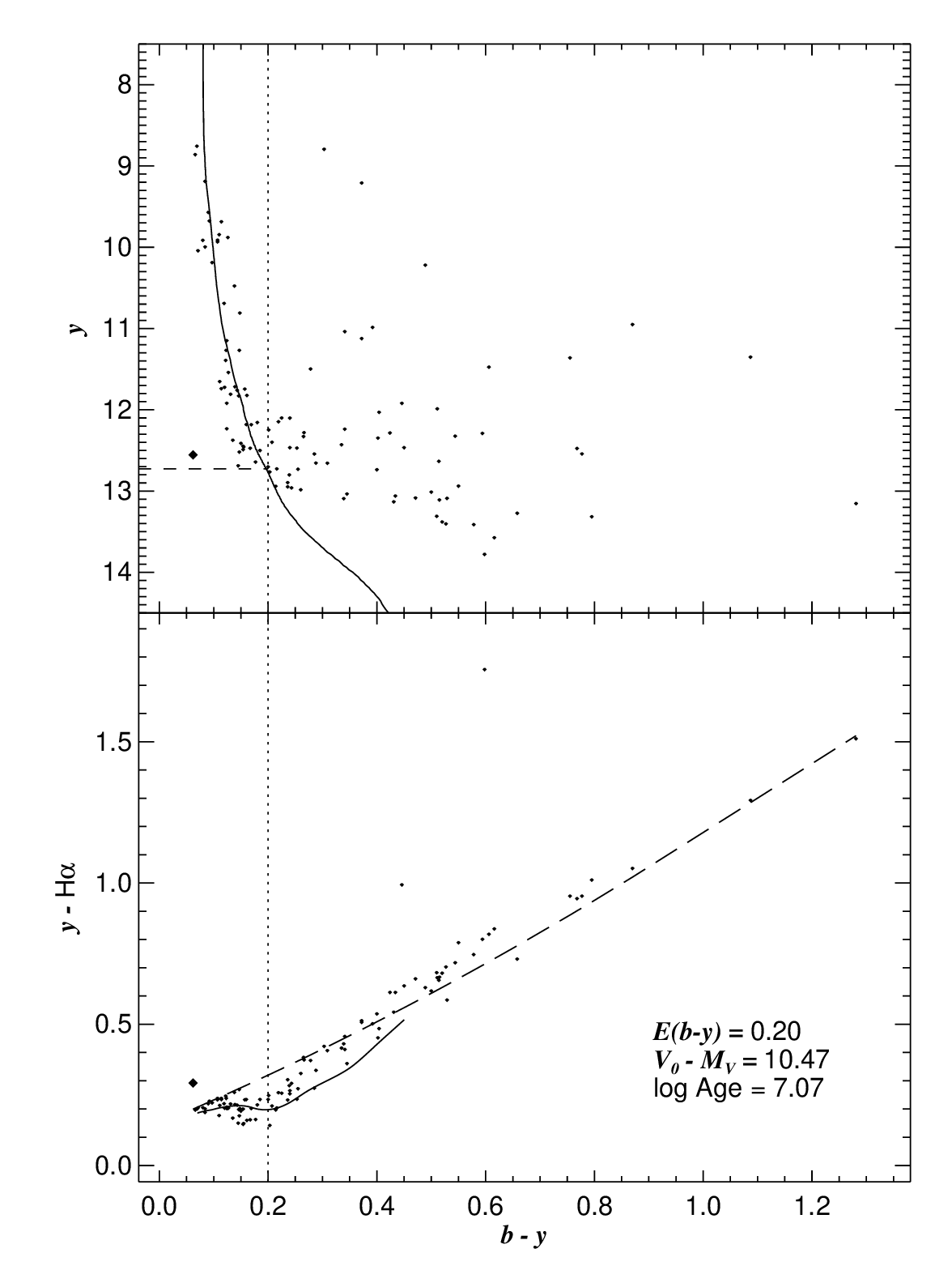}
\caption{
Color-magnitude (\textit{top}) and color-color 
(\textit{bottom}) diagrams of the cluster NGC 6531
in the same format as Figure \ref{Basel1}.
\label{NGC6531}
}
\end{figure}

\begin{figure}
\includegraphics[angle=0,scale=0.4]{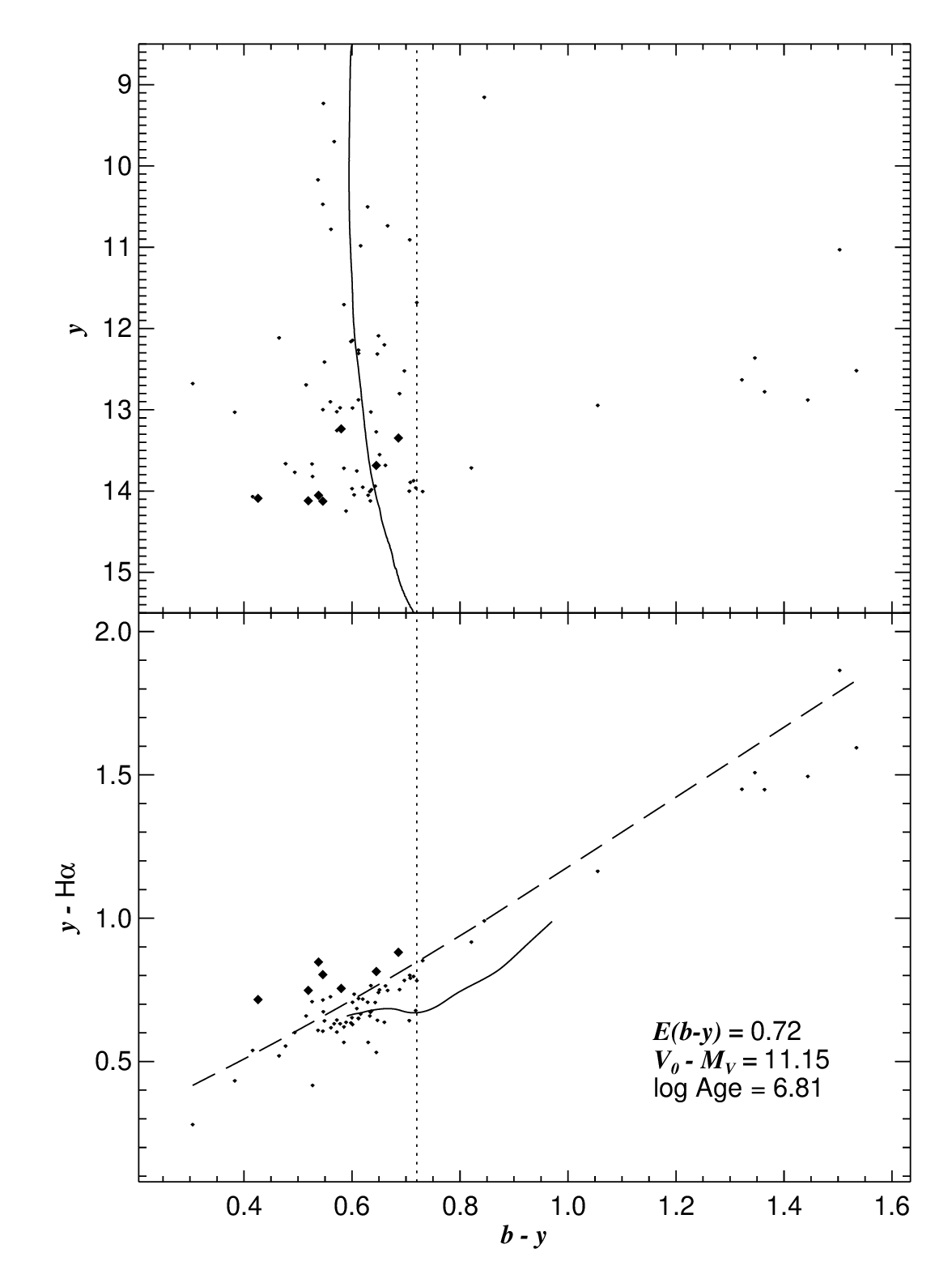}
\caption{
Color-magnitude (\textit{top}) and color-color 
(\textit{bottom}) diagrams of the cluster NGC 6604
in the same format as Figure \ref{Basel1}.
\label{NGC6604}
}
\end{figure}

\begin{figure}
\includegraphics[angle=0,scale=0.4]{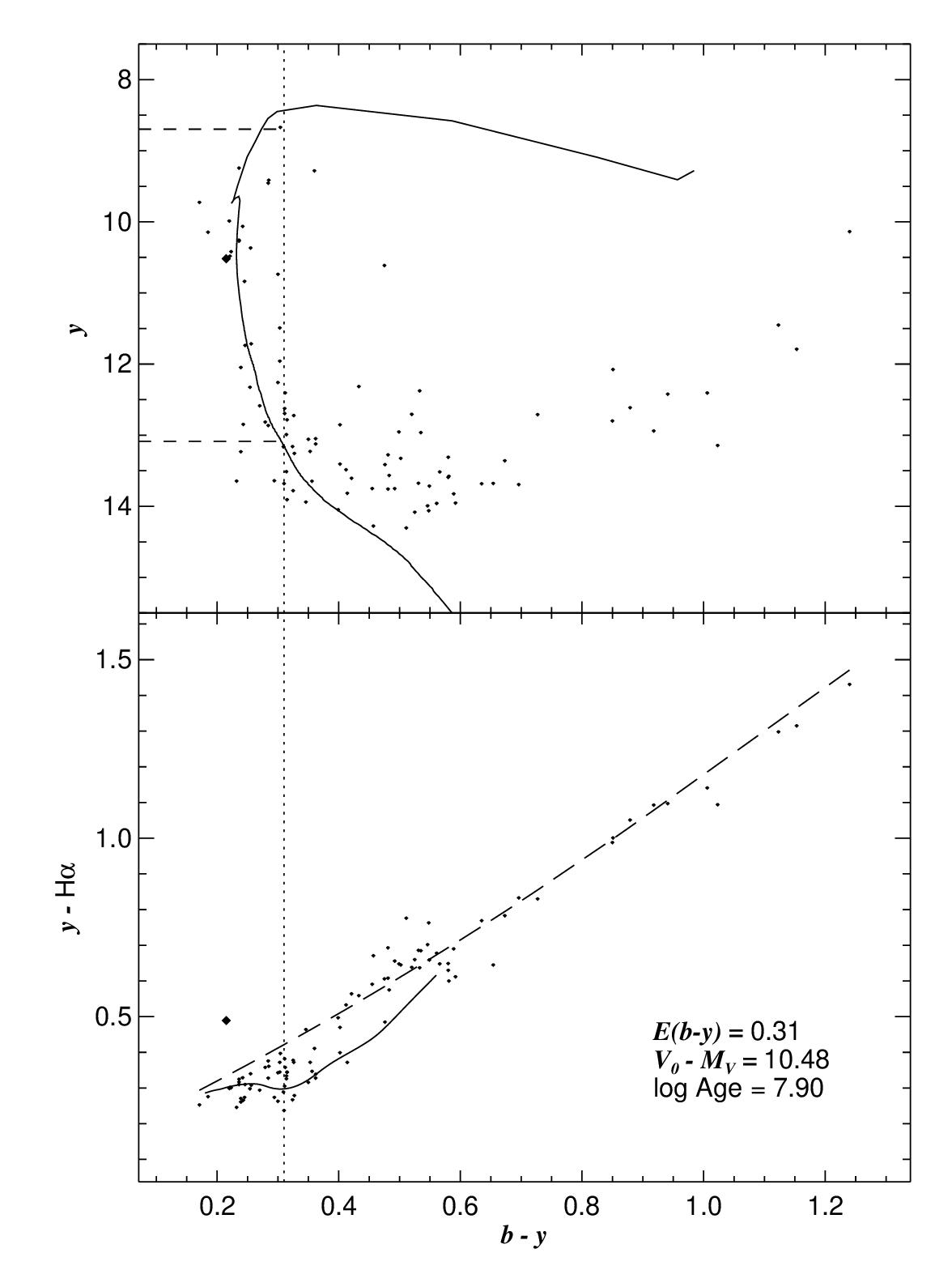}
\caption{
Color-magnitude (\textit{top}) and color-color 
(\textit{bottom}) diagrams of the cluster NGC 6613
in the same format as Figure \ref{Basel1}.
\label{NGC6613}
}
\end{figure}

\begin{figure}
\includegraphics[angle=0,scale=0.4]{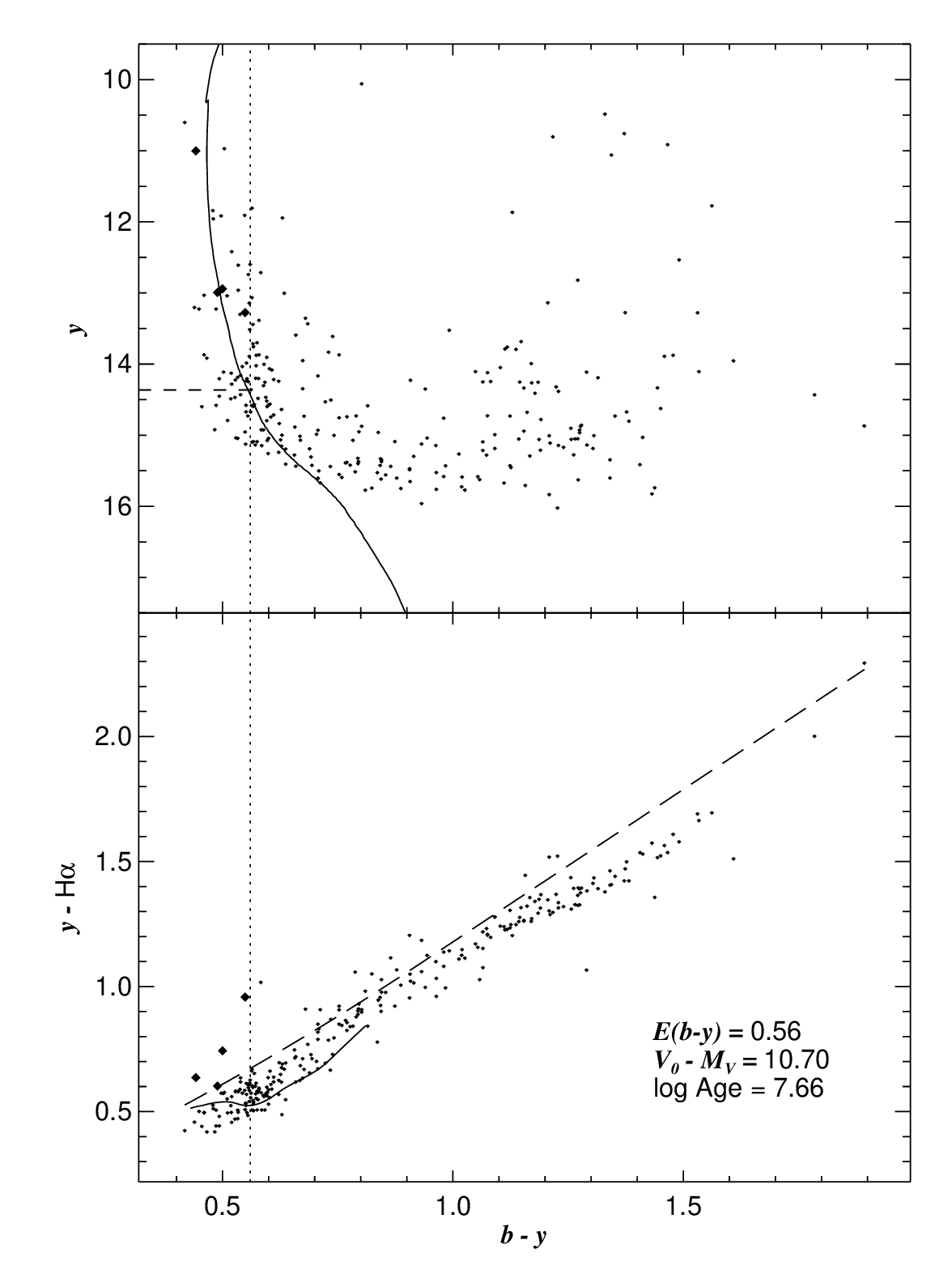}
\caption{
Color-magnitude (\textit{top}) and color-color 
(\textit{bottom}) diagrams of the cluster NGC 6664
in the same format as Figure \ref{Basel1}.
\label{NGC6664}
}
\end{figure}

\begin{figure}
\includegraphics[angle=0,scale=0.4]{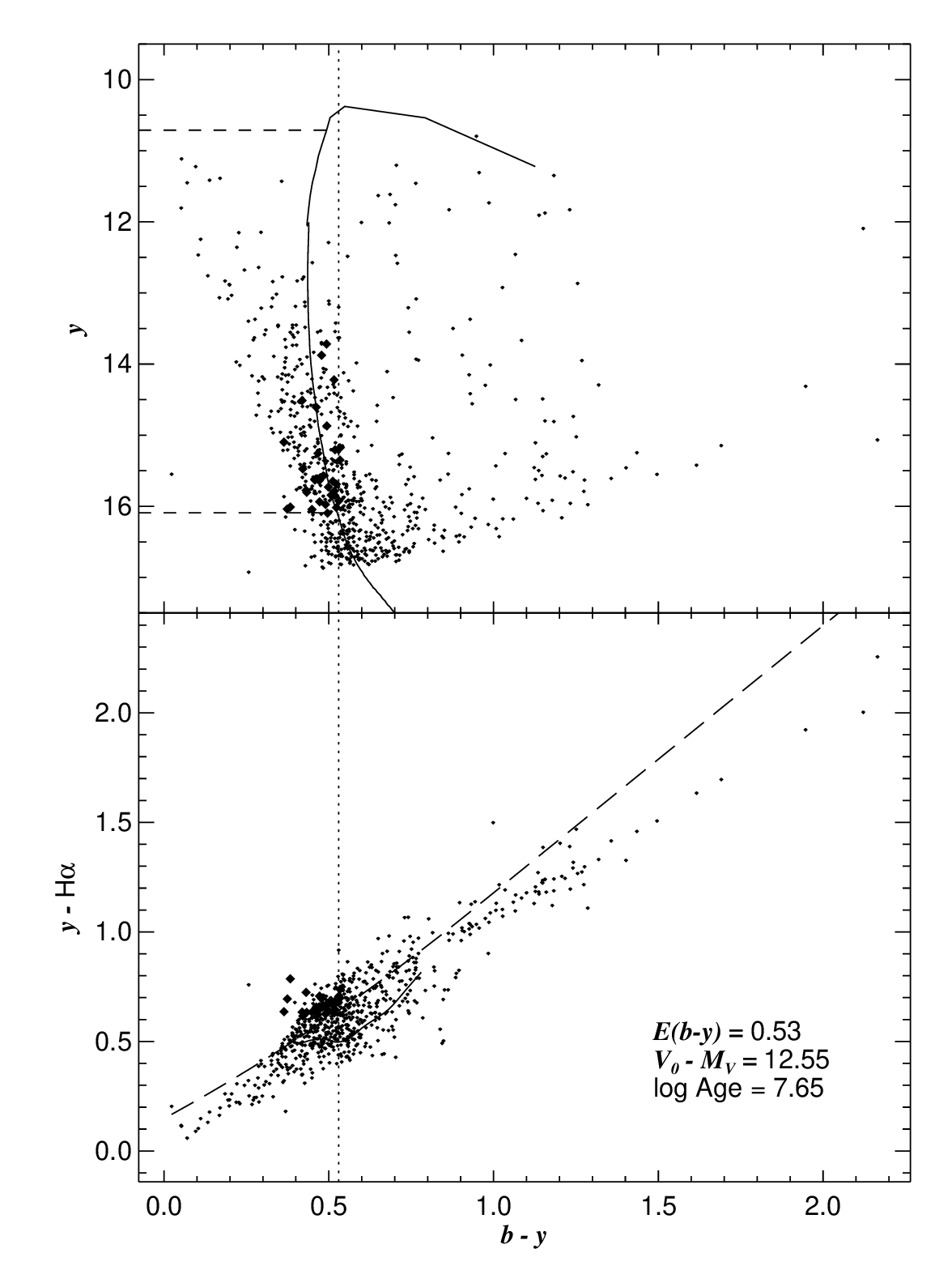}
\caption{
Color-magnitude (\textit{top}) and color-color 
(\textit{bottom}) diagrams of the cluster Ruprecht 79
in the same format as Figure \ref{Basel1}.
\label{Ruprecht79}
}
\end{figure}

\begin{figure}
\includegraphics[angle=0,scale=0.4]{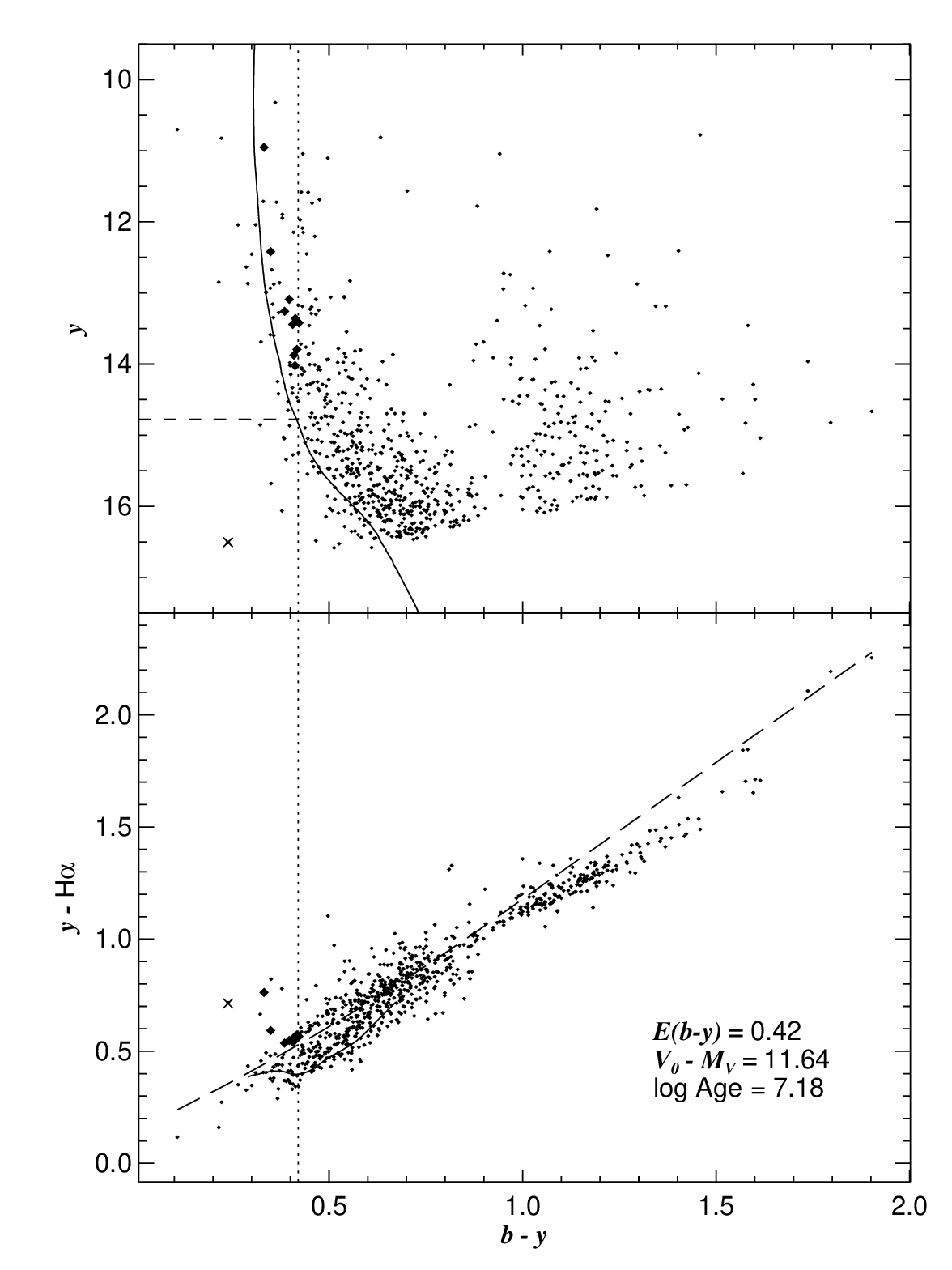}
\caption{
Color-magnitude (\textit{top}) and color-color 
(\textit{bottom}) diagrams of the cluster Ruprecht 119
in the same format as Figure \ref{Basel1}.
\label{Ruprecht119}
}
\end{figure}

\begin{figure}
\includegraphics[angle=0,scale=0.4]{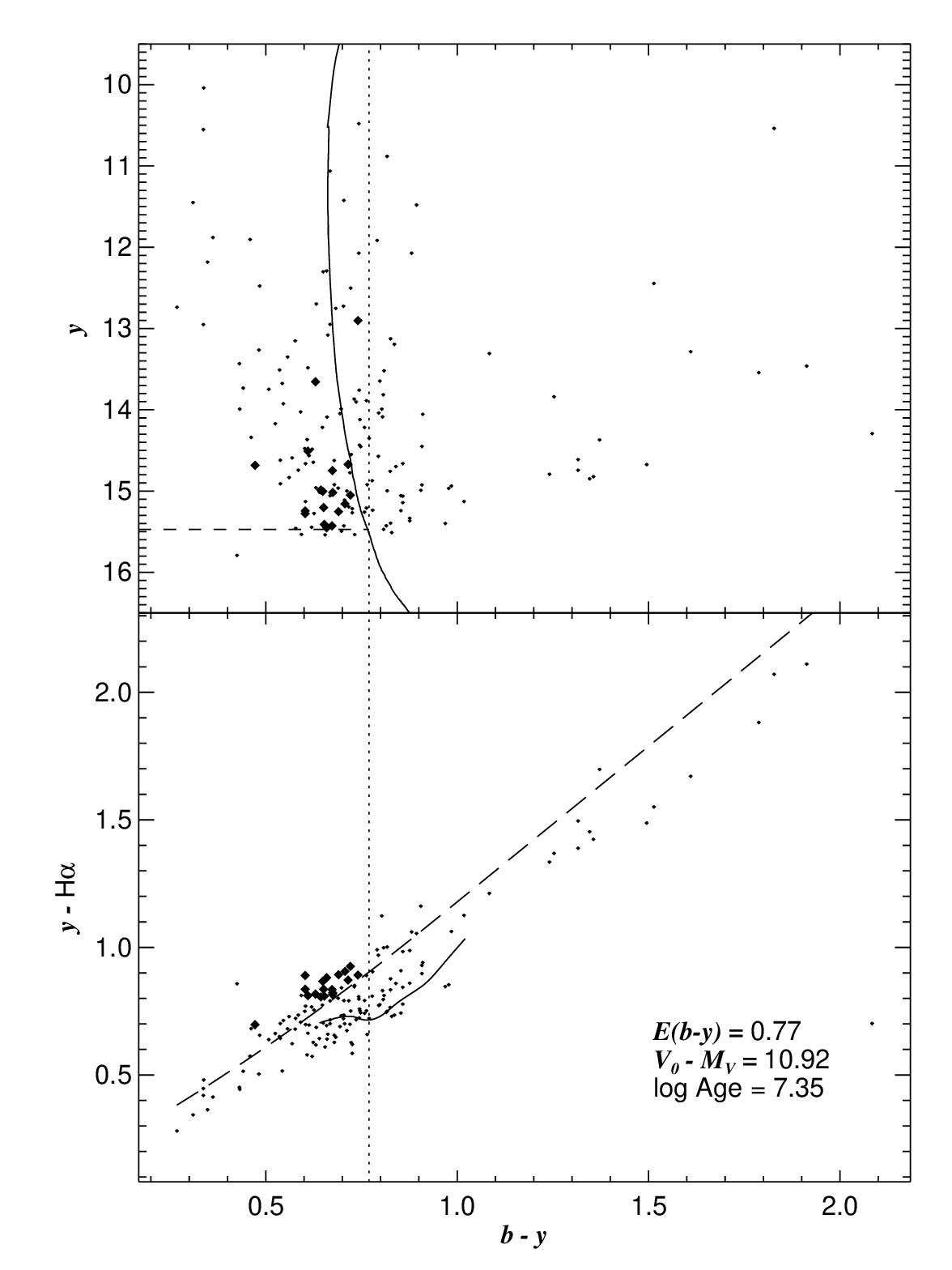}
\caption{
Color-magnitude (\textit{top}) and color-color 
(\textit{bottom}) diagrams of the cluster Ruprecht 127
in the same format as Figure \ref{Basel1}.
\label{Ruprecht127}
}
\end{figure}

\begin{figure}
\includegraphics[angle=0,scale=0.4]{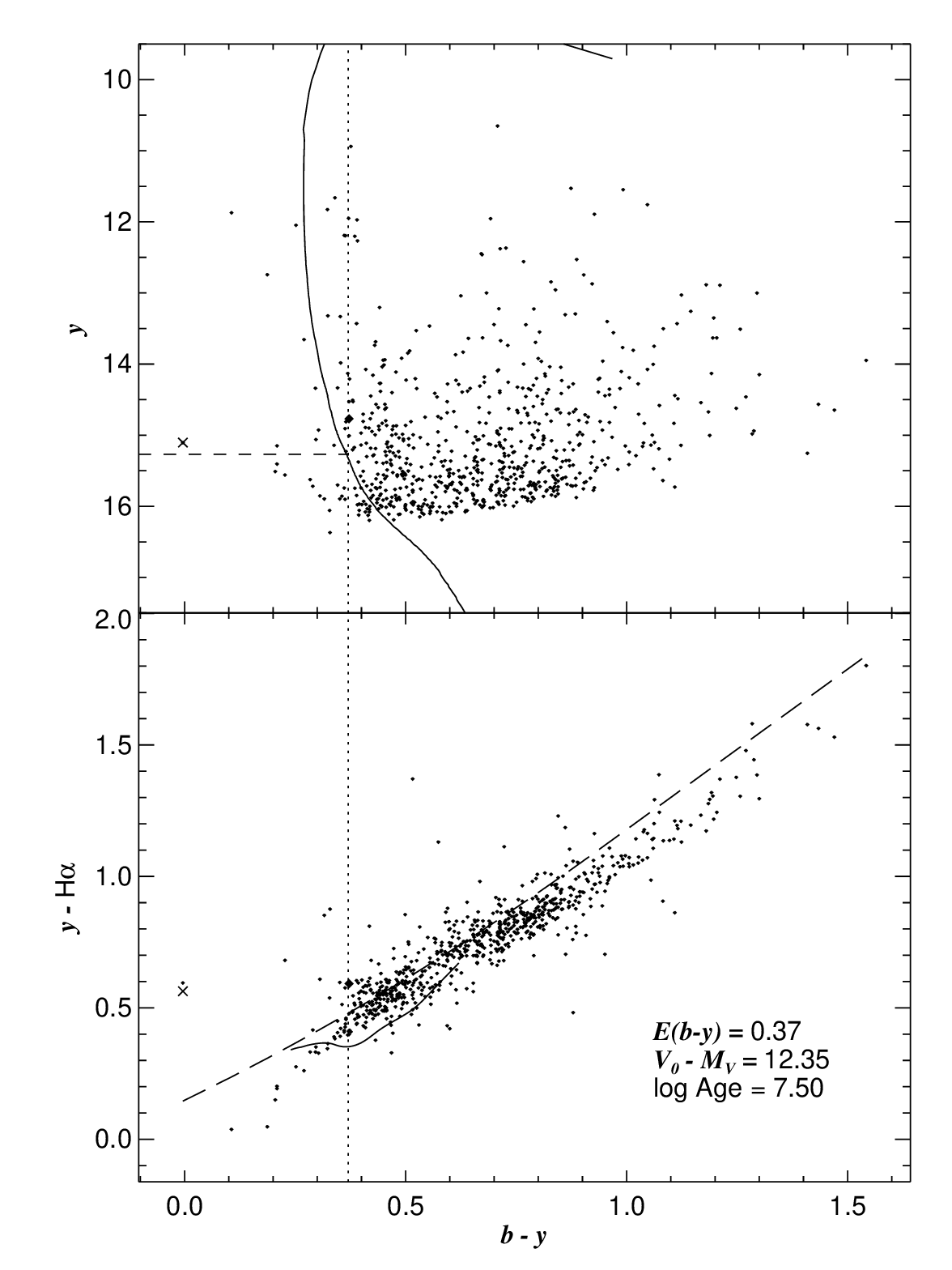}
\caption{
Color-magnitude (\textit{top}) and color-color 
(\textit{bottom}) diagrams of the cluster Ruprecht 140
in the same format as Figure \ref{Basel1}.
\label{Ruprecht140}
}
\end{figure}

\begin{figure}
\includegraphics[angle=0,scale=0.4]{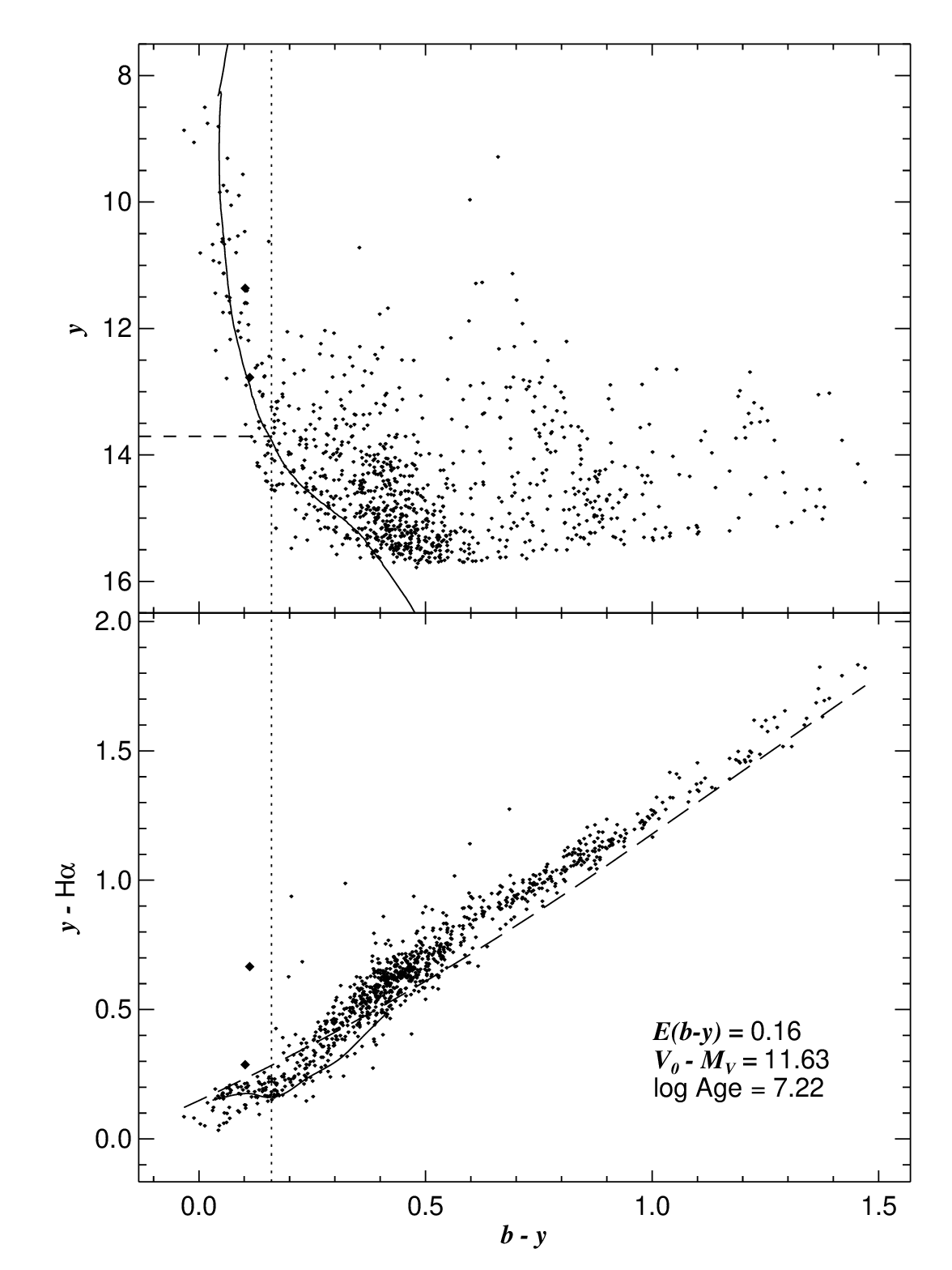}
\caption{
Color-magnitude (\textit{top}) and color-color 
(\textit{bottom}) diagrams of the cluster Stock 13
in the same format as Figure \ref{Basel1}.
\label{Stock13}
}
\end{figure}

\begin{figure}
\includegraphics[angle=0,scale=0.4]{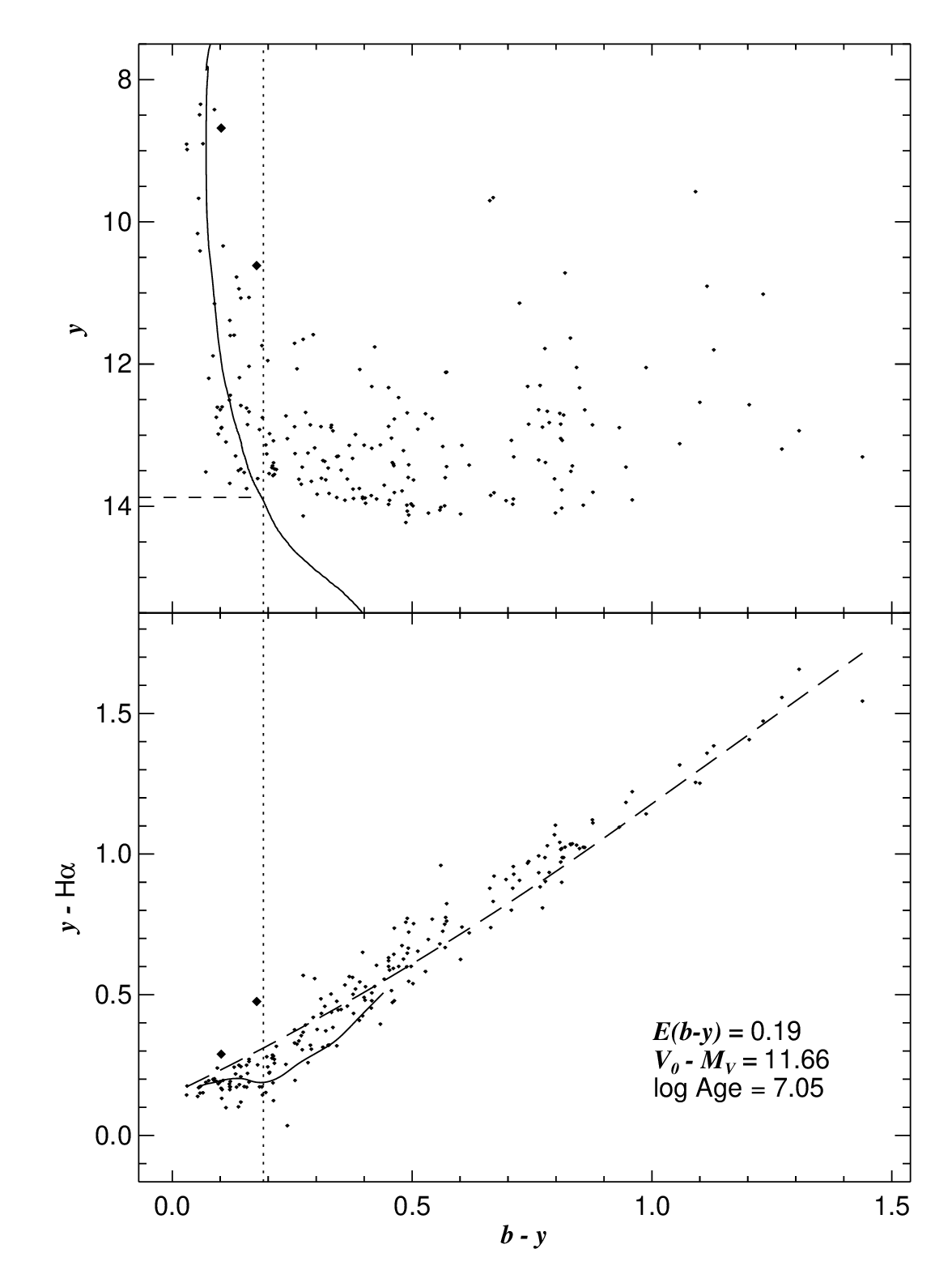}
\caption{
Color-magnitude (\textit{top}) and color-color 
(\textit{bottom}) diagrams of the cluster Stock 14
in the same format as Figure \ref{Basel1}.
\label{Stock14}
}
\end{figure}

\begin{figure}
\includegraphics[angle=0,scale=0.4]{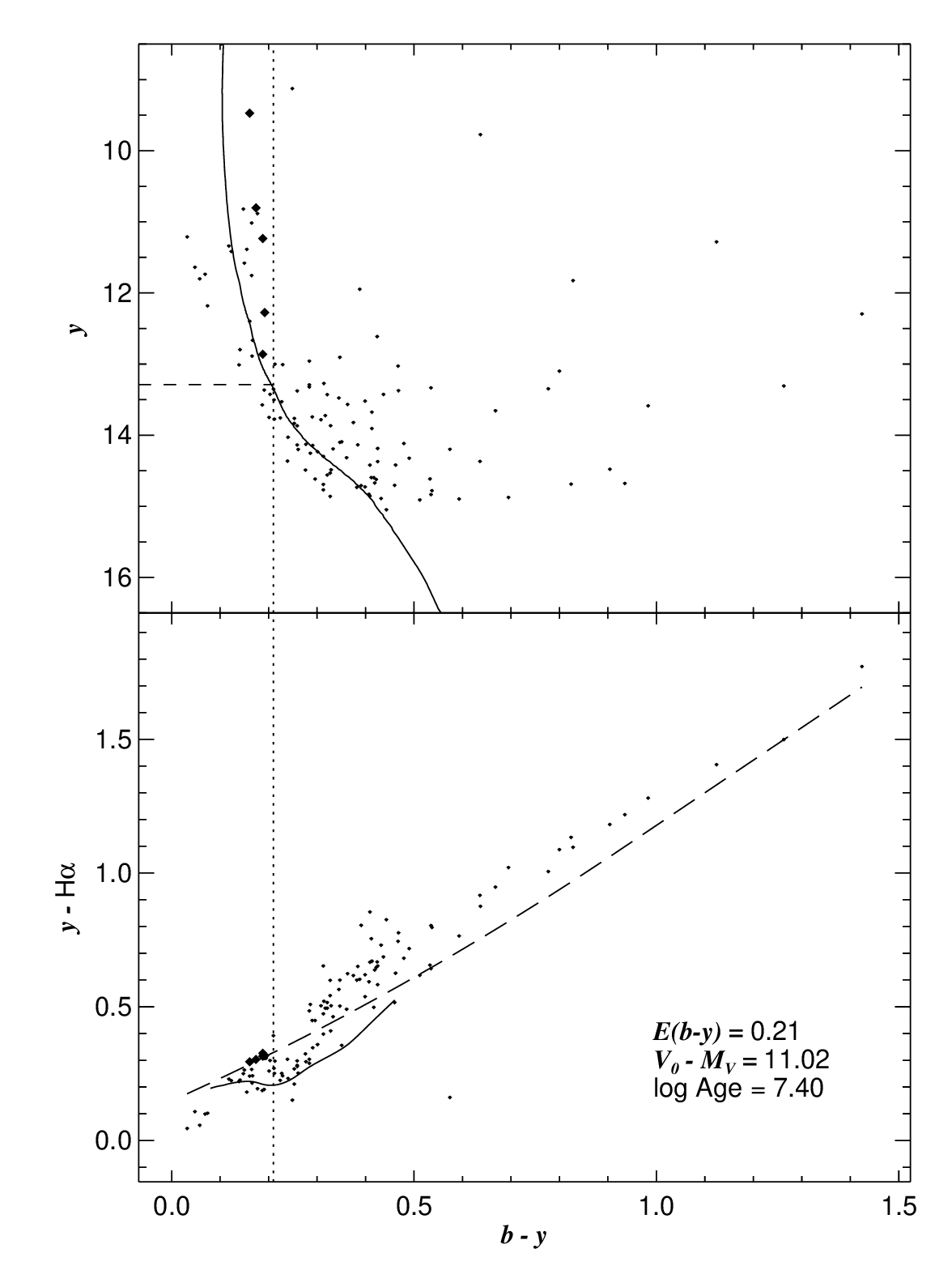}
\caption{
Color-magnitude (\textit{top}) and color-color 
(\textit{bottom}) diagrams of the cluster Trumpler 7
in the same format as Figure \ref{Basel1}.
\label{Trumpler7}
}
\end{figure}

\begin{figure}
\includegraphics[angle=0,scale=0.4]{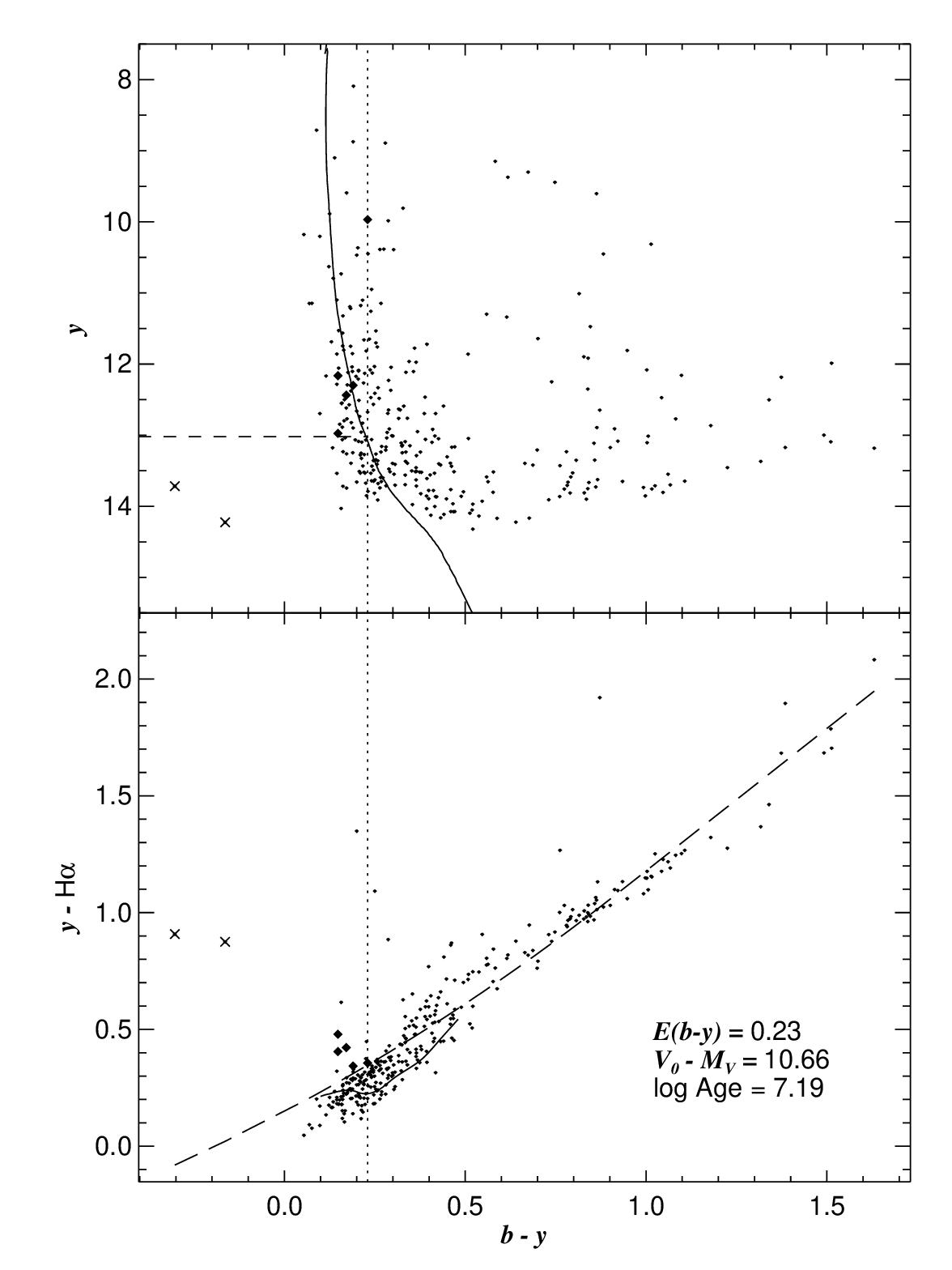}
\caption{
Color-magnitude (\textit{top}) and color-color 
(\textit{bottom}) diagrams of the cluster Trumpler 18
in the same format as Figure \ref{Basel1}.
\label{Trumpler18}
}
\end{figure}

\begin{figure}
\includegraphics[angle=0,scale=0.4]{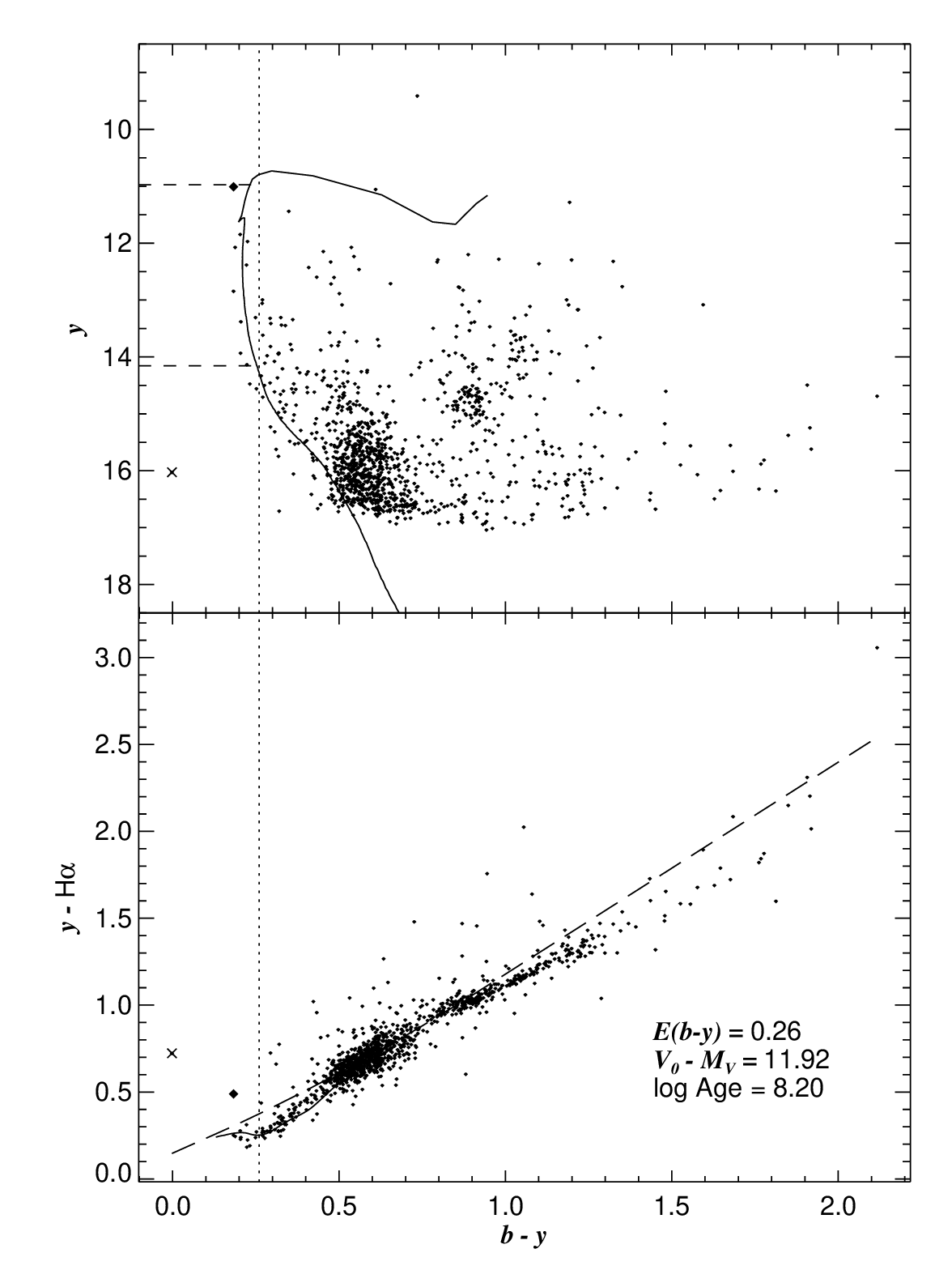}
\caption{
Color-magnitude (\textit{top}) and color-color 
(\textit{bottom}) diagrams of the cluster Trumpler 20
in the same format as Figure \ref{Basel1}.
\label{Trumpler20}
}
\end{figure}

\begin{figure}
\includegraphics[angle=0,scale=0.4]{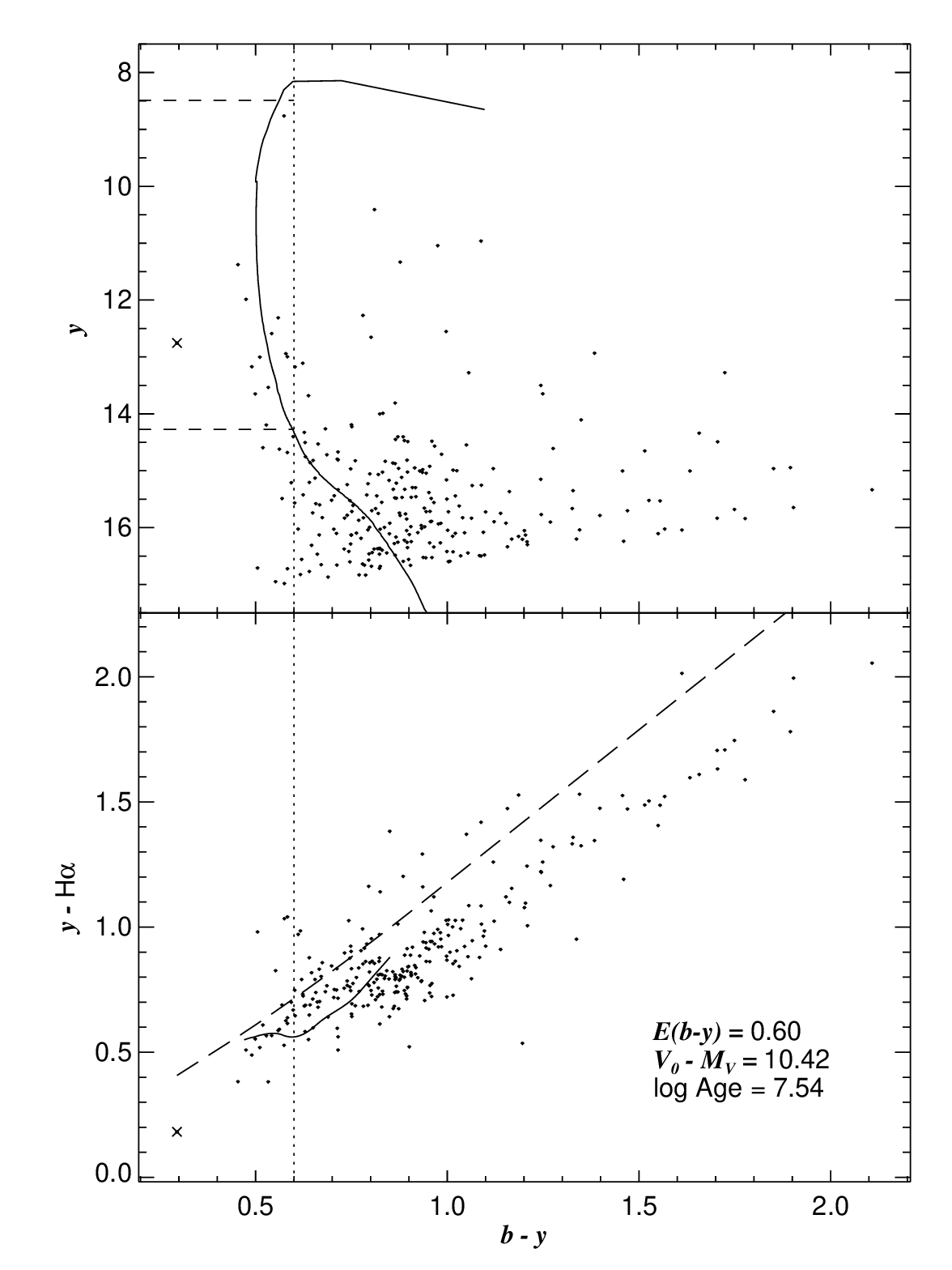}
\caption{
Color-magnitude (\textit{top}) and color-color 
(\textit{bottom}) diagrams of the cluster vdB-Hagen 217
in the same format as Figure \ref{Basel1}.
\label{vdB-Hagen217}
}
\end{figure}

The ($b-y,y$) color-magnitude and ($b-y,y-\rm H\alpha$) color-color
diagrams for each cluster are presented in Figures \ref{Basel1} $-$
\ref{vdB-Hagen217}.  In each color-magnitude diagram, the closest matching
theoretical isochrone, assuming solar metallicity, from 
\citet{lejeune2001} is plotted as a solid line.  The limiting magnitudes
of B-type stars, $y_{Bmin}$ and $y_{Bmax}$, are indicated as dashed lines
when they are within the plotting range.  In each color-color diagram, the
theoretical color-color curve, derived from \citet{kurucz1979} model
spectra, is shown as a solid line.  Be stars are identified by their
excess H$\alpha$ brightness that shifts their $y-\rm{H}\alpha$ color above
this reference line (Paper 1).  We also made a parabolic fit of the color
indices derived from dereddened spectra of 161 stars from the atlas of
\citet*{jacoby1984}, shown as a dashed curve, and unreddened foreground
stars will probably be found close to this curve in the observed
color-color diagrams.  In both diagrams, the massive O- and B-type stars
all lie to the left of the vertical dashed line that represents $E(b-y)$.  
The definite and possible Be stars are represented with large diamonds in
each plot, while all other stars are represented with small diamonds.  In
a few cases, potential foreground stars or white dwarfs that lie to the
left of the reddening line have been excluded from the Be sample, and
these stars are marked with an X.

\begin{figure}
\includegraphics[angle=0,scale=0.4]{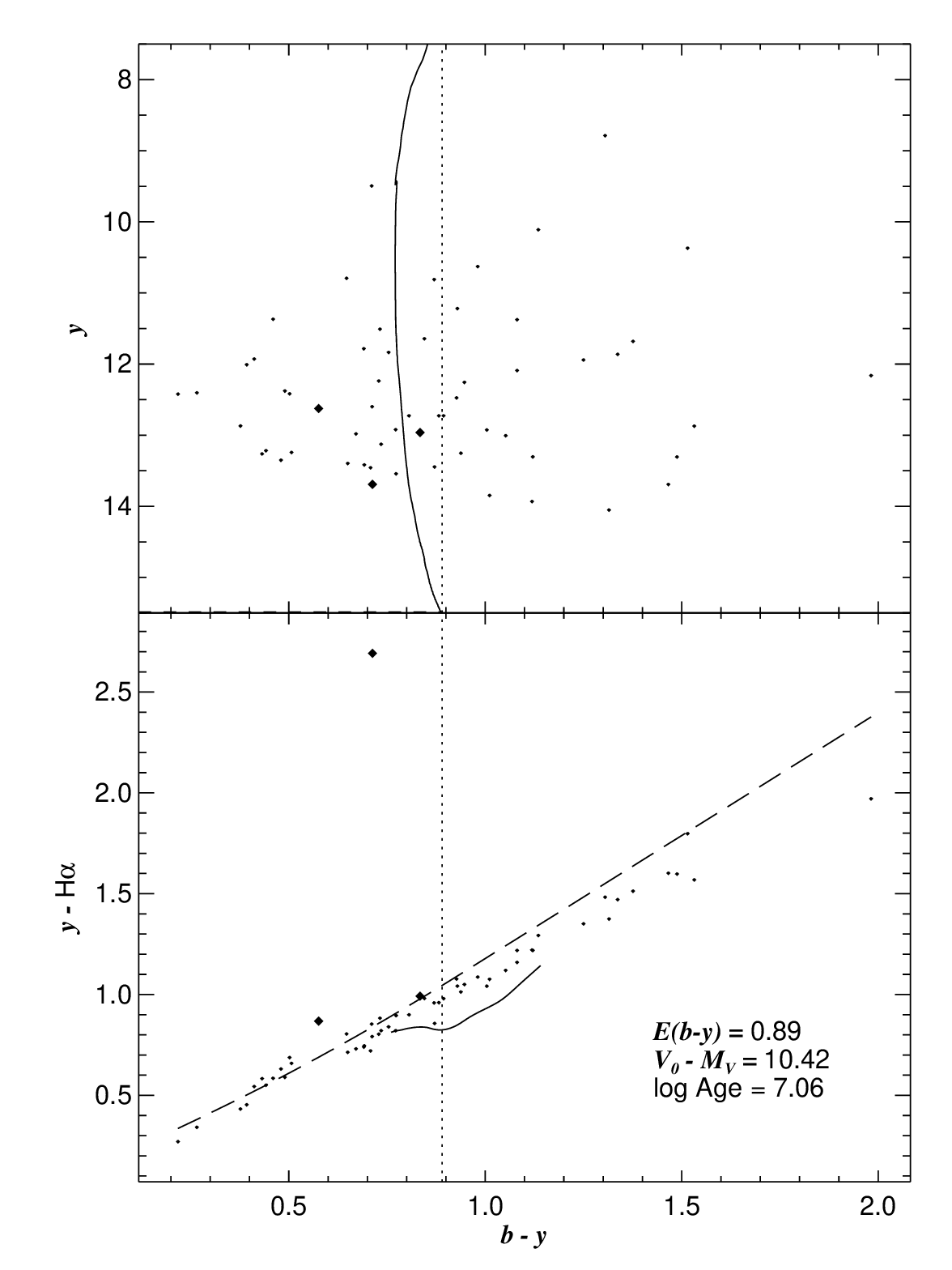}
\caption{
Color-magnitude (\textit{top}) and color-color 
(\textit{bottom}) diagrams of the cluster Trumpler 27
in the same format as Figure \ref{Basel1}.
\label{Trumpler27}
}
\end{figure}

\begin{figure}
\includegraphics[angle=0,scale=0.4]{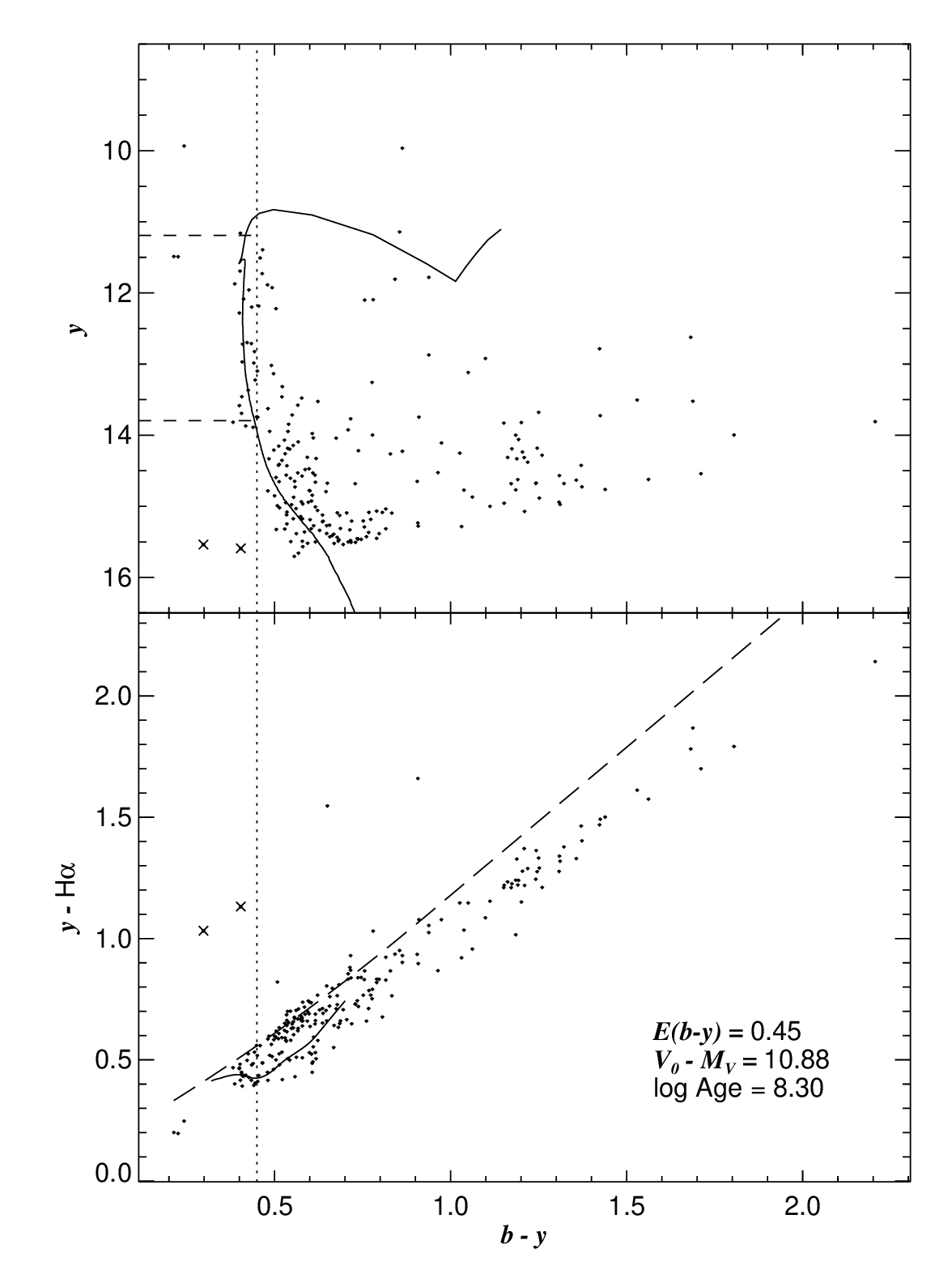}
\caption{
Color-magnitude (\textit{top}) and color-color 
(\textit{bottom}) diagrams of the cluster Trumpler 28
in the same format as Figure \ref{Basel1}.
\label{Trumpler28}
}
\end{figure}

\begin{figure}
\includegraphics[angle=0,scale=0.4]{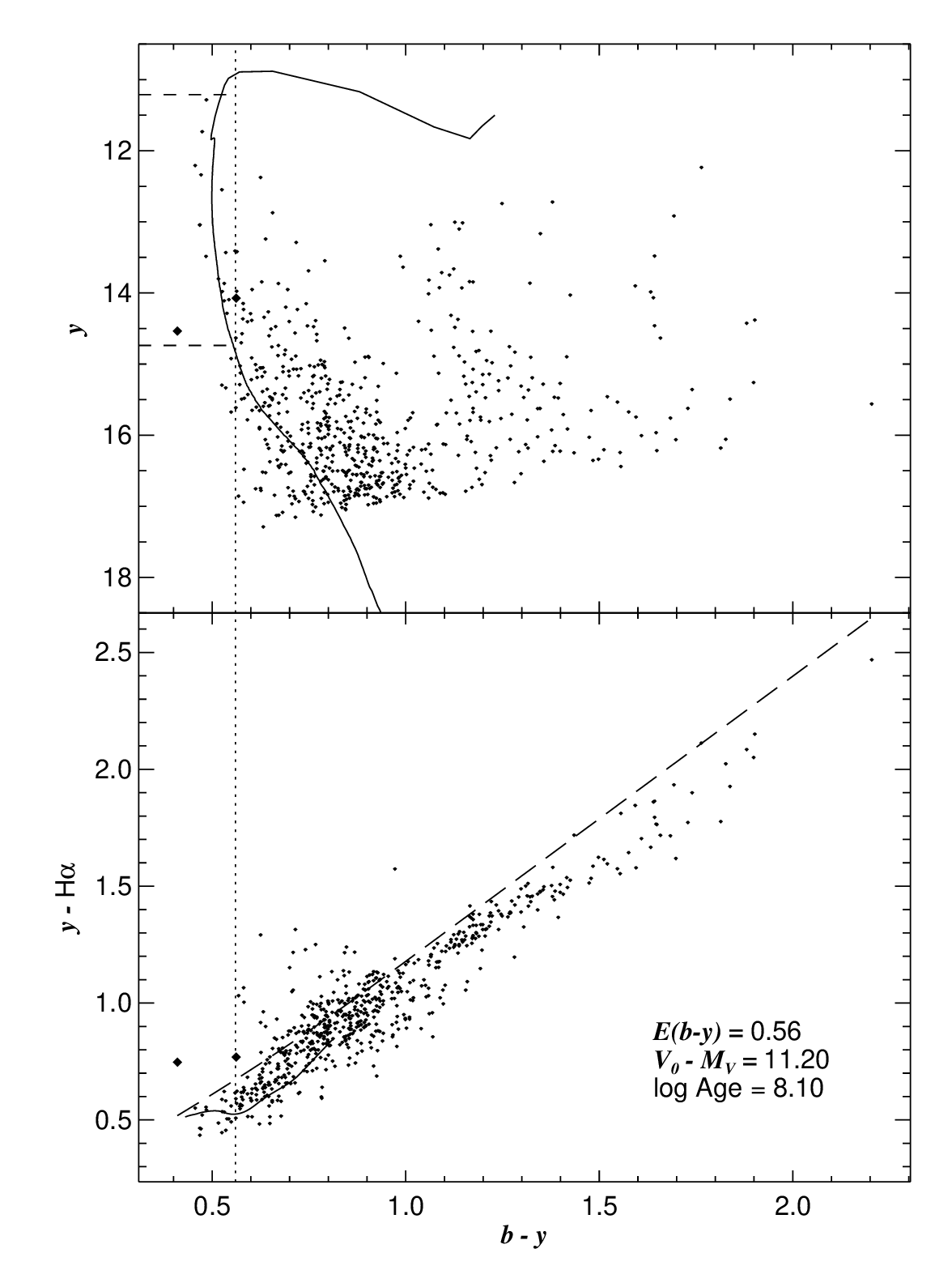}
\caption{
Color-magnitude (\textit{top}) and color-color 
(\textit{bottom}) diagrams of the cluster Trumpler 34
in the same format as Figure \ref{Basel1}.
\label{Trumpler34}
}
\end{figure}

\placefigure{Basel1}
\placefigure{Bochum13}
\placefigure{Collinder272}
\placefigure{Haffner16}
\placefigure{Hogg16}
\placefigure{Hogg22}
\placefigure{IC2395}
\placefigure{IC2581}
\placefigure{IC2944}
\placefigure{NGC2343}
\placefigure{NGC2362}
\placefigure{NGC2367}
\placefigure{NGC2383}
\placefigure{NGC2384}
\placefigure{NGC2414}
\placefigure{NGC2421}
\placefigure{NGC2439}
\placefigure{NGC2483}
\placefigure{NGC2489}
\placefigure{NGC2571}
\placefigure{NGC2659}
\placefigure{NGC3293}
\placefigure{NGC3766}
\placefigure{NGC4103}
\placefigure{NGC4755}
\placefigure{NGC5281}
\placefigure{NGC5593}
\placefigure{NGC6178}
\placefigure{NGC6193}
\placefigure{NGC6200}
\placefigure{NGC6204}
\placefigure{NGC6231}
\placefigure{NGC6249}
\placefigure{NGC6250}
\placefigure{NGC6268}
\placefigure{NGC6322}
\placefigure{NGC6425}
\placefigure{NGC6530}
\placefigure{NGC6531}
\placefigure{NGC6604}
\placefigure{NGC6613}
\placefigure{NGC6664}
\placefigure{Ruprecht79}
\placefigure{Ruprecht119}
\placefigure{Ruprecht127}
\placefigure{Ruprecht140}
\placefigure{Stock13}
\placefigure{Stock14}
\placefigure{Trumpler7}
\placefigure{Trumpler18}
\placefigure{Trumpler20}
\placefigure{Trumpler27}
\placefigure{Trumpler28}
\placefigure{Trumpler34}
\placefigure{vdB-Hagen217}

\begin{figure}
\includegraphics[angle=0,scale=0.4]{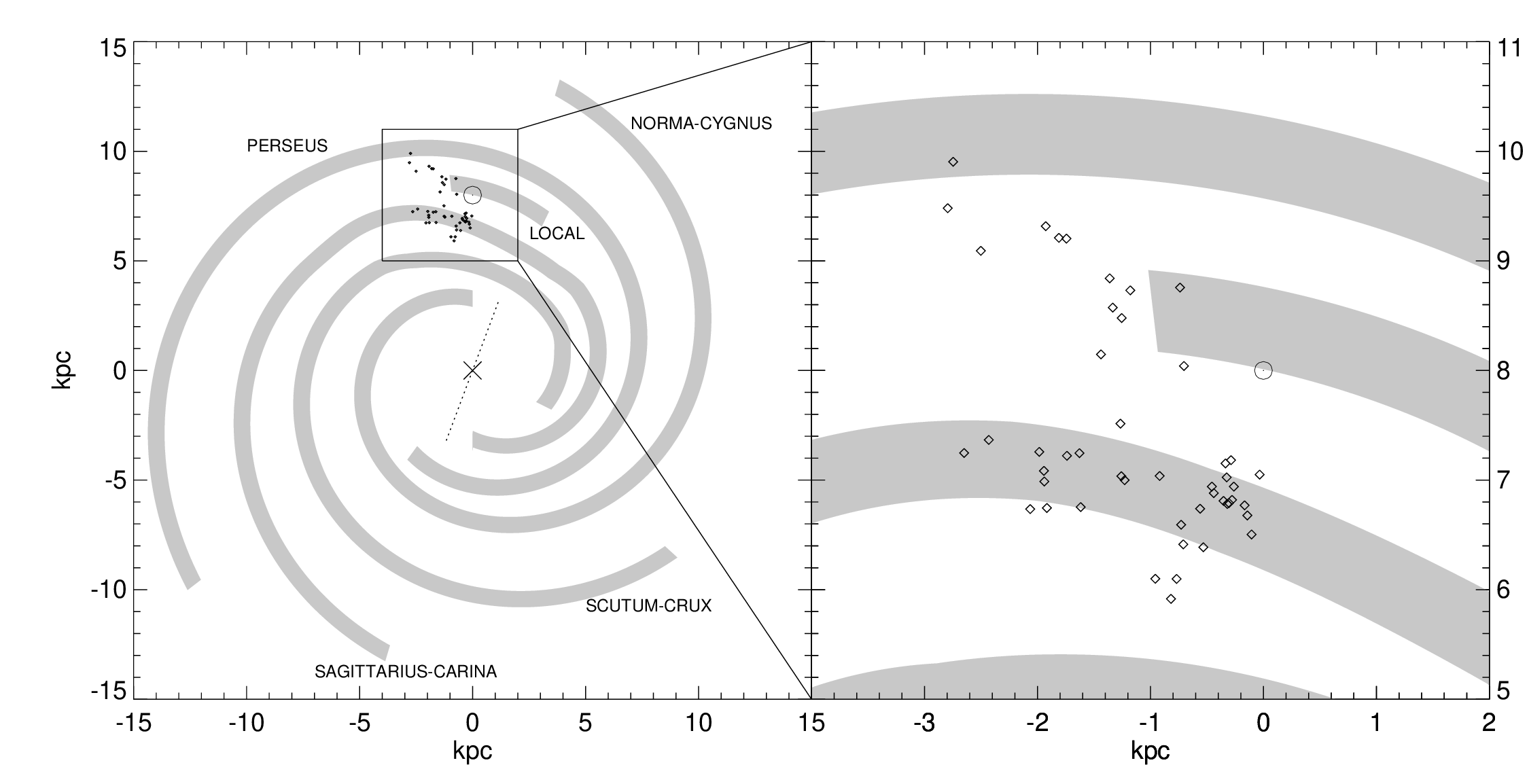}
\caption{
The distribution of the true open clusters is shown relative to the Sun
($\odot$) and the Galactic center ($\times$).  The positions of the spiral
arms (\textit{gray}) are from \citet{cordes2003}, and the orientation of
the central bar (\textit{dotted line}) is from \citet*{bissantz2003}.
\label{galaxy}
}
\end{figure}

\begin{itemize}

\item \textbf{Basel 1:} 
Only one star, GSC $05126-02385$, is saturated in our images of Basel 1,
and we detect no Be stars in this cluster.

\item \textbf{Bochum 13:}
Our images of Bochum 13 contain two bright, saturated stars: HD 156134 and
HD 156154.  The former is an O+ star, and the latter is a B0 Iab star.  
We identify Nos.\ 16 and 30 as definite Be stars, and No.\ 22 is a 
possible Be star candidate.  

Aside from the two bright stars that are saturated, Bochum 13 is a sparse
cluster that contains mostly fainter members.  The MS is not well defined,
and we find it difficult to determine accurate estimates of $E(b-y)$,
$V_0-M_V$, and log age for this cluster.  Therefore we consider our
results less reliable, and we ignore this cluster in our statistics.

\item \textbf{Collinder 272:} 
There are no saturated stars in our images of the cluster Collinder 272.  
The three Be stars in this cluster are Nos.\ 508, 565, and 767.  We also
find five possible Be stars: Nos.\ 345, 513, 560, 562, and 774.

\item \textbf{Haffner 16:} 
In our images of Haffner 16, the cluster appears to have a small diameter
of approximately 4\arcmin, and many field stars are visible in the region.  
However, to create the color-magnitude and color-color diagrams of the
cluster, we used the stars from the entire $13\farcm5 \times 13\farcm5$
field of view of the CCD.  From our color-magnitude diagram shown in
Figure \ref{Haffner16}, the region of sky appears to contain two distinct
stellar populations.  Table \ref{distribution} and Figure \ref{galaxy}
reveal that Haffner 16 is located within the Perseus arm of the Galaxy and
is viewed through the local spur, causing significant contamination from
foreground stars along the line of sight.

Four stars in this cluster show H$\alpha$ emission along with a color $b-y
< 0.123$.  One of the stars, No.\ 756, is within the dense cluster region
and we classify it as a possible Be star.  The others have a $y$ magnitude
fainter than the B star limit for the cluster, so they are likely
foreground stars.  There are three stars in our images of Haffner 16 that
are saturated, but none are identified in the SIMBAD database.  

\item \textbf{Hogg 16:} 
Two stars are saturated in Hogg 16: HD 116887, a K0 III star, and HD
116875, a B8 V star.  We find one possible Be star in Hogg 16, No.\ 195.  
The SIMBAD database also identifies No.\ 199 as a B2e star, and while 
we do observe H$\alpha$ emission from this star, its $b-y$ color is too 
red to classify it a B-type star in the cluster.

The cluster Hogg 16 lies about 16\arcmin~west of Collinder 272.  We find 
slightly different $E(b-y)$, $V_0-M_V$, and ages for these two clusters, 
although they may be physically associated \citep{vazquez1997}.

\item \textbf{Hogg 22:} 
The only saturated star in our images of Hogg 22 is HD 150958, an O type
star and a speckle binary with angular separation of $0\farcs3$
\citep{mcalister1990}.  We find one possible Be star, No.\ 9, in Hogg 22.

Hogg 22 is only 6\arcmin~away from the cluster NGC 6204.  While the two
appear almost as the same group on the sky, Hogg 22 is the more distant
yet less evolved cluster \citep{forbes1996}.  Because the two clusters
share the same FOV in our images, we used the finder chart of
\citet{forbes1996} to consider the photometry of Hogg 22 members
separately.  However, the known members of Hogg 22 do not include the full
B-type MS, and as a result the cluster parameters have a large error.  A
more detailed analysis will be required to identify more cluster members.

\item \textbf{IC 2395:}
There are three saturated stars in our photometry of IC 2395: HD 74234, a
B2 IV star; HD 74455, a B1.5 Vn star; and HD 74531, a B2 V star in a
double system.  We identify only one possible Be star candidate in this
cluster, No.\ 92.

Our images of IC 2395 also contain members of the cluster BH 47, although
both \citet{claria2003} and \citet{jorgensen1988} suggest that IC 2395 and
BH 47 are really a single open cluster within the Vela OB1C association.  
Therefore we consider the entire field without discriminating between the
two clusters.

\item \textbf{IC 2581:} 
The very bright, saturated star that dominates images of this cluster is
V399 Car, a Cepheid variable.  Two other bright stars, HD 90706 and HD
90707, were defocused slightly during our observations to avoid
saturation.  HD 90706 is a B1/B2 Iab variable star (SIMBAD) and HD 90707 
is a B1~III eclipsing binary \citep{lloydevans1969}.

Our photometry led to the detection of two definite Be stars in IC 2581.  
The first, No.\ 158, is a B2 Ve star, and the other is No.\ 243, a B1 Ve 
star.  These two Be stars were also identified as Be stars by
\citet{lloydevans1969, lloydevans1980}.  \citet{lloydevans1980} identifies
No.\ 148 as a third Be star in IC 2581; however, it does not exhibit
H$\alpha$ emission in our results. We identify an additional 8 possible Be
stars in IC 2581: Nos.\ 10, 123, 151, 157, 219, 267, 270, and 323.

\item \textbf{IC 2944:}
There are three saturated stars in our images of IC 2944.  The first is HD
101131, an O6.5 V((f)) + O8.5 V spectroscopic binary \citep{gies2002}.  
In addition, HD 101205 is an O8 star and eclipsing binary of $\beta$ Lyr
type, and HD 101298 is an O6 V star.  We identify No.\ 58 as a definite Be
star in IC 2944, and Nos.\ 21 and 51 are possible Be stars.

IC 2944 has a very distinctive appearance in our images; in addition to a
slight amount of H$\alpha$ nebulosity, several dark, sharp-edged Thackeray
globules \citep{reipurth1997} are visible in the H$\alpha$ images.  
Although these provide clear signs of active star formation in the field
of IC 2944, there is disagreement within the literature about whether this
is a significant physical cluster.  While \citet{walborn1987} argues that
the O stars of IC 2944 contribute to the ionization of an extensive HII
region nearby, \citet{perry1986} point out that both IC 2944 and IC 2948
are associated with this HII region in the Centaurus OB2 association, so
the hot stars of IC 2944 are not required to ionize the region.  They
argue instead that IC 2944 is a chance superposition of early type field
stars along the Carina arm of the Galaxy.  Although the true nature of IC
2944 is not resolved, Figure \ref{IC2944} appears to represent a
physically bound group, so we do include the cluster with our final
statistics on Be stars in open clusters.

\item \textbf{NGC 2343:}
No stars are saturated in our images of NGC 2343, and we do not identify
any Be stars in this cluster.  Although \citet{claria1972} identified our 
Nos.\ 19, 21, and 159 as probable foreground stars, they appear to lie on 
the cluster's MS and we classify these stars as B-type stars within the 
cluster.  Also, he found that our No.\ 160 is probably not a cluster 
member, although we find it close to the MS and barely too red to classify 
as another B star.  Finally, \citet{claria1972} identified No.\ 149 as a 
probable non-member of the cluster, but this star may be the only member 
of this sparse cluster to appear in the MS turnoff region, and 
confirmation of its membership in NGC 2343 could provide a better estimate 
of the cluster age.

\item \textbf{NGC 2362:} 
Two stars in this cluster were saturated in our images, HD 57061 and HD
57192.  HD 57061 is a quadruple star system containing a visual double,
two O-type stars with a separation of $0\farcs151$.  The brighter of the
two O stars is a 154.9 day period spectroscopic binary that also contains
an eclipsing binary with a 1.3 day period \citep{vanleeuwen1997}.  The
other saturated star in our images of NGC 2362 is HD 57192, a B2 V
eclipsing binary \citep{kazarovets1999}.  We do not identify any definite
Be stars in this cluster, and there is only one star, No.\ 19, that we 
classify as a possible detection.

\item \textbf{NGC 2367:} 
There are no definite Be stars in this cluster, and only three possible
candidates: Nos.\ 55, 72, and 138.  The two stars HD 57370 and
BD~$-21^{\circ}1881$ are saturated in our images of NGC 2367.  HD 57370 is 
a B2/B3 III star, and BD~$-21^{\circ}1881$ is a B1 V star.  Although 
\citet{ahumada1995} classify HD 57370 as a blue straggler, we find several 
bright B-type stars in the cluster, and in fact NGC 2367 is one of the 
youngest clusters in this study.

\item \textbf{NGC 2383:} 
Many of the stars in the nearby cluster NGC 2384 are in the same field of
view as NGC 2383, so we removed the 13 known members of the neighboring
cluster from this set of photometry.  The only saturated star in our
images of NGC 2383 is the star HD 58509, but because it is a member of NGC
2384, its photometry is included with that cluster's data set.  Two of the
brightest stars in the field of NGC 2383 are not cluster members according
to \citet{subramaniam1999}.  They classify the first nonmember, No.\ 269,
as an A3 I star.  The other nonmember is No.\ 329, an M1 I star.  Two
definite Be stars are found in NGC 2383, Nos.\ 11 and 341.

\item \textbf{NGC 2384:}
Three known members of the nearby cluster NGC 2383 fell into the field of 
view of this cluster, so they were removed from the photometry set.  None 
of the stars in NGC 2384 are saturated, and we found two possible Be stars 
in this cluster, Nos.\ 176 and 194.

Although the two clusters NGC 2383 and NGC 2384 are separated by only
about 5\arcmin, they are not associated.  \citet{vogt1972} and
\citet{subramaniam1999} have shown that NGC 2384 is both younger and at a
farther distance than its neighbor, indicating a chance superposition in
the plane of the Galaxy.  According to \citet{vogt1972}, the two clusters
have nearly the same reddening because NGC 2383 lies on the outer edge of
the local spiral arm, so there is little dust between it and the more
distant NGC 2384.

\item \textbf{NGC 2414:}
The star HD 60308, a B2 Iab star, is saturated in our images of NGC 2414.  
There are five definite Be stars in this cluster (Nos.\ 12, 65, 73, 87,
and 97).  In addition, we found 13 possible Be star candidates (Nos.\ 2, 
3, 4, 14, 17, 19, 34, 37, 40, 106, 114, 119, and 122).  However, there are 
a large number of apparent foreground stars contaminating the MS of this
cluster, so we doubt that the Be candidates are true H$\alpha$ emitters.

In fact, we doubt that this is a true, physically associated cluster.  
Although the group appears somewhat prominent against the field, we see
two possible main-sequences in the color-magnitude diagram that we attempt
to fit with different values for $E(b-y)$ and $V_0-M_V$.  Neither value of
$E(b-y)$ provides a good match between the observed color-color sequence
and the theoretical color-color curve, which suggests that the group 
either has a highly variable reddening or is not associated.  We see no 
evidence for nebulosity or dust gradients in our H$\alpha$ images of the 
cluster.

\item \textbf{NGC 2421:}
There are four stars in NGC 2421 with definite H$\alpha$ emission and $b-y
< 0.350$, the cutoff for OB stars based on the cluster's reddening.  
These definte Be stars are Nos.\ 311, 356, 378, and 396.  
\citet{slettebak1985} observed H$\gamma$ emission from No.\ 356 and
classified its spectral type as B3$-$B5 (V:).  He also observed weak
H$\gamma$ central emission in No.\ 311, which he also classified as a
B3$-$B5 (V:)  star.  Also, \citet{stephenson1977} identified No.\ 396 as a
B9 star with weak emission lines.  No.\ 378 is a newly identified Be star.

We also find six possible Be stars in NGC 2421: Nos.\ 83, 197, 256, 341,
483, and 530.  \citet{stephenson1977} also observed weak emission in No.\
341.  In addition, \citet{stephenson1977} and \citet{slettebak1985} saw
weak emission in No.\ 264, although this star is both too red and too
faint in H$\alpha$ to be a Be star in our results.

No stars in our images of NGC 2421 were saturated.

\item \textbf{NGC 2439:}
In our images of NGC 2439, there were two stars that had to be defocused
to avoid saturation.  The brightest is R Pup, an F8/G0 Ia star.  
\citet{eggen1983} describes R Pup as a ``pseudocepheid'', a non-periodic
variable supergiant whose low dispersion spectrum resembles a Cepheid.  
The second brightest star in the cluster is V384 Pup, an M3 Ia0-Ia
pulsating variable star.

Our photometry of NGC 2439 revealed six definite Be stars, Nos.\ 22, 24,
58, 101, 193, and 196.  No.\ 24 was classified as a B1 V: star with weak
central emission in the H$\gamma$ line by \citet{slettebak1985}.  He also
observed double peaked H$\alpha$ emission and sharp central absorption
cores in the H$\gamma$ and H$\delta$ lines of No.\ 58.  
\citet{slettebak1985} also noted that the star No.\ 101 had been
previously identified as a Be star, presumably in an unpublished catalog
by Sanduleak \& Robertson (1977).  According to Slettebak, Sanduleak \&
Robertson list a total of five Be stars in NGC 2439, although the
remaining two are not identified by Slettebak.  In addition to the six
definite Be stars found from our photometry, we detect seven possible
cases: Nos.\ 13, 16, 43, 91, 116, 147, and 199.

NGC 2439 may not be a true open cluster (\citealt{eggen1983};  
\citealt*{kaltcheva2001}).  \citet{kaltcheva2001} found many localized
clumps of OB stars within the cluster as well as highly variable
reddening.  In addition, they found a large scatter in distances, proper
motions, and radial velocities among their sample of early type stars in
the field.  While NGC 2439 may simply be a collection of field stars
revealed by an absorption hole in the galactic plane, its color-magnitude
and color-color diagrams in Figure \ref{NGC2439} appear to represent a
true physical cluster, so we do include the cluster in our analysis in \S
3 and \S 4.

\item \textbf{NGC 2483:}
There were no stars that were saturated in our shortest exposures of this
cluster, so there is no chance of any Be stars being overlooked.  One of
the definite Be stars in this cluster, No.\ 186, was classified as an A0 I
emission line star by \citet{stephenson1977}, but they noted an
uncertainty in the spectral type and a Be classification is likely.  In
addition, we identify No.\ 87 as a Be star, and there are ten possible Be
stars in this cluster: Nos.\ 7, 27, 40, 42, 53, 90, 97, 179, 290, and 292.

NGC 2483 is not a true, physically associated cluster, and we ignore it
in our analysis.  \citet{fitzgerald1975} noted that the region has only a
slight increase in the number density of stars over the background field,
and they identified three distinct groups of stars, each at different
distances, that are superimposed on the field.  \citet{havlen1976}
confirmed these results and recommended that NGC 2483 no longer be
classified as a galactic cluster.  Havlen also showed that the four bluest
stars in our photometry are a group of foreground B6 $-$ B8 stars that are
somewhat less reddened (mean $E(B-V) = 0.11$) than the population of 9 O9
$-$ B5 stars that we used to fit the ``cluster's'' isochrone.

\item \textbf{NGC 2489:}
There are no Be stars in this cluster, and no stars are saturated.

\item \textbf{NGC 2571:} 
In our images of NGC 2571, no stars were saturated.  The single Be star in
this cluster is No.\ 76, and it is classified as a B4 star in SIMBAD.

\item \textbf{NGC 2659:} 
None of the stars in NGC 2659 were saturated in our images.  One definite
Be star emerged from our color-color plot of this cluster, No.\ 233.  In 
addition to this Be star, there are 4 possible Be stars (Nos.\ 3, 42, 89, 
and 222).

\item \textbf{NGC 3293:}
NGC 3293 is a young cluster with ongoing star formation around the
cluster's periphery \citep{baume2003}, although its nuclear age is used in
this study.  Baume et al.\ found many pre-MS members in this cluster, as
well as several massive stars that have already begun to move off of the
ZAMS.  Both they and \citet{turner1980} observed variable reddening in
this cluster, and our own H$\alpha$ image of NGC 3293 reveals thin, wispy
clouds of gas across the region.  The differential reddening is a likely
cause for error in the isochrone fit, and the background H$\alpha$
emission as well as emission from pre-MS stars is also likely to
contaminate the number of Be stars detected in this survey.

To further amplify the problem of detecting Be stars in NGC 3293, a
large number of stars are saturated in our images of this bright 
cluster.  They are: 
HD 91943, a B0.5/B1 II/III variable star;
HD 91969, a B0 Ib star;
HD 92007, a B1 III variable star of $\lambda$ Eri type and a probable Be 
star \citep{balona1994}; 
HD 92044, a B1 II star;
CD~$-57^{\circ}$3348, a B0.5 V star;
CPD~$-57^{\circ}$3506B, a B1 III star that is a suspected eclipsing binary 
and a variable star of $\beta$ Cep type \citep{balona1994};
CPD~$-57^{\circ}$3521, a B1 III star;
CPD~$-57^{\circ}$3523, a B1 III star;
CPD~$-57^{\circ}$3526B, a B0.5 III star; and
V361 Car, an M0 star.
Clearly, the Be star population is probably not well sampled in this
cluster due to the large number of B-type giants that are omitted from the
photometry, including at least one probable Be star.

We find three definite Be stars in NGC 3293: Nos.\ 6, 42, and 46.  In 
addition, there are eight possible Be stars with
somewhat weaker emission: Nos.\ 4, 8, 27, 51, 54, 84, 99, and 116.  No.\ 
99 was noted to be a Be star by \citet{shobbrook1980},
and No.\ 51 was also identified as a Be star by \citet{feast1958}.

\item \textbf{NGC 3766:}
The cluster NGC 3766 was the test case for our photometric search for Be
stars, and it was discussed in detail in Paper 1.  The five definite Be
stars that we identified with our photometry are (in our numbering 
scheme) Nos.\ 47, 127, 154, 198, and 200.  Our spectra confirmed these 
five and also identified three other Be stars with weak H$\alpha$ 
emission: Nos.\ 31, 83, and 92.  From our photometry, we classified Nos.\ 
61, 73, 83, 92, 94, 126, 130, and 139 as possible Be detections.  Our 
spectroscopy proves that at least some of these possible Be stars are 
indeed weak emission line stars.

The five saturated stars in our images of NGC 3766 are V910 Cen, HD
100840, HD 100943, HD 306794, and HD 306799.  One of these saturated
stars, HD 100943, was identified as a Be star by \citet{shobbrook1987} and
confirmed by our spectra.  However, for consistency in the detection
technique, we only include Be stars detected with our photometry in this 
study.

\item \textbf{NGC 4103:} 
No stars in our images of NGC 4103 were saturated.  One star, No.\ 231,
clearly exhibits H$\alpha$ emission in the color-color diagram of this
cluster.  It was classified as a B2 IVe star by \citet{wesselink1969}, and
we classify it as a definite Be star detection.  Two other possible Be
stars (Nos.\ 13 and 209) were also found in this cluster.

\item \textbf{NGC 4755:}
In our images of NGC 4755, six bright stars are saturated.  They are:  HD
111904, a B9 Ia variable star; HD 111934, a B2 Ib variable star;  
HD~111973, a B5 Ia star; HD 111990, a B1/B2 Ib star in an optical double
system; SAO 252073, an M2 Iab variable star of irregular type; and SAO
252075, a B0.5~Vn ellipsoidal variable star.

Of the remaining unsaturated stars, five are definitely Be stars.  They
are: Nos.\ 43, 129 (B1.5 Vpne; \citealt{schild1970}), 171 (B2 IVne;  
\citealt{knoechel1980}), 180, and 187.  Our photometry detected 7
additional Be star candidates (Nos.\ 20, 53, 131, 168, 189, 222, and 223).

\item \textbf{NGC 5281:} 
One star, HD 119699, is saturated in our images of this cluster, and it is
classified as an A1 II star.  The only definite Be star detected in our
study is No.\ 74, and five additional Be candidates were found: Nos.\ 11,
13, 29, 66, and 79.  The bright star No.\ 80 also exhibits H$\alpha$
emission that distinguishes it in the color-color diagram, and it was
previously classified as a Be star by \citet{humphreys1975} and as a blue
straggler by \citet{ahumada1995}. However, its ($b-y$) color places it
slightly redder than the B-type stars in our study so we do not classify
it as a Be star, nor does it appear to be a blue straggler in the cluster.

\item \textbf{NGC 5593:}
There are no saturated stars in our images of NGC 5593.  The star No.\ 204
is a definite Be star in this cluster, and No.\ 180 is a possible Be star.

\item \textbf{NGC 6178:}
The only saturated star in our images of NGC 6178 is HD 149277, a B2 IV/V
star, and we do not identify any Be stars in this cluster.

\item \textbf{NGC 6193:}
This young cluster is rich in spectroscopic binaries, and
\citet{arnal1988} found a binary fraction as high as 72\%.  Two of the
saturated stars in NGC 6193 are HD 150135 and HD 150136, an O6.5 V + O5
visual binary and the only two O stars in the Ara OB1 association
\citep{herbst1977}.  In addition, the B1/B2 Ib supergiant HD 150041 is
saturated in our images.  NGC 6193 is an extremely young cluster, about
the same age as the Orion nebula cluster \citep{herbst1977}, with pre-MS B
stars moving onto the ZAMS at magnitudes $V > 12.4$ \citep{moffat1973}.  
Therefore the possible Be star candidate in NGC 6193, No.\ 38, may be a
pre-MS Herbig Be star.

\item \textbf{NGC 6200:}
Although the cluster NGC 6200 barely stands out against the background 
field, it contains a large number of B-type stars and Be star candidates.  
We identify Nos.\ 2, 38, 112, 213, 239, 247, 259, and 305 as possible Be 
stars and Nos.\ 115 and 153 as definite H$\alpha$ emission stars.  NGC 
6200 appears close to NGC 6204 and Hogg 22 on the sky, although this is a 
chance superposition since the cluster is younger and more distant than 
its apparent neighbors.

\item \textbf{NGC 6204:}
NGC 6204 is only 6\arcmin~away from Hogg 22 on the sky (see the notes for
that cluster).  Members of Hogg 22 were removed from our results for NGC 
6204, and no members of this cluster are saturated in our images.  
While we identify no definite Be stars in NGC 6204, there are two
possible Be stars: Nos.\ 46 and 47.  No.\ 4 was also identified as an
H$\alpha$ emitter by \citet{vega1980}, but it is barely too red to be a
B-type star according to our results.

\item \textbf{NGC 6231:}
NGC 6231 is a very bright cluster, and a large number of massive stars are 
saturated in our images.  They are:
HD 152233, an O6 III variable star;
HD 152234, a B0.5 Ia star;
HD 152248, an O7 I $+$ O7 I spectroscopic binary \citep*{penny1999};
HD 152270 (WR 79), a WC+ Wolf-Rayet star;
HD 326331, a B star;
HD 152314, an O+ variable star;
HD 152219, an O9 IV spectroscopic binary;
SAO 227385, an O9 III spectroscopic binary;
HD 152249, an O9 Ib supergiant; and 
HD 152218, an O9 V spectroscopic binary.
The cluster has an extremely high binary frequency; \citet{garcia2001}
found that 82\% of the members earlier than B1.5 V are binary.  It also
contains a large number of pulsating variable stars, and
\citet{arentoft2001} identify at least 14 stars with $\beta$ Cephei,
$\delta$ Scuti, $\gamma$ Doradus, and other types of pulsations.  
We identify two definite Be stars, Nos.\ 67 and 79, and three possible 
Be stars, Nos.\ 2, 35, and 221, but none of these are identified as 
pulsators by \citet{arentoft2001}.  \citet*{sung1998} found 19 pre-MS 
stars or candidates in NGC 6231, including our Be star No.\ 79.

\item \textbf{NGC 6249:}
No stars are saturated in our images on NGC 6249, and we identify one 
definite Be star, No.\ 207.

\item \textbf{NGC 6250:}
There are two saturated stars in our images of NGC 6250.  They are HD 
152917, an A6 III/IV star, and HD 152853, a B2 III star.  The cluster is 
sparse, with only seven B-type stars, and we find no Be stars in NGC 6250.

\item \textbf{NGC 6268:}
NGC 6268 is an evolved cluster with very few B stars remaining.  We 
identify one of them, No.\ 51, as a definite Be star.  The only saturated 
star in our images of this cluter is HD 322550, an F0 star and likely 
foreground to the cluster.

\item \textbf{NGC 6322:}
Three stars in NGC 6322 are saturated in our images, HD 156292, a B1 Ib/II
supergiant, HD 156234, a B0 III star, and HD 156189, an A3 III/IV star.  
The definite Be star in this cluster is No.\ 12, and Nos.\ 7, 26, 50, 51,
76, 79, 84, and 85 are possible candidates.  Most of these are likely
Herbig Be stars as pre-MS contraction is still occuring in NGC 6322 among
stars with $V > 12.5$ \citep{moffat1975c}.

\item \textbf{NGC 6425:}
The only saturated star in our images of NGC 6425 is HD 161508, a G6/G8 V
star and likely a foreground star.  We find no Be stars in this cluster.

\item \textbf{NGC 6530:}
NGC 6530 is part of a complex star-forming region, the Lagoon Nebula, and
many of the bright B stars in the cluster are pre-MS stars moving onto the
ZAMS \citep*{sung2000}.  The B0 IVpne star HD 164906 \citep*{hiltner1965} 
is
saturated in our images of NGC 6530.  We find one definite Be star, No.\
50, and two possible Be stars, Nos.\ 20 and 31, in the cluster.  No. 50
was classified as a B2 Ve star by \citet{hiltner1965} and it was 
identified as a Herbig Be star by \citet*{boesono1987}.

\item \textbf{NGC 6531:}
The two stars HD 164863 and HD 164844 are saturated in our images of NGC
6531.  They are O+ and B1/B2 III stars, respectively.  Although
\citet{schild1976} found no Be stars in this cluster, we see H$\alpha$
emission from No.\ 12.  Even though we classify this as a definite Be star
because its $y-\rm H\alpha$ color is more than 5 standard deviations
above the other B star colors, the photometry errors for this faint star
are comparable to the strength of the emission so it may be a false Be
star detection.

\item \textbf{NGC 6604:}
NGC 6604 appears faint without a well-defined MS in our images due to 
its location at the core of the HII region S54 \citep{barbon2000}.  The
saturated star in our images is HD 167971, an O5$-$8 V + O5$-$8 V + O8 Ib 
triple system \citep{leitherer1987} and eclipsing binary of $\beta$ Lyr 
type.  The cluster may have some faint pre-MS members with $V \ge 16$ 
\citep{barbon2000}.  There are seven possible Be stars in NGC 6604, Nos.\ 
2, 3, 9, 11, 21, 66, and 73, although they all lie at the faint end of the 
MS and the errors in their $y-\rm{H \alpha}$ colors are relatively high.  
We believe that the apparent emission is more likely due to photometric 
scatter.

\item \textbf{NGC 6613:}
NGC 6613 = M18 is a sparse cluster in Sagittarius.  No stars are saturated
in our images of this cluster, and we find one definite Be star, No.\ 1,
also discovered as an H$\alpha$ emission star by \citet{macconnell1981}.

\item \textbf{NGC 6664:}
No stars are saturated in our images of NGC 6664, and we find several Be
star candidates.  Nos.\ 177 and 221 have definite H$\alpha$ emission, and
Nos.\ 137 and 182 are possibly Be stars as well.  Star No.\ 176 also
appears to have H$\alpha$ emission, but it is not classified as a B-type
star in our results.

\item \textbf{Ruprecht 79:}
There is one star that is barely saturated in our $b$ images of Ruprecht
79, but not much is known about it.  The SIMBAD database identifies it as
GSC 08585-01823 with $B$ = 10.9 and $V$ = 10.28.  Its color, $B-V$ = 0.6,
suggests that it is either a massive OB star in the cluster or an
unreddened, foreground star.  The latter is more likely, given the
presence of many stars to the left of the MS which must be foreground
stars.  \citet{walker1987} points out that unreddened G and K stars fall
into the same region as the reddened B and A stars in his color-color
diagram of this cluster.

In a color-magnitude diagram of Ruprecht 79, the MS is not clearly defined
due to the overlap of many foreground G and K stars \citep{walker1987}.  
They also contaminate the color-color plot, making it difficult to fit a
theoretical reddening curve to our data.  Because of the contamination,
there are a large number of stars that lie more than $2\sigma$ above the
theoretical reddening curve in the Be star region of the color-color plot,
but they are not likely to be Be stars since there is no clear gap in the
stellar population indicating a group of H$\alpha$ emitters.  Therefore
the Be star detections in this cluster are likely incorrect, and we
disregard these detections in the analysis below.

The cluster Ruprecht 79 has been studied a handful of times in the
literature \citep{moffat1975a, harris1976, walker1987} because the Cepheid
variable CS Velorum is a member of the cluster.  However,
\citet{harris1976} note that the cluster is ``rather loose'', and they
point out that no large-scale photographs of the region were available at
the time of their observing program.  That is unfortunate, because a
$60\arcmin \times 60\arcmin$ image of Ruprecht 79 from the STScI Digitized
Sky Survey reveals that Ruprecht 79 is not at all prominent against the
field.  In fact, the eastern half of the image is more crowded with stars
than the west, suggesting a gradient in the dust absorption across the
field.  From our eyeball estimates, there are many other regions in the
field with comparable stellar density to this cluster.

In addition, the galactic coordinates of Ruprecht 79 and its distance (see
Table \ref{distribution}) place it within the Sagittarius-Carina spiral
arm of the Milky Way \citep{binney1998}.  The cluster is viewed through
the tangential edge of the arm \citep{dame1987, binney1998}, so a gradient
in absorption across the field of view is likely.  Therefore we conclude
that Ruprecht 79 simply corresponds to a hole in the dust in this region
of the Galaxy, and it is not a true open cluster.

\item \textbf{Ruprecht 119:}
Although \citet{moffat1973} classified Ruprecht 119 as a non-cluster,
\citet{piatti2000b} showed that it is in fact physically bound.  We find
a distinct MS for the cluster, although we confirm a slight gap in the MS
around $y = 13.5$, and possibly at $y=16$, both identified by
\citet{piatti2000b}.  No stars are saturated in our results for Ruprecht
119, and we find one definite Be star (No.\ 332) and nine possible Be
stars (Nos.\ 267, 301, 412, 427, 487, 549, 574, 674, and 706).

\item \textbf{Ruprecht 127:}
There are no saturated stars in our photometry of Ruprecht 127.  We find
no clear MS visible from our data, and any true MS is somewhat
contaminated by field stars.  We experimented with fitting the MS using
the range of $E(b-y)$, $V_0-M_V$, and ages available in the literature
(see Table \ref{clusters}) as well as our own values for a less reddened,
closer group of stars.  Each isochrone fit illustrated that multiple,
sparse stellar populations are in the FOV, and there appear to be two
stellar populations, with distinctly different reddening, in the
color-color sequence as well.  The Be star candidates found using the
parameters in Table \ref{clusters} are likely foreground stars rather than
true emission stars.  We find no Be stars when we fit the more nearby 
population.  We doubt that this is a physical cluster, and instead we 
believe it is a chance physical alignment of several bright stars.

\item \textbf{Ruprecht 140:}
Only the star GSC $07397-00784$ is saturated in our images of Ruprecht
140.  We find no clear MS in this cluster, and we agree with
\citet{piatti2001} who concluded that Ruprecht 140 is not a physical
cluster but rather a density fluctuation over the background field.  We 
find one possible Be star in this FOV (No. 468).

\item \textbf{Stock 13:}
We find one definite Be star, No.\ 831 , and one possible Be star, No.\
271, in our results from Stock 13.  No stars are saturated in our data.  
There may be some pre-MS stars in this cluster, as \cite{moffat1975b}
identified both Nos.\ 376 and 452 as possible pre-MS stars.  We see a
distinct cluster MS, although the color-color diagram possibly indicates a
small population of nearly unreddened, late B- and early A-type foreground
stars that are also present.  \citet{steppe1977} also found a group of 20
foreground red giants in the field of Stock 13.  The presence of
foreground stars is not surprising considering the relatively large
distance to this cluster (see Table \ref{distribution}).

\item \textbf{Stock 14:}
In our images of Stock 14, the F8 Ia supergiant V810 Cen is saturated, 
although \citet{kienzle1998} found that this star is not a member of the 
cluster.  We find one definite Be star (No.\ 164) and a possible Be 
star (No.\ 105) in Stock 14, and both of these stars are also identified 
as Be or emission line stars in SIMBAD.

\item \textbf{Trumpler 7:}
No definite Be stars are found in this cluster, but 5 possible Be stars
have H$\alpha$ emission more than $2\sigma$ above the color-color fit.  
These Be star candidates are Nos.\ 13, 55, 57, 59, and 82.  Only the A1 IV 
star HD~59163 is saturated in our images of Trumpler~7.

\item \textbf{Trumpler 18:}
No stars are saturated in our results from Trumpler 18, and we identify 
five possible Be stars (Nos.\ 10, 29, 55, 137, and 219).  Like 
\citet{vazquez1990}, we see a clear MS for this cluster.  However, 
\citet{vazquez1990} questioned the physical association of Trumpler 18 
because they found a peculiar luminosity function for the cluster.  We 
compute the mass function for Trumpler 18 based on the method described 
in \S 3, and we find a normal power law relation with $\Gamma = -1.5 \pm 
0.16$, depending on the number of $\log M$ bins used.  With this value of 
$\Gamma$ we should expect $75 \pm 4$ B-type stars in the cluster, and we 
detect 71 (see Table 2).  We conclude that Trumpler 18 is a true open 
cluster.

\item \textbf{Trumpler 20:}
There are no saturated stars in our images of Trumpler 20.  We find very 
few B-type stars in this cluster, and only one possible Be star (No.\ 
1039).

\item \textbf{Trumpler 27:}
We find two saturated stars in our images of Trumpler 27: HD 318014, a B8
Iab supergiant, and CD $-33^\circ 12241$, an M0 Ia binary.  The cluster
contains one star with very strong emission, No.\ 15 (WR 95).  In this
case, the emission more likely originates from the He II $\lambda$ 6527
line than H$\alpha$ in this Wolf-Rayet star. We also find two possible Be
stars, Nos.\ 49 and 52.  We do not see a well defined MS for Trumpler 27,
although we believe this is due to the cluster's significant differential
reddening.  \citet{the1970} found that the cluster lies behind the dust
lane B217 that partially encircles the cluster, and \citet{feinstein2000}
used polarization measurements to determine that two separate dust
components lie in the field of Trumpler 27.  We believe that our estimates
of the cluster parameters are unreliable, and we disregard this cluster in
our statistical analysis of Be stars (\S 3).

\item \textbf{Trumpler 28:}
The B3 star HD 317971 is saturated in our images of Trumpler 28, and we 
find no Be stars in this cluster.

The cluster Trumpler 28 is only 14\arcmin~from NGC 6383, and the two may
form a binary open cluster system \citep{subramaniam1995}.  However, the
large angular size of NGC 6383 (20\arcmin; WEBDA) excluded it from our
target list, and we do not present photometry of that cluster in this
study.  According to \citet{subramaniam1995}, NGC 6383 is significantly
younger than Trumpler 28, and the three apparent blue stragglers near the 
MS of Trumpler 28 (Nos.\ 139, 161, and 193) are probably members of NGC 
6383 instead.

\item \textbf{Trumpler 34:}
No stars are saturated in our images of Trumpler 34.  We find one
definite Be star, No.\ 355, and one possible Be star, No.\ 297, in this 
cluster.  No.\ 355 is very blue for a late B-type star, but its 
photometric errors are small enough that we trust its position in Figure 
\ref{Trumpler34}.  However, it may possibly be a foreground star.

\item \textbf{vdB-Hagen 217:}
There are no saturated stars in our images of vdB-Hagen 217.  We find no 
Be stars in this cluster.

\end{itemize}



\clearpage

\clearpage

\begin{figure}
\includegraphics[angle=90,scale=0.3]{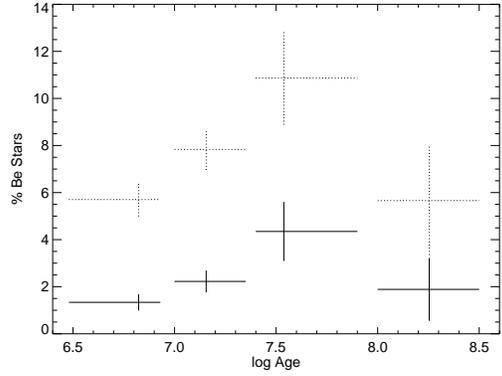}
\caption{
The average definite Be frequency (\textit{solid lines}) and total 
possible Be frequency (\textit{dashed lines}) are plotted for four age 
bins corresponding to $\log \rm age < 7.0$, $7.0 \le \log \rm age < 7.4$, 
$7.4 \le \log \rm age < 8.0$, and $\log \rm age \ge 8.0$.  The errors in 
Be frequency represent the Poisson statistics for our sample, and the age 
error bars represent the maximum and minimum ages observed in each bin.
\label{befreq}
}
\end{figure}


\clearpage
 
\clearpage
 
\clearpage
 
\clearpage
 
\clearpage
 
\clearpage
 
\clearpage
 
\clearpage
 
\clearpage
 
\clearpage
 
\clearpage
 
\clearpage
 
\clearpage
 
\clearpage
\clearpage
\begin{deluxetable}{rrrrrcrcrcccc}
\rotate
\tablewidth{0pc}
\tabletypesize{\scriptsize}
\tablecaption{Photometry of Open Clusters \label{table1}}
\tablehead{
\colhead{Cluster} &
\colhead{No.} &
\colhead{RA (2000)} &
\colhead{Dec (2000)} &
\colhead{$y$~~} &
\colhead{$\delta y$} &
\colhead{$b-y$} &
\colhead{$\delta (b-y)$} &
\colhead{$y-{\rm H}\alpha$} &
\colhead{$\delta (y-{\rm H}\alpha)$} &
\colhead{Code} &
\colhead{WEBDA} &
\colhead{Identifier}
}
\startdata
Basel1  \dotfill & 1 &  18 47 56.08 & --5 45 51.0  & 16.445  & 0.066  &
0.988  & 0.103  &  0.917  & 0.103   &   O &  \nodata & \nodata \\
Basel1  \dotfill & 2 &  18 48 01.11 & --5 45 50.7  & 15.981  & 0.056  &
0.744  & 0.084  &  0.770  & 0.094   &   O &  \nodata & \nodata \\
Basel1  \dotfill & 3 &  18 48 23.49 & --5 45 50.9  & 16.137  & 0.060  &
0.868  & 0.092  &  0.876  & 0.098   &   O &  \nodata & \nodata \\
Basel1  \dotfill & 4 &  18 48 33.34 & --5 45 53.7  & 16.570  & 0.069  &
0.770  & 0.102  &  0.898  & 0.114   &   O &  \nodata & \nodata \\
Basel1  \dotfill & 5 &  18 48 06.99 & --5 45 55.7  & 14.437  & 0.050  &
1.396  & 0.079  &  1.537  & 0.064   &   O &  \nodata & \nodata \\
Basel1  \dotfill & 6 &  18 48 37.87 & --5 45 53.8  & 15.479  & 0.048  &
0.657  & 0.070  &  0.693  & 0.081   &   O &  \nodata & \nodata \\
Basel1  \dotfill & 7 &  18 47 52.85 & --5 45 56.7  & 16.315  & 0.064  &
0.868  & 0.096  &  1.037  & 0.096   &   O &  \nodata & \nodata \\
Basel1  \dotfill & 8 &  18 48 01.30 & --5 45 57.1  & 14.802  & 0.047  &
1.111  & 0.074  &  1.181  & 0.066   &   O &  \nodata & \nodata \\
Basel1  \dotfill & 9 &  18 48 31.17 & --5 45 56.8  & 15.899  & 0.054  &
0.812  & 0.084  &  0.683  & 0.093   &   O &  \nodata & \nodata \\
Basel1 \dotfill & 10 &  18 48 34.64 & --5 45 56.7  & 13.519  & 0.031  &
0.429  & 0.046  &  0.395  & 0.052   &   O &  \nodata & \nodata \\
\enddata

\tablecomments{The complete version of this table is in the electronic
edition of the Journal.  For each star observed, we give the cluster name
and our identifying number, the right ascension (RA), and declination
(Dec) for the epoch 2000.  We also provide the $y$ magnitude, the $b-y$
and $y-\rm H\alpha$ colors, and the error for each.  The code is used to
label definite Be stars (Be), possible Be stars (Be?), B-type stars (B),
probable foreground stars (F), and other stars in the field (O).
Finally, the WEBDA number and other identifiers are provided where
available.}

\notetoeditor{Only the first few lines of Table 1 should be included
in the printed version; the full table should be published in
electronic format only.}

\end{deluxetable}


\clearpage
\clearpage
\begin{deluxetable}{lcccccccrc}
\tablewidth{0pc}
\tabletypesize{\scriptsize}
\tablecaption{Summary of Open Clusters \label{clusters}} 
\tablehead{
\colhead{ } &
\colhead{ } &
\colhead{ } &
\colhead{ } &
\colhead{ } &
\colhead{ } &
\colhead{$N_{Be}$} &
\colhead{$N_{Be}$} &
\colhead{ } &
\colhead{ } \\
\colhead{Cluster} &
\colhead{RA (2000)} &
\colhead{Dec (2000)} &
\colhead{$E(b-y)$} &
\colhead{ $V_0-M_V$} &
\colhead{log Age} &
\colhead{(Definite)} &
\colhead{(Possible)} &
\colhead{$N_{B+Be}$} &
\colhead{References}
}
\startdata
\object[cl Basel 1]{Basel 1}	                   \dotfill   &  18 48 12  &  $-$05 51 00  &  0.33  &  11.00  &  8.45  &   0  &   0  &      3  &  6  \\
\object[cl Bochum 13]{Bochum 13}\tablenotemark{a} \dotfill   &  17 17 24  &  $-$35 33 00  &  0.64  &  10.16  &  6.82  &   2  &   1  & $>$ 23  &  1  \\
\object[cl Collinder 272]{Collinder 272}              \dotfill   &  13 30 26  &  $-$61 19 00  &  0.34  &  11.55  &  7.11  &   3  &   5  &     91  &  1,30  \\
\object[cl Haffner 16]{Haffner 16}                 \dotfill   &  07 50 20  &  $-$25 28 00  &  0.13  &  12.50  &  7.08  &   0  &   1  &     19  &  1,7,14  \\
\object[cl Hogg 16]{Hogg 16}                    \dotfill   &  13 29 18  &  $-$61 12 00  &  0.31  &  11.00  &  7.05  &   0  &   1  &     35  &  1  \\
\object[cl Hogg 22]{Hogg 22}                    \dotfill   &  16 46 37  &  $-$47 05 00  &  0.47  &  11.75  &  6.70  &   0  &   1  & $>$ 22  &  8,14  \\
\object{IC 2395}                    \dotfill   &  08 42 37  &  $-$48 06 48  &  0.05  & \phn 9.24  &  7.22  &   0  &   1  &     14  &  1  \\
\object{IC 2581}                    \dotfill   &  10 27 24  &  $-$57 38 00  &  0.31  &  12.00  &  7.14  &   2  &   8  &    117  &  1,13,14  \\
\object{IC 2944}\tablenotemark{b}   \dotfill   &  11 38 20  &  $-$63 22 22  &  0.24  &  11.27  &  6.82  &   1  &   2  &     50  &  1  \\
\object{NGC 2343}                   \dotfill   &  07 08 18  &  $-$10 39 00  &  0.11  &  10.12  &  8.00  &   0  &   0  &     11  &  1,11,14  \\
\object{NGC 2362}                   \dotfill   &  07 18 48  &  $-$24 57 00  &  0.07  &  10.71  &  6.91  &   0  &   1  &     41  &  1  \\
\object{NGC 2367}                   \dotfill   &  07 20 06  &  $-$21 53 00  &  0.26  &  11.84  &  6.74  &   0  &   3  &     16  &  1,7,31  \\
\object{NGC 2383}                   \dotfill   &  07 24 41  &  $-$20 56 42  &  0.20  &  12.62  &  7.40  &   2  &   0  &     20  &  14,26,31  \\
\object{NGC 2384}                   \dotfill   &  07 25 12  &  $-$21 01 24  &  0.22  &  11.63  &  6.90  &   0  &   2  &     33  &  1,31  \\
\object{NGC 2414}\tablenotemark{b}  \dotfill   &  07 33 11  &  $-$15 27 12  &  0.41  &  11.99  &  6.80  &   5  &  13  &     86  &  14,31  \\
\object{NGC 2421}                   \dotfill   &  07 36 18  &  $-$20 37 00  &  0.35  &  11.69  &  7.40  &   4  &   6  &     65  &  1,14,16  \\
\object{NGC 2439}\tablenotemark{b}  \dotfill   &  07 40 48  &  $-$31 39 00  &  0.30  &  12.18  &  7.25  &   6  &   7  &    111  &  1,14  \\
\object{NGC 2483}\tablenotemark{b}  \dotfill   &  07 55 54  &  $-$27 56 00  &  0.30  &  12.31  &  7.00  &   2  &  10  &     49  &  14  \\
\object{NGC 2489}                   \dotfill   &  07 56 12  &  $-$30 04 00  &  0.30  &  10.81  &  8.38  &   0  &   0  &      7  &  14,19  \\
\object{NGC 2571}                   \dotfill   &  08 18 54  &  $-$29 44 00  &  0.10  &  10.64  &  7.49  &   1  &   0  &     24  &  1  \\
\object{NGC 2659}                   \dotfill   &  08 42 36  &  $-$44 57 00  &  0.38  &  10.80  &  6.89  &   1  &   4  &     32  &  1,27  \\
\object{NGC 3293}                   \dotfill   &  10 35 49  &  $-$58 13 00  &  0.22  &  12.20  &  6.90  &   3  &   8  & $>$ 91  &  4  \\
\object{NGC 3766}                   \dotfill   &  11 36 13  &  $-$61 36 55  &  0.15  &  11.40  &  7.16  &   5  &   8  & $>$ 146 &  1,25  \\
\object{NGC 4103}                   \dotfill   &  12 06 43  &  $-$61 15 21  &  0.22  &  11.70  &  7.35  &   1  &   2  &     92  &  23,33  \\
\object{NGC 4755}                   \dotfill   &  12 53 42  &  $-$60 22 00  &  0.28  &  11.80  &  6.85  &   5  &   7  & $>$ 178 &  3  \\
\object{NGC 5281}                   \dotfill   &  13 46 30  &  $-$62 54 54  &  0.19  &  11.00  &  7.71  &   1  &   5  &     40  &  14,15  \\
\object{NGC 5593}                   \dotfill   &  14 25 54  &  $-$54 49 00  &  0.26  &  10.62  &  7.00  &   1  &   1  &     15  &  34  \\
\object{NGC 6178}                   \dotfill   &  16 35 42  &  $-$45 38 00  &  0.18  & \phn 9.80  &  7.60  &   0  &   0  &      6  &  15,21  \\
\object{NGC 6193}\tablenotemark{c}  \dotfill   &  16 41 18  &  $-$48 46 00  &  0.33  &  10.31  &  6.48  &   0  &   1  &     31  &  1,10,15  \\
\object{NGC 6200}                   \dotfill   &  16 44 12  &  $-$47 29 00  &  0.43  &  11.56  &  6.93  &   2  &   8  &    112  &  1  \\
\object{NGC 6204}                   \dotfill   &  16 46 08  &  $-$47 00 44  &  0.35  &  10.40  &  8.10  &   0  &   2  &     34  &  5,8  \\
\object{NGC 6231}\tablenotemark{c}  \dotfill   &  16 54 09  &  $-$41 49 36  &  0.33  &  10.47  &  6.90  &   2  &   3  &    129  &  1,20  \\
\object{NGC 6249}                   \dotfill   &  16 57 36  &  $-$44 47 00  &  0.34  &  10.06  &  7.40  &   1  &   0  &     13  &  14,15  \\
\object{NGC 6250}                   \dotfill   &  16 57 58  &  $-$45 56 36  &  0.26  & \phn 9.69  &  7.41  &   0  &   0  &      7  &  1  \\
\object{NGC 6268}                   \dotfill   &  17 02 40  &  $-$39 44 18  &  0.31  &  10.19  &  8.50  &   1  &   0  &      2  &  22,34  \\
\object{NGC 6322}\tablenotemark{c}  \dotfill   &  17 18 30  &  $-$42 56 00  &  0.50  &  10.50  &  7.06  &   1  &   8  &     43  &  1,18  \\
\object{NGC 6425}                   \dotfill   &  17 47 02  &  $-$31 30 00  &  0.30  & \phn 9.89  &  7.35  &   0  &   0  &     25  &  1,29  \\
\object{NGC 6530}\tablenotemark{c}  \dotfill   &  18 04 48  &  $-$24 20 00  &  0.25  &  10.62  &  6.87  &   1  &   2  & $>$ 41  &  1  \\
\object{NGC 6531}                   \dotfill   &  18 04 36  &  $-$22 30 00  &  0.20  &  10.47  &  7.07  &   1  &   0  &     49  &  1,9  \\
\object{NGC 6604}\tablenotemark{c}  \dotfill   &  18 18 06  &  $-$12 14 00  &  0.72  &  11.15  &  6.81  &   0  &   7  & $>$ 63  &  1  \\
\object{NGC 6613}                   \dotfill   &  18 19 54  &  $-$17 08 00  &  0.31  &  10.48  &  7.90  &   1  &   0  &     29  &  12,14,34  \\
\object{NGC 6664}                   \dotfill   &  18 36 42  &  $-$08 13 00  &  0.56  &  10.70  &  7.66  &   2  &   2  &     36  &  24  \\
\object{Ruprecht 79}\tablenotemark{b} \dotfill &  09 40 59  &  $-$53 51 00  &  0.53  &  12.55  &  7.65  &   1  &  31  &    288  &  1,28,32  \\
\object{Ruprecht 119}                 \dotfill &  16 28 15  &  $-$51 30 00  &  0.42  &  11.64  &  7.18  &   1  &   9  &     50  &  1,22,34  \\
\object{Ruprecht 127}\tablenotemark{b} \dotfill &  17 37 51  & $-$36 18 00  &  0.77  &  10.92  &  7.35  &   0  &  18  &     97  &  1,18  \\
\object{Ruprecht 140}\tablenotemark{b} \dotfill &  18 21 51  & $-$33 13 00  &  0.37  &  12.35  &  7.50  &   0  &   1  &     25  &  34  \\
\object[cl Stock 13]{Stock 13}\tablenotemark{c}    \dotfill &  11 13 05  &  $-$58 53 00  &  0.16  &  11.63  &  7.22  &   1  &   1  &     64  &  1,17  \\
\object[cl Stock 14]{Stock 14}                   \dotfill   &  11 43 48  &  $-$62 31 00  &  0.19  &  11.66  &  7.06  &   1  &   1  &     47  &  1,14  \\
\object[cl Trumpler 7]{Trumpler 7}                 \dotfill   &  07 27 22  &  $-$23 57 00  &  0.21  &  11.02  &  7.40  &   0  &   5  &     24  &  14  \\
\object[cl Trumpler 18]{Trumpler 18}                \dotfill   &  11 11 28  &  $-$60 40 00  &  0.23  &  10.66  &  7.19  &   0  &   5  &     71  &  1  \\
\object[cl Trumpler 20]{Trumpler 20}                \dotfill   &  12 39 34  &  $-$60 37 00  &  0.26  &  11.92  &  8.20  &   0  &   1  &     10  &  34  \\
\object[cl Trumpler 27]{Trumpler 27}\tablenotemark{a} \dotfill &  17 36 20 &   $-$33 31 00  &  0.89  &  10.42  &  7.06  &   1  &   2  & $>$ 34  &  1,29  \\
\object[cl Trumpler 28]{Trumpler 28}                \dotfill   &  17 37 00  &  $-$32 29 00  &  0.45  &  10.88  &  8.30  &   0  &   0  &     19  &  14,34  \\
\object[cl Trumpler 34]{Trumpler 34}                \dotfill   &  18 39 48  &  $-$08 25 00  &  0.56  &  11.20  &  8.10  &   1  &   1  &     18  &  34  \\
\object[cl VDBH 217]{vdB-Hagen 217}              \dotfill   &  17 16 06  &  $-$40 49 00  &  0.60  &  10.42  &  7.54  &   0  &   0  &     12  &  2,34  \\
\enddata
\tablenotetext{a}{Cluster parameters unreliable.}
\tablenotetext{b}{May not be a true cluster.}
\tablenotetext{c}{May contain pre-MS Herbig Be stars.}
\tablerefs{
(1) WEBDA;
(2) \citealt{ahumada2000}; 
(3) \citealt{alfaro1991}; 
(4) \citealt{baume2003};
(5) \citealt{carraro2004}; 
(6) \citealt*{delgado1997}; 
(7) using mean values from \citealt{fitzgerald1979};
(8) \citealt{forbes1996}; 
(9) \citealt{hagen1970}; 
(10) \citealt{herbst1977}; 
(11) \citealt{kalirai2001}; 
(12) \citealt{lindoff1971}; 
(13) \citealt{lloydevans1969}; 
(14) \citealt{lynga1987};
(15) \citealt{moffat1973}; 
(16) \citealt{moffat1975a};
(17) \citealt{moffat1975b};
(18) \citealt{moffat1975c}; 
(19) \citealt{paunzen2001};
(20) \citealt*{perry1991}; 
(21) \citealt*{piatti2000a}; 
(22) \citealt*{piatti2000b};  
(23) \citealt{sanner2001}
(24) \citealt{schmidt1982}; 
(25) \citealt{shobbrook1985}; 
(26) \citealt{subramaniam1999};
(27) \citealt{stetson1981};
(28) \citealt{tadross2001};
(29) \citealt{the1970}; 
(30) \citealt{vazquez1997};  
(31) \citealt{vogt1972};
(32) \citealt{walker1987};
(33) \citealt{wesselink1969};
(34) this work.
}
\end{deluxetable}

\clearpage

\clearpage
\begin{deluxetable}{lcccc}
\tablewidth{0pc}
\tablecaption{
Cluster mass distribution
\label{massfcn}
}
\tablehead{
\colhead{Cluster} &
\colhead{$\Gamma$} &
\colhead{$N_{tot}$} &
\colhead{$N_{missing}$} &
\colhead{$\Delta M$\tablenotemark{a} $(M_\odot)$}
}
\startdata
 Hogg 22        \dotfill  &  $-1.11 \pm 0.27$  &  $\phn 49 \pm     10$  &  27  &  3.9 $-$ 24.8  \\
 NGC 3293       \dotfill  &  $-1.00 \pm 0.14$  &  $    189 \pm     21$  &  98  &  4.2 $-$ 18.4  \\
 NGC 3766       \dotfill  &  $-1.67 \pm 0.10$  &  $    191 \pm \phn 9$  &  45  &  2.7 $-$ 12.2  \\
 NGC 4755       \dotfill  &  $-1.04 \pm 0.13$  &  $    247 \pm     17$  &  69  &  2.7 $-$ 20.0  \\
 NGC 6530       \dotfill  &  $-0.68 \pm 0.34$  &  $\phn 66 \pm \phn 7$  &  25  &  2.7 $-$ 13.8  \\
 NGC 6604       \dotfill  &  $-1.27 \pm 0.19$  &  $    127 \pm     18$  &  64  &  3.8 $-$ 20.8  \\
\enddata
\tablenotetext{a}{$\Delta M$ represents the observed range of masses for 
the B stars in each cluster.}
\end{deluxetable}

\clearpage

\clearpage
\begin{deluxetable}{lrrcc}
\tablewidth{0pc}
\tablecaption{
Distribution of open clusters
\label{distribution}
}
\tablehead{
\colhead{ } &
\colhead{$\ell$} &
\colhead{$b$} &
\colhead{$d_\odot$} &
\colhead{$d_{cent}$} \\
\colhead{Cluster} &
\colhead{(deg)} &
\colhead{(deg)} &
\colhead{(kpc)} &
\colhead{(kpc)}
}
\startdata
 Basel 1        \dotfill  &   27.36  &  $-1.95$  &  1.58  &   6.63  \\
 Collinder 272  \dotfill  &  307.62  &  $ 1.25$  &  2.04  &   6.94  \\
 Haffner 16     \dotfill  &  242.08  &  $ 0.48$  &  3.16  &   9.88  \\
 Hogg 16        \dotfill  &  307.48  &  $ 1.34$  &  1.58  &   7.15  \\
 Hogg 22        \dotfill  &  338.56  &  $-1.14$  &  2.24  &   5.97  \\
 IC 2395        \dotfill  &  266.65  &  $-3.58$  &  0.70  &   8.07  \\
 IC 2581        \dotfill  &  284.60  &  $ 0.01$  &  2.51  &   7.76  \\
 IC 2944        \dotfill  &  294.85  &  $-1.65$  &  1.79  &   7.43  \\
 NGC 2343       \dotfill  &  224.32  &  $-1.14$  &  1.06  &   8.79  \\
 NGC 2362       \dotfill  &  238.18  &  $-5.53$  &  1.39  &   8.81  \\
 NGC 2367       \dotfill  &  235.65  &  $-3.84$  &  2.33  &   9.51  \\
 NGC 2383       \dotfill  &  235.26  &  $-2.44$  &  3.34  &  10.28  \\
 NGC 2384       \dotfill  &  235.39  &  $-2.41$  &  2.12  &   9.37  \\
 NGC 2421       \dotfill  &  236.24  &  $ 0.08$  &  2.18  &   9.39  \\
 NGC 2439       \dotfill  &  246.42  &  $-4.42$  &  2.73  &   9.43  \\
 NGC 2489       \dotfill  &  246.71  &  $-0.78$  &  1.45  &   8.68  \\
 NGC 2571       \dotfill  &  249.11  &  $ 3.54$  &  1.34  &   8.57  \\
 NGC 2659       \dotfill  &  264.16  &  $-1.63$  &  1.45  &   8.27  \\
 NGC 3293       \dotfill  &  285.86  &  $ 0.07$  &  2.75  &   7.72  \\
 NGC 3766       \dotfill  &  294.11  &  $-0.03$  &  1.91  &   7.43  \\
 NGC 4103       \dotfill  &  297.58  &  $ 1.15$  &  2.19  &   7.25  \\
 NGC 4755       \dotfill  &  303.21  &  $ 2.53$  &  2.29  &   7.01  \\
 NGC 5281       \dotfill  &  309.17  &  $-0.70$  &  1.58  &   7.11  \\
 NGC 5593       \dotfill  &  316.34  &  $ 5.58$  &  1.33  &   7.10  \\
 NGC 6178       \dotfill  &  338.40  &  $ 1.23$  &  0.91  &   7.16  \\
 NGC 6193       \dotfill  &  336.70  &  $-1.57$  &  1.15  &   6.96  \\
 NGC 6200       \dotfill  &  337.99  &  $-1.09$  &  2.05  &   6.15  \\
 NGC 6204       \dotfill  &  338.56  &  $-1.03$  &  1.20  &   6.89  \\
 NGC 6231       \dotfill  &  343.46  &  $ 1.19$  &  1.24  &   6.82  \\
 NGC 6249       \dotfill  &  341.55  &  $-1.16$  &  1.03  &   7.03  \\
 NGC 6250       \dotfill  &  340.68  &  $-1.93$  &  0.87  &   7.19  \\
 NGC 6268       \dotfill  &  346.10  &  $ 1.22$  &  1.09  &   6.95  \\
 NGC 6322       \dotfill  &  345.29  &  $-3.07$  &  1.26  &   6.79  \\
 NGC 6425       \dotfill  &  357.97  &  $-1.59$  &  0.95  &   7.05  \\
 NGC 6530       \dotfill  &    6.14  &  $-1.38$  &  1.33  &   6.68  \\
 NGC 6531       \dotfill  &    7.71  &  $-0.44$  &  1.24  &   6.77  \\
 NGC 6604       \dotfill  &   18.26  &  $ 1.69$  &  1.70  &   6.41  \\
 NGC 6613       \dotfill  &   14.15  &  $-1.01$  &  1.25  &   6.80  \\
 NGC 6664       \dotfill  &   23.95  &  $-0.49$  &  1.38  &   6.76  \\
 Ruprecht 119   \dotfill  &  333.27  &  $-1.90$  &  2.13  &   6.17  \\
 Stock 13       \dotfill  &  290.52  &  $ 1.57$  &  2.12  &   7.52  \\
 Stock 14       \dotfill  &  295.24  &  $-0.64$  &  2.15  &   7.35  \\
 Trumpler 7     \dotfill  &  238.28  &  $-3.38$  &  1.60  &   8.95  \\
 Trumpler 18    \dotfill  &  290.98  &  $-0.14$  &  1.36  &   7.62  \\
 Trumpler 20    \dotfill  &  301.49  &  $ 2.24$  &  2.42  &   7.04  \\
 Trumpler 28    \dotfill  &  355.99  &  $-0.27$  &  1.50  &   6.50  \\
 Trumpler 34    \dotfill  &   24.09  &  $-1.28$  &  1.74  &   6.45  \\
 vdB-Hagen 217  \dotfill  &  346.76  &  $-1.48$  &  1.21  &   6.82  \\
\enddata
\end{deluxetable}

\clearpage

\clearpage
\begin{deluxetable}{lccccc}
\tablewidth{0pc}
\tablecaption{
B Star Rotational Velocities
\label{critrot}
}
\tablehead{
\colhead{ } &
\colhead{$M_B$} &
\colhead{$R_B$} &
\colhead{$v_{crit}$} &
\colhead{$\langle v \sin i \rangle$} &
\colhead{ } \\
\colhead{Spectral Type} &
\colhead{$(M_\odot)$} &
\colhead{$(R_\odot)$} &
\colhead{(km~s$^{-1}$)} &
\colhead{(km~s$^{-1}$)} &
\colhead{$\langle v \sin i \rangle/v_{crit}$}
}
\startdata
B1 \dotfill  &  13.5  &  5.9  &  560  &  127  &  0.23 \\
B4 \dotfill  &   5.2  &  3.2  &  470  &  108  &  0.23 \\
B7 \dotfill  &   3.5  &  2.5  &  440  &  152  &  0.35 \\
B9 \dotfill  &   2.4  &  2.1  &  390  &  134  &  0.34 \\
\enddata
\end{deluxetable}

\end{document}